\documentclass[useAMS,usenatbib]{mn2e}

\usepackage{graphicx}
\usepackage{amssymb}
\usepackage[authoryear]{natbib}

\bibpunct{(}{)}{;}{a}{}{,}
\title[Stellar population gradients in the centres of nearby field E+A galaxies]{Stellar population gradients in the cores of nearby field E+A galaxies} 
\author[Michael.~ B.~ Pracy et al.]{
\parbox[t]{\textwidth}{
       Michael B.~Pracy$^{1,2}$, Matt S.~Owers$^{1}$, Warrick J.~Couch$^{1}$,  Harald Kuntschner$^{3}$, Kenji Bekki$^{4}$, Frank Briggs$^{5}$, Philip Lah$^{6,5}$, and Martin Zwaan$^{3}$}
\vspace*{6pt}\\
$^1$Center for Astrophysics \& Supercomputing, Swinburne University of Technology, P.O. Box 218, Hawthorn, Vic, Australia\\
$^2$Sydney Institute for Astronomy, School of Physics, University of Sydney, NSW 2006, Australia\\
$^3$European Southern Observatory, Karl-Schwarzschild Strasse 2, 85748, Garching, Germany\\
$^4$ICRAR, M468, The University of Western Australia, 35 Stirling Highway, Crawley, Western Australia 6009, Australia \\
$^5$Mount Stromlo Observatory, The Australian National University, Weston Creek, ACT 2611, Australia \\
$^6$National Centre for Radio Astrophysics, Post Bag 3, Ganeshkhind, Pune 411 007, India
}

\begin{document}

\date{Received 0000; Accepted 0000}

\pagerange{\pageref{firstpage}--\pageref{lastpage}} \pubyear{2010}

\maketitle

\label{firstpage}

\begin{abstract}
We have selected a sample of local E+A galaxies from the Sloan Digital Sky Survey (SDSS) Data Release 7  for follow
up integral field spectroscopy with the Wide Field Spectrograph (WiFeS) on the ANU 2.3-m telescope. 
The sample was selected using the H$\alpha$ line in place of the [OII]$\lambda 3727$
line as the indicator of on-going star formation (or lack thereof). This allowed us to select a lower redshift sample of galaxies than available in 
the literature since the [OII]$\lambda 3727$ falls off the blue end of the wavelength coverage in the SDSS for the 
very lowest redshift objects. This low redshift selection means that the galaxies have a large angular to physical scale which allows
us to resolve the central $\sim$1\,kpc region of the galaxies; the region where stellar population gradients are expected.
Such observations have been difficult to make using other higher redshift samples because even at
redshifts $z\sim$0.1 the angular to physical scale is similar to the resolution provided by ground based seeing.
Our integral field spectroscopy has enabled us to make the first 
robust detections of Balmer line gradients in the centres of E+A galaxies. 
Six out of our sample of seven, and all the galaxies with regular morphologies, are observed to have 
compact and centrally--concentrated Balmer line absorption. 
This is evidence for compact young cores and stellar population gradients which are predicted from models of mergers and tidal interactions which funnel gas
into the galaxy core. Given the generally isolated nature of our sample this argues for the galaxies being seen in the late
stage of a merger where the progenitors have already coalesced. 
\end{abstract}


\section{Introduction}
E+A galaxies, their optical spectra characterized by strong Balmer absorption lines and little or no emission, represent
galaxies undergoing rapid evolution in their star formation properties. 
The strong Balmer lines are indicative of a substantial population of A-type stars
which must have formed within the last $\sim$1\,Gyr. However,  the lack of optical emission lines 
imply that the episode of star formation in which they formed has ended and star formation is not ongoing.
The normal interpretation of this spectral signature is that the galaxy is being observed in a post starburst phase 
\citep[e.g.][]{couch87,poggianti99}
although it can also be produced by the abrupt truncation of star formation in a disk, without necessarily requiring a starburst
\citep[e.g.][]{shioya04,bekki05,pracy05}. This spectroscopic evolution may be accompanied by a morphological one where a star--forming
disk galaxy is transformed into a quiescent spheroidal system \citep{caldwell96,zabludoff96}.

Much attention has been paid to the environmental influences which give rise to the abrupt changes in star formation rate within
E+A galaxies and several processes have been suggested. These include major and unequal mass galaxy mergers \citep{mihos96,bekki05}, galaxy--galaxy
interactions with mass ratios close to unity \citep{bekki05} and, 
in the cluster environment, interactions with the cluster tidal field \citep{bekki01}, galaxy harassment \citep{moore98} 
and interaction with the intra--cluster medium \citep{gunn72,dressler83,bothun86}. 
While the situation in galaxy clusters remains somewhat unclear \citep{pracy10}, 
there is growing consensus that galaxy mergers are the mechanism responsible for producing the local field E+A population as
seen in the analysis of E+A samples from large redshift surveys.
Ground based imaging and HST follow up of 
a sample of 21 E+A galaxies selected from the Las Campanas Redshift Survey (LCRS) revealed an increased rate of tidal features implying
galaxy mergers had taken place \citep{zabludoff96,yang08}. Using a sample selected from the Two 
Degree Field Galaxy Redshift Survey (2dFGRS), \citet{blake04} concluded that the significant incidence of tidal disruption, 
the nature of the luminosity function, and the results of a near--neighbor analysis, were consistent with them being
in the late stage of a merger. \citet{goto05} found a higher incidence of near neighbors for E+A galaxies in the SDSS suggesting 
galaxy--galaxy interactions may also be an important mechanism.

Another, more direct, way of discerning the physical cause of the E+A phase is to examine the internal
properties of the galaxy itself. The internal kinematics and spatial distribution of the young and old stellar populations are expected to differ
between different formation scenarios. For example, in a major merger the young stellar population should quickly become dynamically
pressure supported \citep{norton01,bekki05}. For more extreme mass ratio mergers and tidal interactions, significant rotation of the young stellar
population is generally expected \citep{bekki05}. The two major spatially resolved spectroscopic studies of low redshift `field'
E+A galaxies \citep{norton01,pracy09} have reached quite different conclusions on the prevalence of rotation in E+As. 
Using follow--up slit spectroscopy of the LCRS sample, \citet{norton01}
found little evidence for rotation in most cases. In contrast, \citet{pracy09}'s two dimensional Integral Field Unit (IFU)
spectroscopy of a sub--sample of the \citet{blake04} 2dFGRS E+A galaxies showed rotation of the young stellar population 
to be ubiquitous. This discrepancy could partly be explained by the rotation axis not always corresponding to
the isophotal major axis, dampening the amount of rotation measured using a one dimensional slit \citep{pracy09}.

If the E+A phase is the result of gas--rich mergers, then evidence of this should be contained in the spatial distribution
of the stellar populations. During a merger, tidal torques transfer angular momentum from the gas to the stars \citep{barnes96},
driving the gas toward the centre and resulting in a compact central starburst \citep{mihos94,mihos96,barnes96,hopkins09}. 
Feedback processes then halt the starburst \citep{springel03,springel05} leaving an old stellar population that is distributed like a normal 
early--type galaxy \citep[e.g.][]{hernquist92,hopkins09a} and a young component that forms a centrally--concentrated cusp \citep{mihos94,hopkins09}.
This young component should be quite compact, with scales of $\lesssim$1\,kpc \citep{barnes96,bekki05,hopkins09}, and is a possible explanation of
the excess light observed in the centres of merging starburst galaxies \citep{hibbard99,rothberg04}. \citet{kohno02} detected centrally concentrated diffuse and gravitationally 
stable molecular gas in the post-starburst galaxy
NGC 5195 which may be the remaining component of the gas used in fueling the central starburst.
In the E+A galaxies the remnant of such a starburst should be observable as a Balmer line absorption enhancement and gradient 
within the central $\sim$1\,kpc \citep{pracy05,bekki05}. These predictions of mergers resulting in a centrally concentrated 
post-starburst region and strong radial gradients have been strengthened by a recent set of detailed fully three--dimensional hydrodynamic
simulations of E+A galaxy production from galaxy merging taking into account a wide range of initial parameters such as mass, gas fraction,
orbital configuration, mass ratio and dust properties \citep{snyder11}.

Attempts to measure the spatial distribution of the post--starburst stellar
population in E+A galaxies has been hindered by the small expected angular scale
of the starburst remnant at the redshifts of E+A galaxies selected from large redshift surveys \citep{blake04,goto07}, where it is smaller or comparable 
to the typical seeing scale. For example, at the median redshift of these surveys ($z\sim$0.1) the 1~kpc expected scale of the starburst projects to
$\sim$0.5\,arcseconds. \citet{norton01} compared the radial distribution of the young and old
populations in the LCRS sample and concluded that the young stellar population was centrally concentrated. The seeing FWHM
of the \citet{norton01} observations corresponded to $\sim$2--3~kpc and in about half the cases they were able to 
spatially resolve the young stellar population by showing it to be broader than the stellar PSF. The implication, 
in these cases at least was that, the starburst population is not confined
to the galaxy core ($\sim$1~kpc). Likewise, follow up of a HI--detected post starburst galaxy from the SDSS by \citet{chilingarian09} found 
the very young stellar population spread over the central $\sim$2.5~kpc. \citet{yagi06} analyzed three early type E+A galaxies selected from the 
SDSS using long--slit spectroscopy. In the one galaxy they could successfully estimate a H$\delta$ equivalent width profile, they found
that the post-starburst signature was as extended as the continuum light, and in the other two galaxies concluded (on statistical grounds)
that the post-starburst region was also significantly extended. \citet{pracy09} were unable to detect any Balmer line gradients in their sample although their
limited spatial coverage coupled with spatial resolution constraints imposed by the delivered image quality meant they could
not robustly rule out compact gradients in the stellar populations.

In this paper we circumvent these physical scale resolution issues by investigating a new, very low redshift sample
of E+A galaxies which have large angular to physical size scales. This is achieved by modifying the standard definition
of an E+A galaxy, using the H$\alpha$ line as the constraint on emission (and hence ongoing star formation) instead
of the traditional [OII]$\lambda 3727$ line. This change allows selection of a local sample of E+As at $z<0.01$ to easily resolve their central $1$\,kpc.

In what follows, Section 2 describes the details of the sample selection; Section 3 describes the observations, data reduction and analysis; in Section 4
we present our results, including spatially resolved Balmer line maps; and in Section 5 we discuss these results. Our conclusions are given in Section 6. 
Throughout, we adopt an $\Omega_{M}=0.3$, $\Omega_{\Lambda}=0.7$ and H$_{0}=70$\,km\,s$^{-1}$\,Mpc$^{-1}$ cosmology. 

\section{The sample}

\subsection{Selection criteria}
E+A galaxies are rare in the local universe \citep{zabludoff96,blake04,goto07} and there are very few 
nearby ($z<0.01$) examples known. For example, from the 2dFGRS \citet{blake04} cataloged 56 E+A galaxies but 
only two of these had redshifts of $z<0.01$. The catalogue of E+A galaxies produced from the much larger
SDSS by \citet{goto07} has no E+As with redshifts of $z<0.03$. The dearth of E+As is the result of the spectral range of 
the SDSS spectra ($\lambda \gtrsim$3800\,\AA) not containing the [OII]$\lambda 3727$ line for the very lowest redshift objects
and not a result of a true lack of local E+A galaxies in the SDSS. Although the E+A fraction does increase rapidly with redshift \citep[e.g.][]{poggianti99,snyder11}.
The use of low or non-existent equivalent widths of the 
[OII]$\lambda 3727$ line as the constraint on ongoing star formation is historical, since E+A galaxies were originally 
discovered and studied in intermediate redshift clusters \citep{dressler83} where the H$\alpha$ line was not available. 
In fact, at low redshift the H$\alpha$ line is a more natural
choice for this since it is more easily relatable to the star formation rate, intrinsically stronger, less prone to dust obscuration
and has been shown to be a more robust constraint on star formation in E+A galaxies \citep{blake04}. By replacing the [OII]$\lambda 3727$
constraint with one defined using the H$\alpha$ line we have selected a $z<0.01$ E+A sample from the SDSS data release 7 spectral line catalogue. 
The selection criteria used were (our definition of equivalent width has emission denoted as negative values):

\begin{list}{$\bullet$}{\itemsep=0.1cm}

\item Little or no H$\alpha$ emission: EW(H$\alpha$)$> -2.5$\,\AA

\item Strong absorption in the H$\delta$ line: EW(H$\delta$)$> 5.0$\,\AA

\item We also demand strong absorption in the H$\gamma$ and H$\beta$ lines since this
has been shown to select a much higher fidelity sample \citep{blake04}: EW(H$\gamma$) \& EW(H$\beta$)$  > 4.0$\,\AA 

\item Small errors on the equivalent width measurements of the Balmer absorption lines: The errors in each of EW(H$\delta$), EW(H$\gamma$), 
EW(H$\beta$)$<1.0$\,\AA

\item High quality redshift: probability the redshift is correct $z_{\rm confidence}>$0.95

\item Low redshift: $0.0005<z<0.01$

\item Sufficiently bright for IFU follow up: $r<$15.5\,mag

\end{list}
In addition to these criteria there were also constraints on the right ascension and declination for follow up observations. 
In total this gave us eight possible targets that could be observed during our scheduled telescope time and we  obtained data for 
all but one.  In the absence of right ascension and declination constraints, twenty galaxies in the SDSS meet these criteria. Imposing
constraints on the H$\beta$ and H$\gamma$ lines in addition to H$\delta$ has a significant effect on the sample size, 
particularly H$\beta$ which is prone to emission line filling. Removing
the  H$\beta$ constraint increases the sample size from twenty to thirty whilst removing the H$\gamma$ constraint increases the
sample by a single object; removing both constraints results in a sample of thirty seven galaxies.
The final observed targets and their properties are given in Table \ref{tab:targets}. 

\begin{figure}
      \includegraphics[width=5.8cm, angle=90, trim=0 0 0 0]{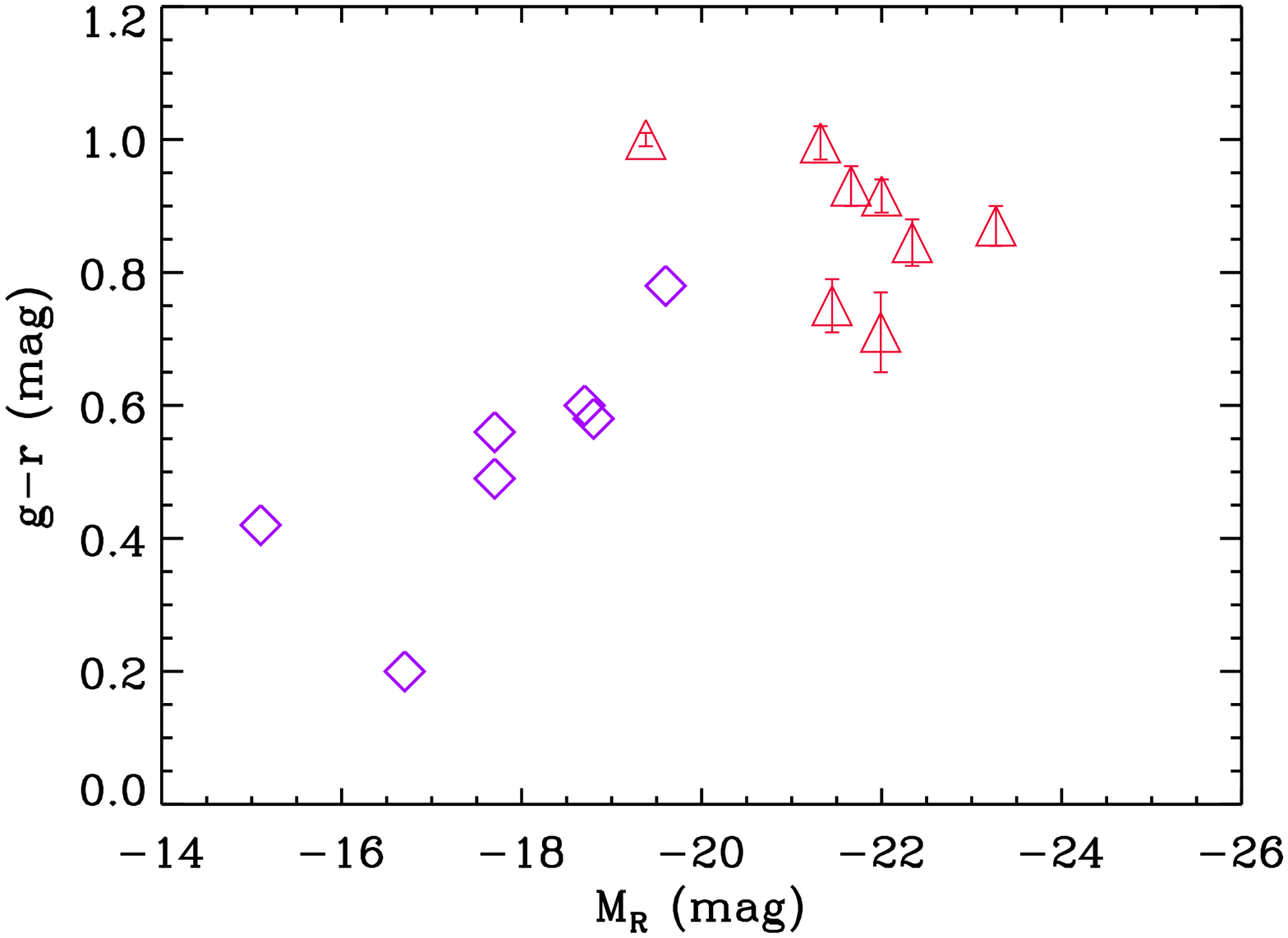}
\caption{\label{fig:colmag}Colour versus absolute magnitude for our current sample ({\it blue diamonds}) and the \citet{pracy09} sample ({\it red triangles}) at
redshifts between 0.04 and 0.20. The error bars on the \citet{pracy09} points give an indication of the uncertainty in the k-corrections.}
\end{figure}

A consequence of selecting an E+A galaxy sample
from the local volume is that our sample contains intrinsically fainter galaxies than those selected at higher redshift. Figure \ref{fig:colmag}
shows our sample in a $M_{R}$ versus $g-r$ color magnitude diagram ({\it blue diamonds}) along with the \citet{pracy09} sample ({\it red triangles}). The brightest 
galaxy in this sample corresponds to the faint--end of the \citet{pracy09} sample in magnitude and colour. There is a strong correlation in
the local sample between colour and luminosity in the sense that the fainter galaxies are bluer.
\begin{table*}
\begin{center}
\caption{\label{tab:targets}Target galaxies and their properties}
\begin{tabular}{llllllccl}\hline
SDSS object ID          &  ID   & $r$    &  $g-r$     &  $z$  &  $\rm{M}_r$& scale               & H$\delta$  & Morphology and environment notes      \\ 
                        &           & (mag)  &            &       &    (mag)  & (kpc/arcsec)    &   \AA           &    \\ \hline
587725073921278177      &  E+A 1    &  14.92 &   0.20         &  0.005 &  -16.7   &        0.10     &  -6.1      & Irregular morphology with lower mass\\
                        &           &        &                &        &           &                 &            & starbursting companion 5\,kpc away.      \\
587726033847451681      &  E+A 2    &  12.35 &   0.58         &  0.004 &  -18.8    &        0.08      &  -5.2     & NGC 3156. Isolated early type.    \\
                        &           &        &                &        &           &                 &            &   Member of the SAURON sample.    \\
588017724397912097      &  E+A 3    &  15.26 &   0.49         &  0.009 &  -17.7   &        0.18      &  -5.8     & NGC 3976A. Late type disk located $\sim$50~kpc from \\
                        &           &        &                &        &           &                  &           & much more massive NGC 3976.      \\
588017730298576932      &  E+A 4    &  14.50 &   0.42         &  0.002 &  -15.1    &        0.04      &  -5.4     & In the Virgo cluster. No evidence of interaction.      \\
588848898841903109      &  E+A 5    &  12.80 &   0.78         &  0.007 &  -19.6     &       0.14      &  -6.0     & NGC 4418. Semi-isolated early type. Starbursting \\
                        &           &        &                &        &            &                 &           & companion $\sim$17\,kpc away.      \\
587729779053887547      &  E+A 6    &  12.98 &   0.60         &  0.005 &  -18.7   &        0.10     &  -6.3      & Isolated early type      \\
587726033341513813      &  E+A 7    &  14.67 &   0.56         &  0.007 &  -17.7    &        0.14      &  -6.4     & Isolated early type      \\ \hline
\end{tabular}
\begin{flushleft}
Notes. Basic properties of our sample. Column 1 is the SDSS object ID and Column 2 is our simplified name. The remainder of the columns are: SDSS $r$--band magnitude;
$g-r$ colour; redshift; absolute $r$ magnitude; angular to physical scale; H$\delta$ absorption line equivalent width from the SDSS integrated spectrum; and a short description
of the morphology and local environment from visual inspection of the SDSS imaging for each galaxy and the surrounding region.
\end{flushleft}
\end{center}
\end{table*}

\subsection{SDSS data}
\begin{figure*}
  \begin{center}
    \begin{minipage}{0.95\textwidth}
      \includegraphics[width=6.0cm, angle=90, trim=0 0 0 0]{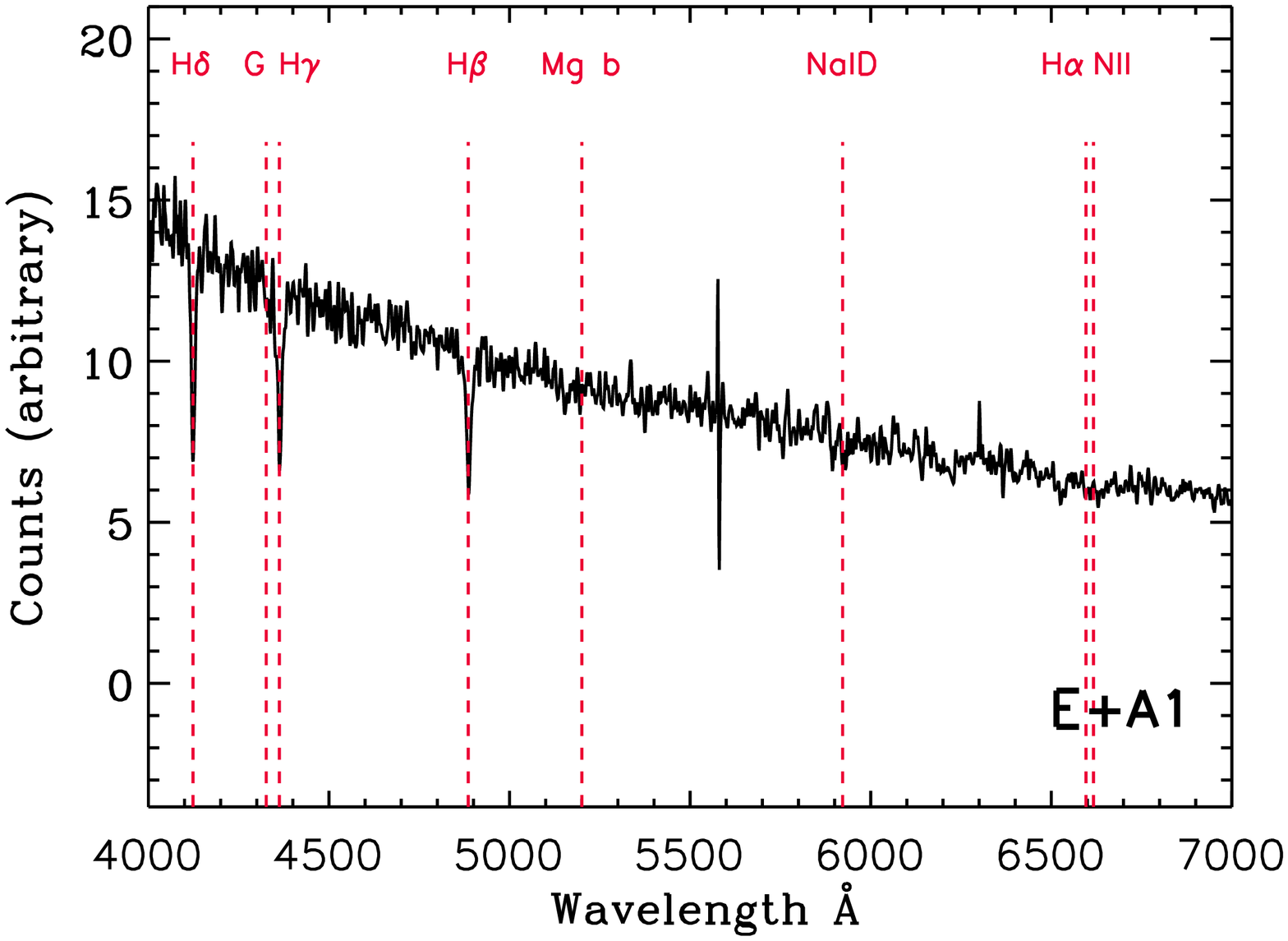}
      \includegraphics[width=6.0cm, angle=90, trim=0 0 0 0]{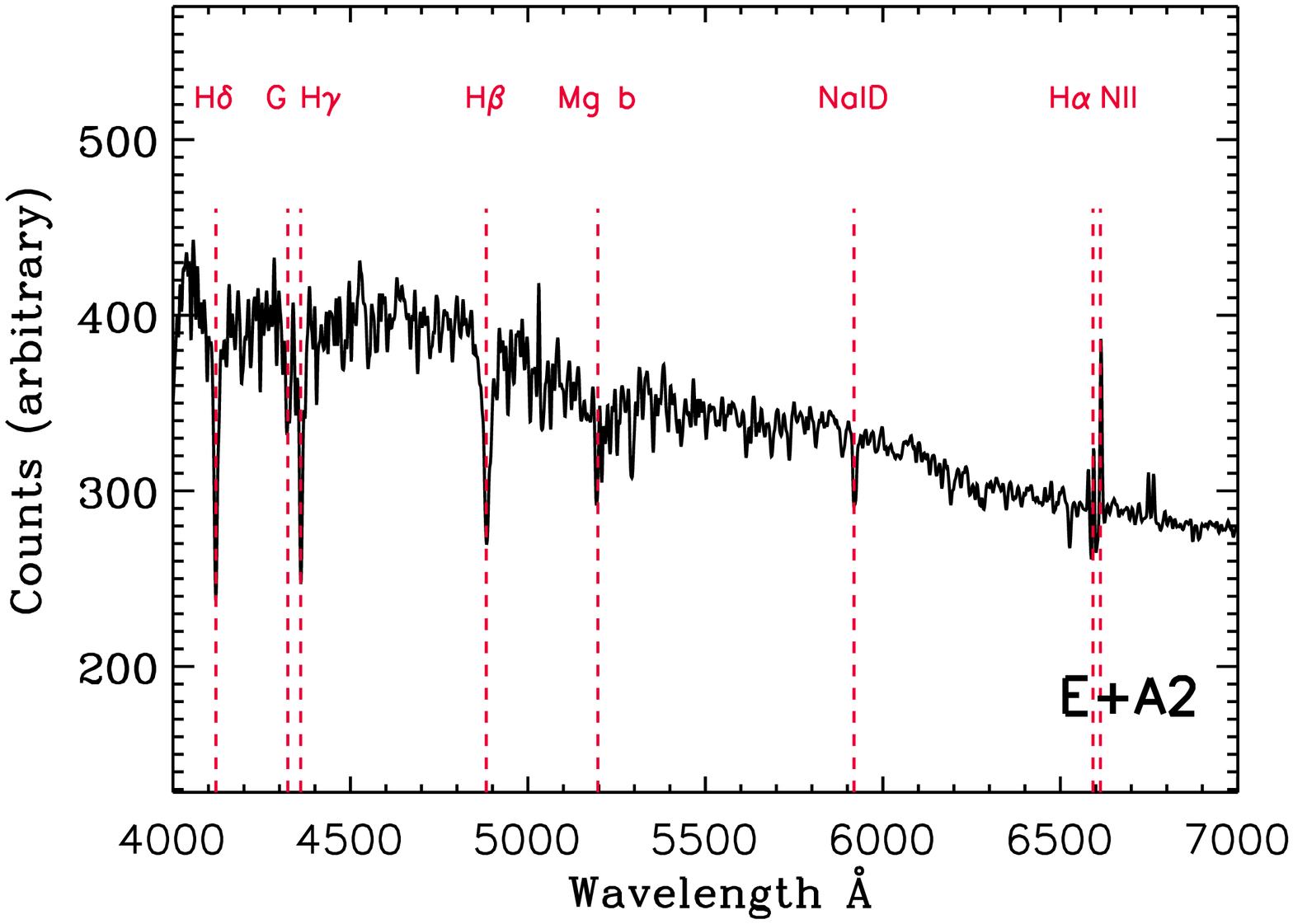}
\vspace{-0.5cm}
    \end{minipage}
    \begin{minipage}{0.95\textwidth}
      \includegraphics[width=6.0cm, angle=90, trim=0 0 0 0]{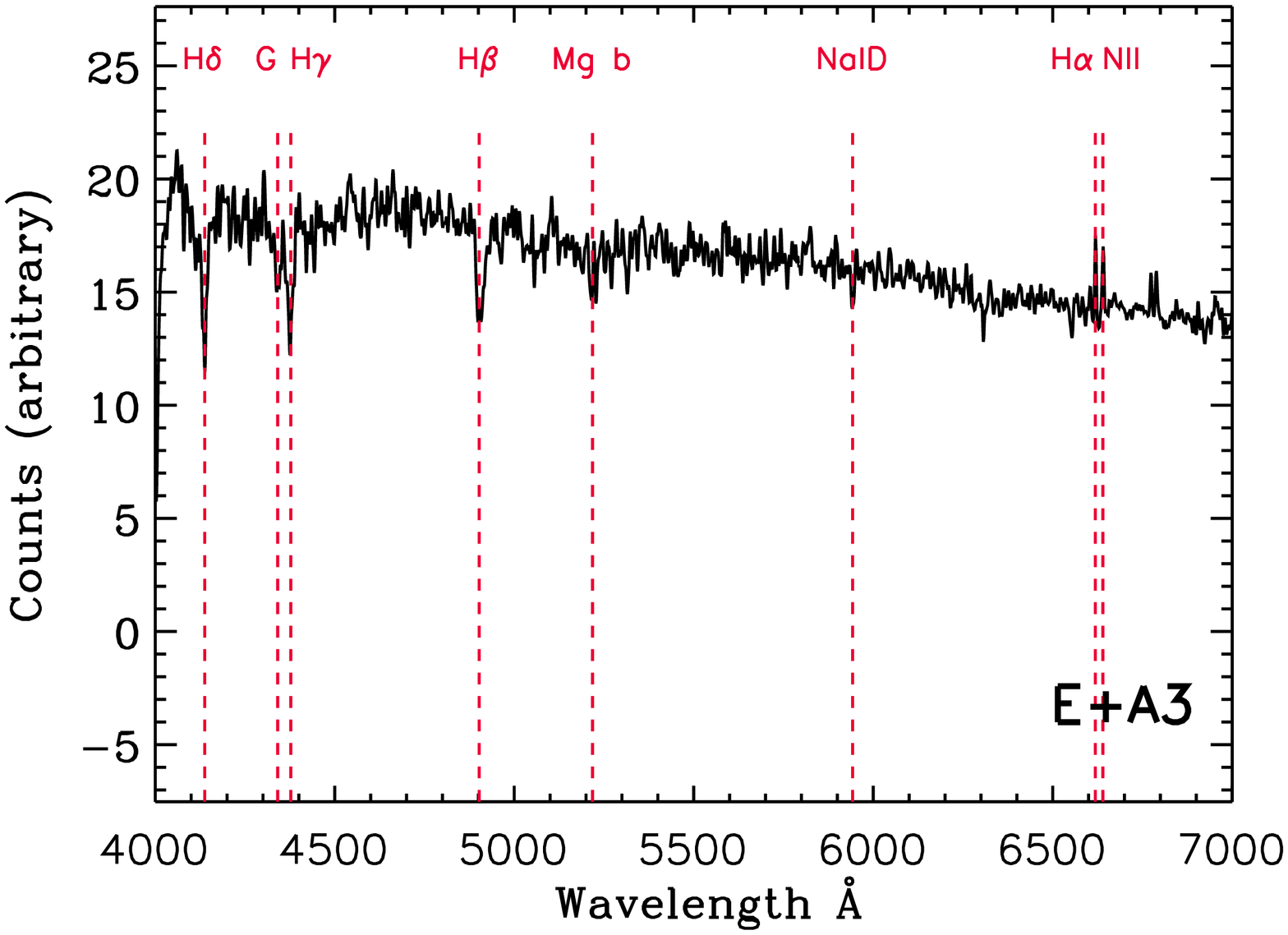}
      \includegraphics[width=6.0cm, angle=90, trim=0 0 0 0]{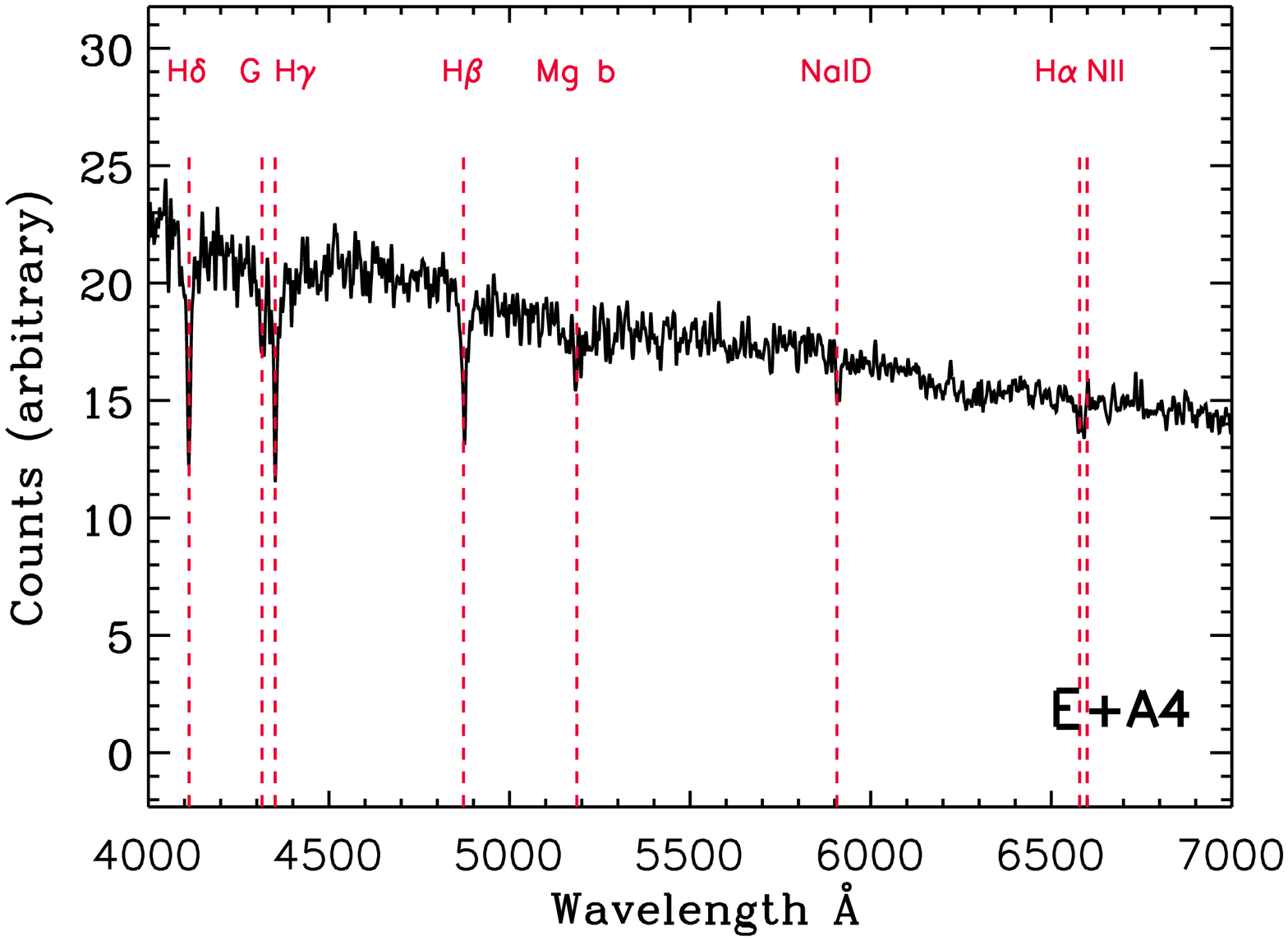}
\vspace{-0.5cm}
    \end{minipage}
    \begin{minipage}{0.95\textwidth}
      \includegraphics[width=6.0cm, angle=90, trim=0 0 0 0]{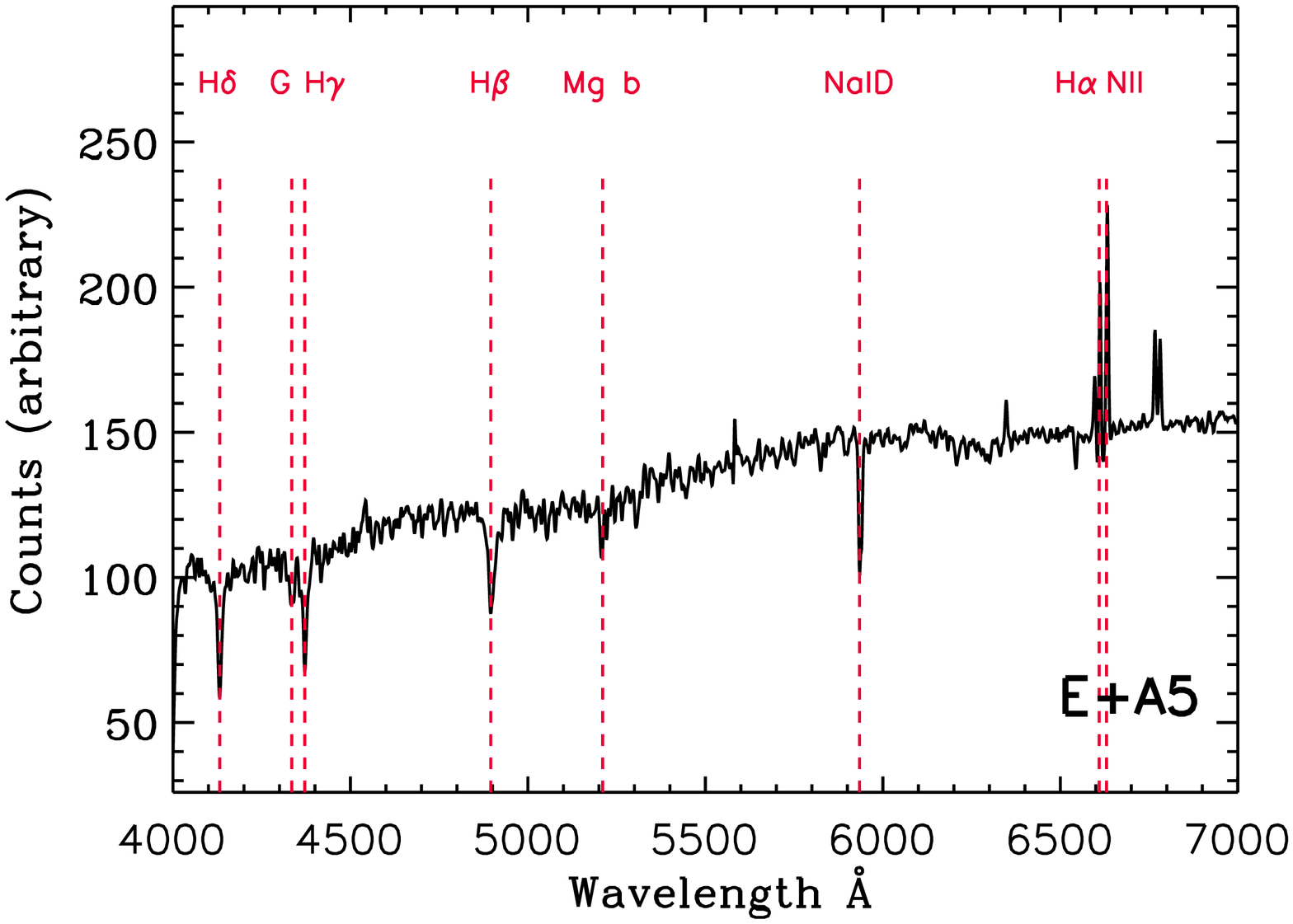}
      \includegraphics[width=6.0cm, angle=90, trim=0 0 0 0]{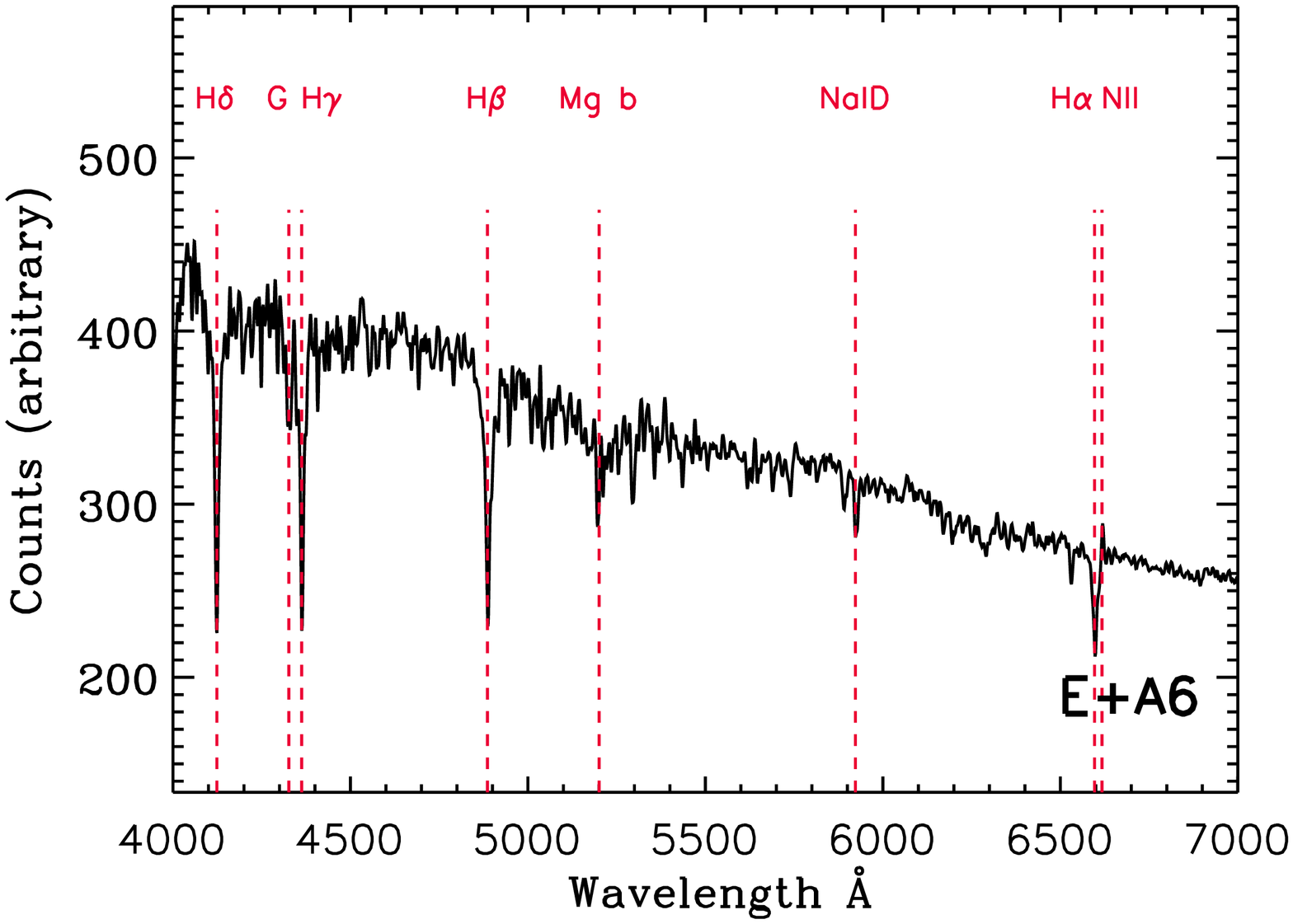}
\vspace{-0.5cm}
    \end{minipage}
    \begin{minipage}{0.95\textwidth}
      \includegraphics[width=6.0cm, angle=90, trim=0 0 0 0]{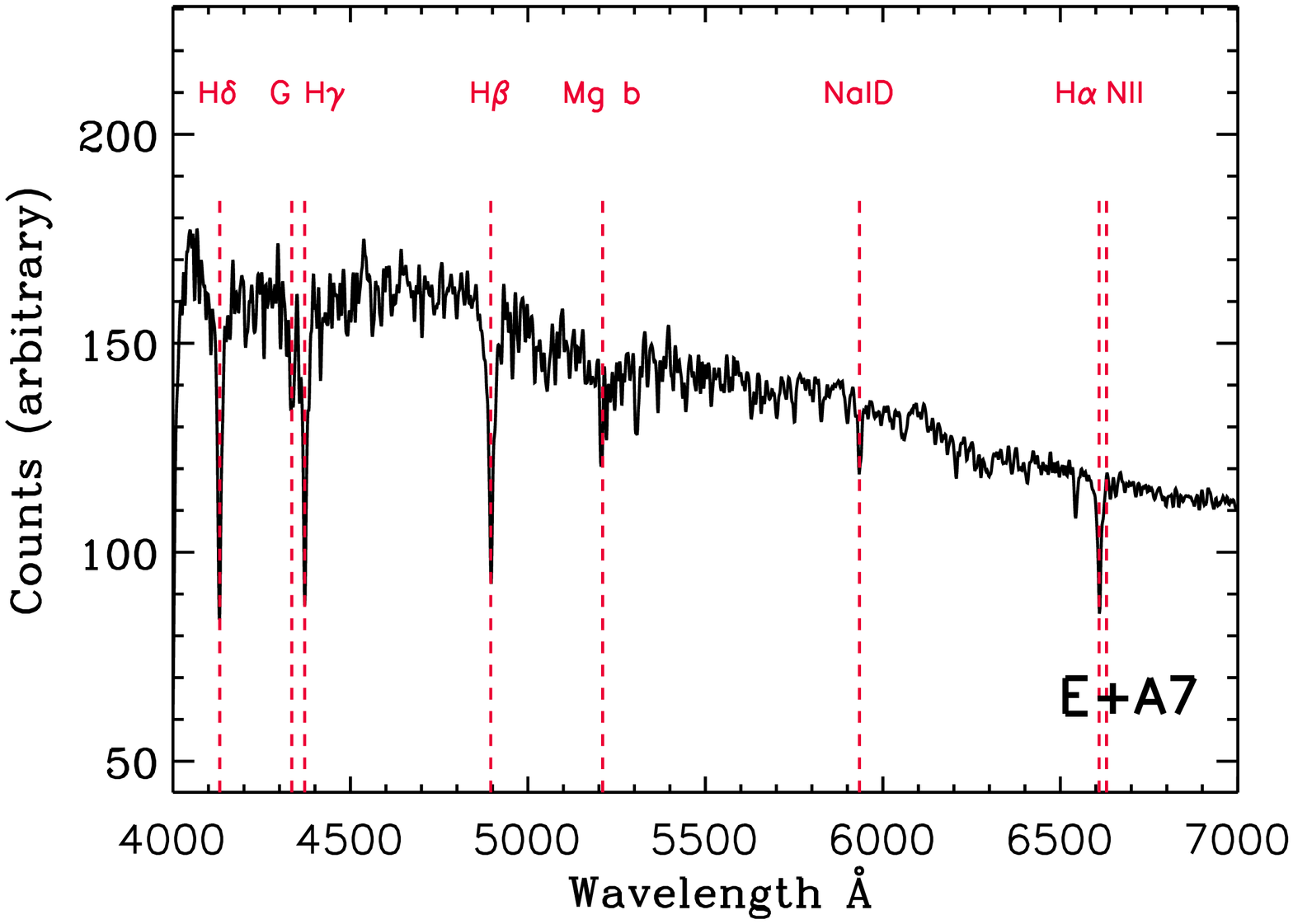}
     \end{minipage}
\end{center}
    \caption{Spectra of our sample of E+A galaxies from the Sloan Digital Sky Survey data release 7.}
    \label{fig:spectra}
\end{figure*}

The SDSS spectra of our sample are shown in Fig. \ref{fig:spectra}. Three galaxies have detectable 
H$\alpha$ emission (E+A 2, E+A 3 and E+A 5). Two of these galaxies have [NII]$\lambda 6583$/{\rm H}$\alpha \lambda 6563 > 1$ with the
other having [NII]$\lambda 6583$/{\rm H}$\alpha \lambda 6563 \sim $1. Such ratios are generally indicative of AGN activity 
\citep{baldwin81}, although, in this case we have selected galaxies with weak
H$\alpha$ emission and underlying stellar populations with strong Balmer line absorption. In this circumstance the 
underlying absorption can have a large effect on the apparent ratio of the lines and we show later that these
spectra are more consistent with a combination of star formation and AGN activity. 
The spectra all have strong higher order Balmer lines (as selected) and blue continua -- E+A 5 has a significantly 
redder continuum than the remainder of the sample but still has an integrated colour that is quite blue with $g-r=0.78$.

Colour SDSS images of the sample are shown in the leftmost column of Fig. \ref{fig:images}. 
The galaxies in the sample generally have early type morphologies which is typical of E+A samples.
There are two exceptions:
E+A 1 which has an irregular morphology and is also the bluest (youngest stellar population) galaxy in our
sample and E+A 3 which is a late type disk galaxy. 

\section{Observations and Data reduction}

\subsection{Observations}
Our data were obtained using the Wide Field Spectrograph \citep[WiFeS;][]{dopita07,dopita10} on the 
Mount Stromlo and Siding Springs Observatory's 2.3-m telescope. The WiFeS is an image--slicing integral 
field spectrograph with a 25\,arcsecond by 38\,arcsecond field-of-view. The spatial pixel size is 1\,arcsecond
square (after on--chip binning) and the spectra have a resolution of R$\sim$3000 and cover from $\sim$3500\,\AA\ to $\sim$9000\,\AA. 
The WiFeS is designed for Nod and Shuffle observations and we used the
instrument in this mode. 
This mode of observation would not normally be used for observations concerned primarily with the blue
end of the optical spectrum. However in this case our objects filled the entire field-of-view so classical sky 
subtraction was not possible.
The nod and shuffle technique
was chosen over observations of offset sky fields (which would have been more efficient) to minimize systematics.
The WiFeS CCD is read out by four separate amplifiers and as such the four quadrants have different
characteristics. We positioned our targets away from the centre of the WiFeS field-of view in the bottom right quadrant (see Fig. \ref{fig:images}).
This was done to avoid the galaxy centre being located where the four read-out sections meet.
The observations were obtained over a 5 night dark--time run in March 2010. Table \ref{tab:obs}
summarizes our observations. 
\begin{table}
\begin{center}
\caption{\label{tab:obs}Summary of observations.}
\begin{tabular}{cccccccc}\hline
 ID   & Exposure time (s)  & N            &  PA  & seeing \\
      & object sky &  (read outs) &      & (arcsec)       \\ \hline
E+A 1 & 2880 2880  &    2         & 155  &  1.7      \\  
E+A 2 & 1200 1200  &    1         & 50   &  1.7      \\
E+A 3 & 8040 8040  &    4         & 140  &  3.0      \\
E+A 4 & 6600 6600  &    4         & 180  &  2.0      \\
E+A 5 & 3600 3600  &    3         & 60   &  1.7      \\
E+A 6 & 1440 1440  &    1         & 170  &  1.6      \\
E+A 7 & 3600 3600  &    3         & 175  &  1.7      \\ \hline
\end{tabular}
\begin{flushleft}
Notes. From left to right: Object ID; total exposure times on object and on sky; the number of readouts; the position angle of the
WiFeS IFU; and an estimate of the seeing during the observation. 
\end{flushleft}
\end{center}
\end{table}

\subsection{Data reduction}
The data were reduced using the WiFeS IRAF data reduction package \citep{dopita10}.
The raw frames were bias--subtracted using the overscan region and trimmed. 
Since the data were taken in nod-and-shuffle mode, sky subtraction is performed
by shifting the image so that the sky regions are moved to the corresponding primary region
and subtracted from the original image. Flat-fielding was performed utilizing quartz-halogen lamp exposures. 
At this point the individual slice regions were extracted into separate image
extensions. The wavelength solution was derived and applied using CuAr arc lamp exposures. 
The slices were then combined into a single three dimensional data cube.

\subsection{Adaptive binning}
Within each data cube there is a large variation of the spectrum signal-to-noise ratio
achieved depending on its spatial position. For the most part the signal-to-noise
ratio drops off rapidly moving away from the bright galaxy centre. In general, the signal-to-noise ratio
of a 1\,arcsecond\, spaxel in the central regions is sufficient for quantitative 
analysis (SNR$\sim$10\,\AA$^{-1}$). However, in the outer regions single 1\,arcsecond\, spaxels 
do not provide sufficient quality spectra. In order to increase the SNR in the outer parts 
and to obtain more consistent spectral quality across the field-of-view we bin the
data spatially using the Voronoi spatial binning method of \citet{cappellari03}.
This method, adaptively bins spatial pixels in such a way as to obtain approximately
the same signal-to-noise ratio in each final spatial bin. This results in smaller spatial
bins in the central region and progressively larger bin sizes moving outward. 
A different final signal-to-noise ratio per adaptive bin was chosen for each galaxy 
depending on the quality of the data; the final signal-to-noise ratios ranged from
$\sim$6 to 12\,\AA$^{-1}$ per spaxel.

\subsection{Recession velocities}
We measure the recession velocity of each spatially binned spectrum by fitting
them with  rest frame template spectra. 
The template spectra used were from the single-age single-metallicity stellar population models
of \citet{vazdekis07} which are constructed from empirical stellar spectra \citep{sanchez06}. The stellar 
population ages used in the models ranged from $\sim$100~Myrs to $\sim$13\,Gyrs. 
The spectral resolution of the science data is degraded 
to match that of the templates. The fitting is then performed using the penalized pixel fitting algorithm
of \citet{cappellari04} which fits the spectrum using a combination of input template spectra
and simultaneously fitting for the recession velocity and velocity dispersion.

\subsection{Line indices}
We measure line equivalent widths from each spectrum using the flux summing technique. This uses
three wavelength bands to define the equivalent width; one centred on the line itself
and two flanking bands on either side of the line to define the expected continuum level by linear extrapolation.
For lines at the blue end of the spectrum, including H$\delta$, H$\gamma$ and H$\beta$, we measure line strength equivalent widths on the Lick/Intermediate Dispersion
Spectrograph system \citep{worthey97,trager98}. Prior to measuring the indices the science spectra are convolved with a wavelength--dependent Gaussian to the
Lick resolution ($\sim$9\,\AA). For lines red-ward of the Lick indices, in particular H$\alpha$ and [NII]$\lambda 6583$ that do not have standard definitions, we
again use the flux--summing technique with appropriate line and continuum bands.

\section{Results}

\subsection{Balmer line gradients}
\begin{figure*}
  \begin{center}
    \begin{minipage}{0.95\textwidth}
\vspace{-0.5cm}
      \includegraphics[width=2.9cm, angle=0, trim=0 -50 0 0]{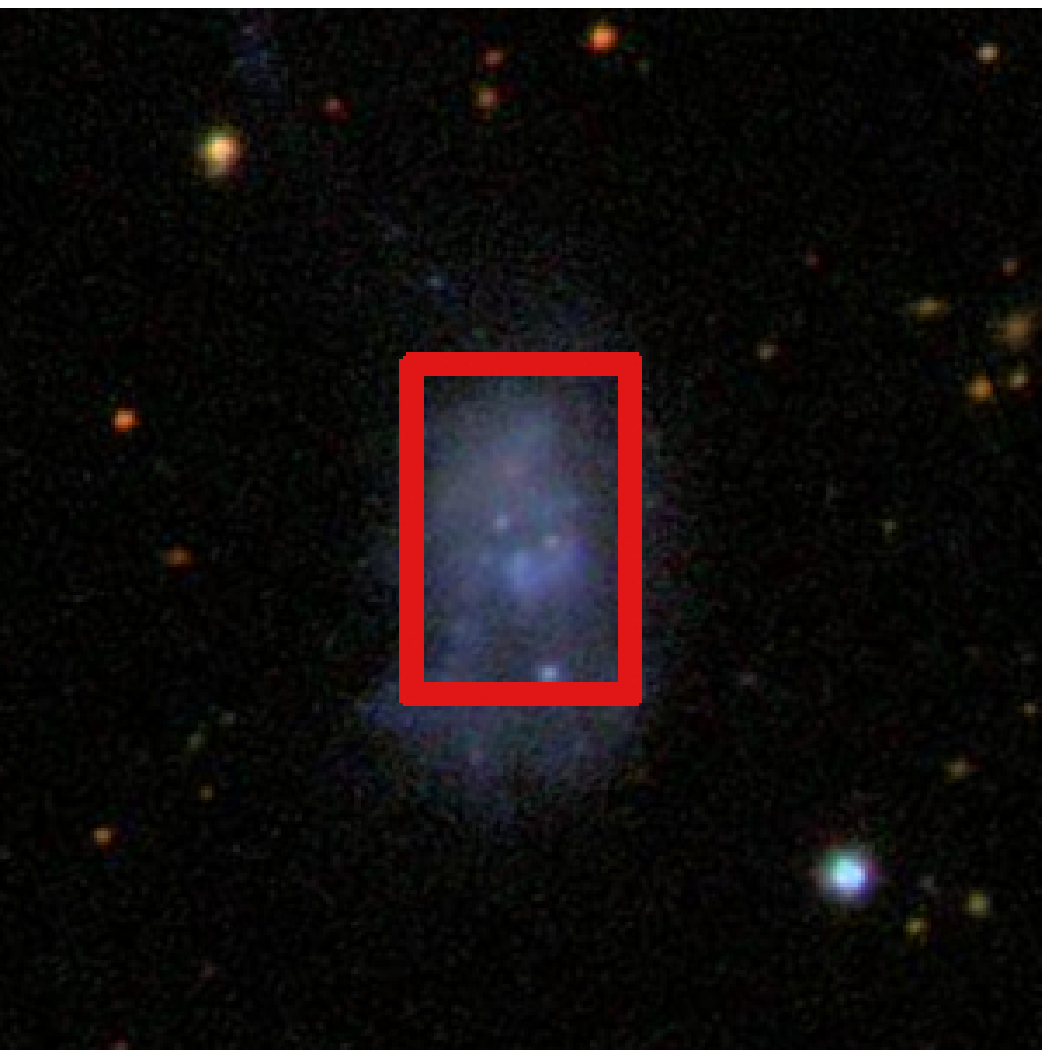}
      \includegraphics[height=4.0cm, angle=0, trim=70 0 0 0]{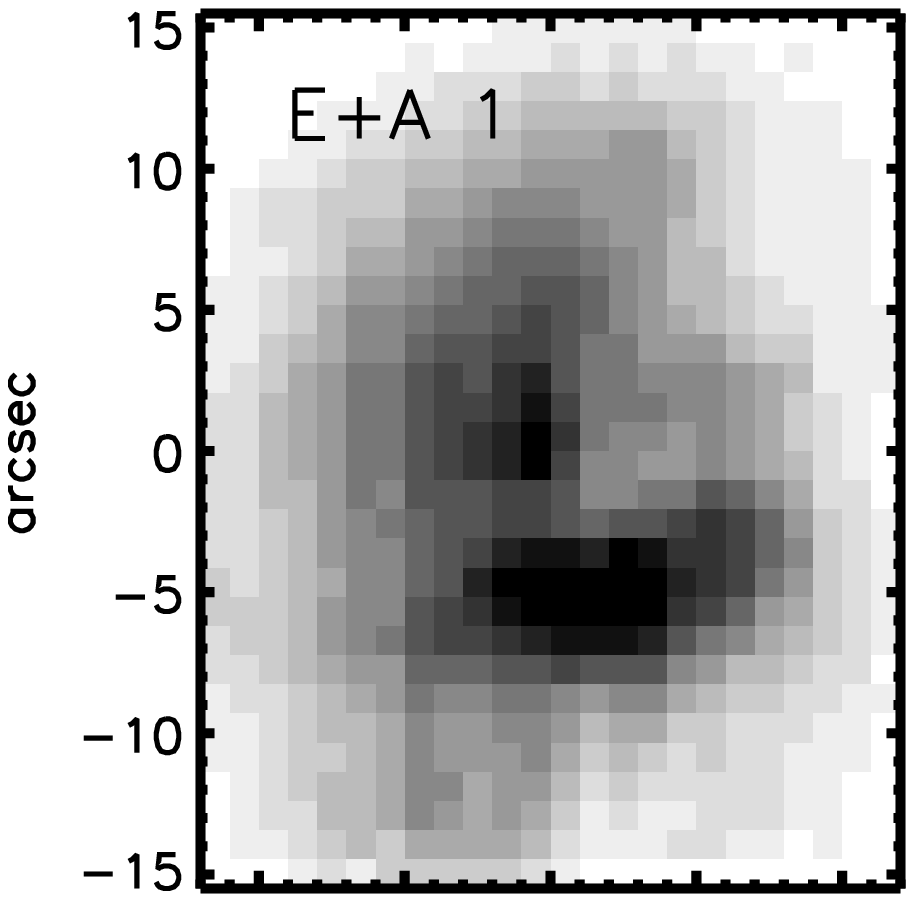}
      \includegraphics[height=3.6cm, angle=0, trim=210 -20 0 0]{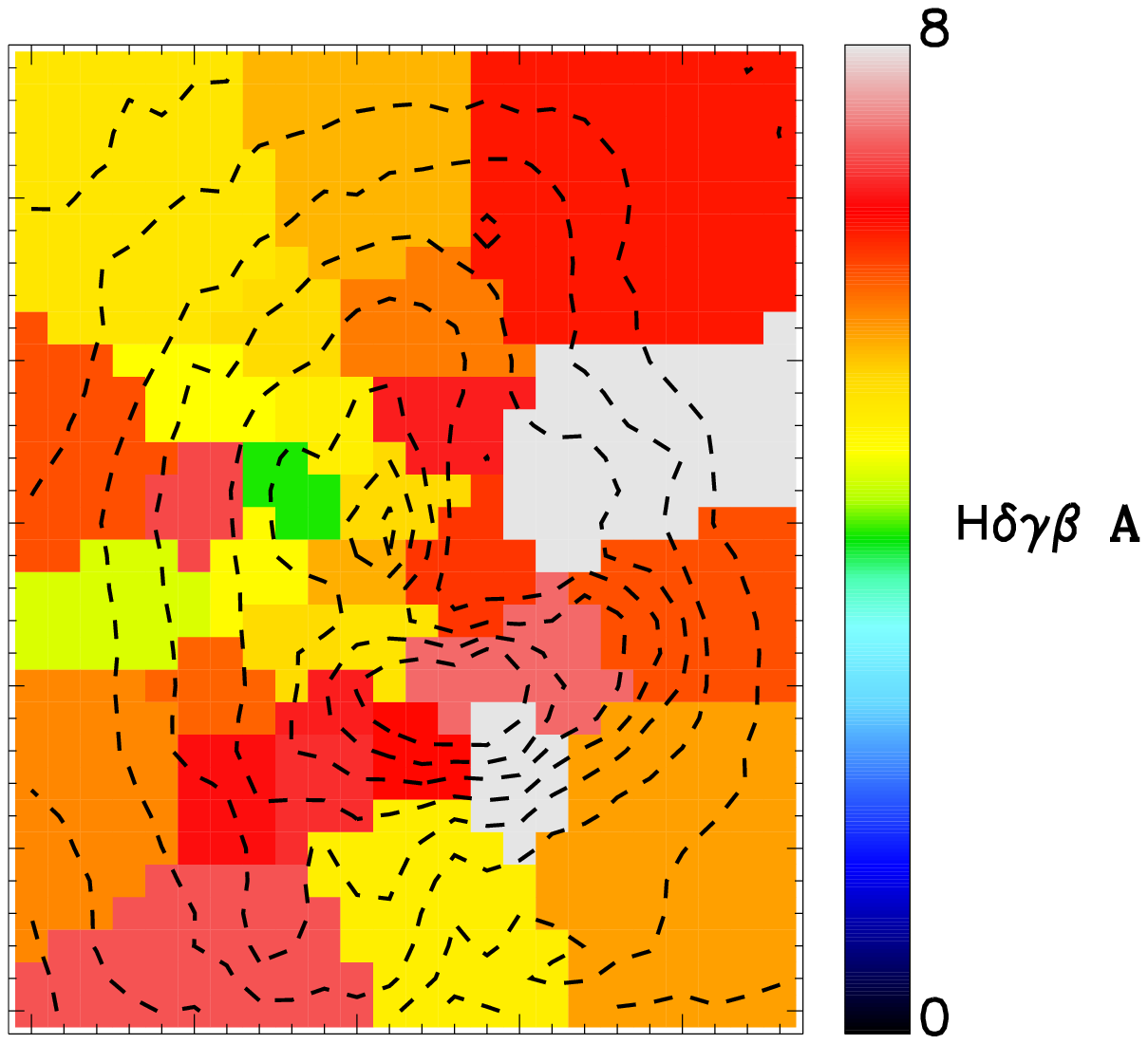}
      \includegraphics[height=4.0cm, angle=90, trim=-60 -30 0 100]{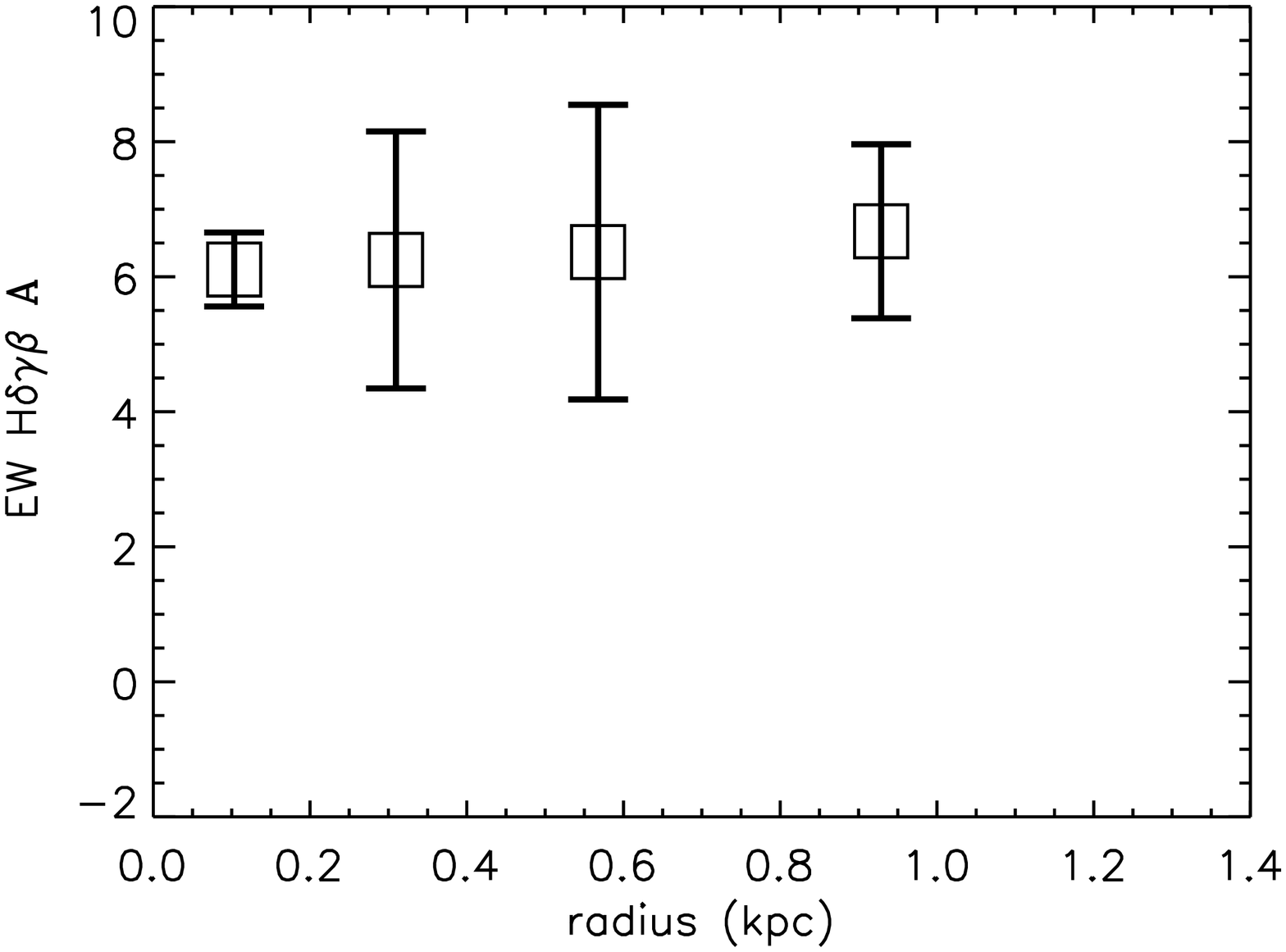}
      \vspace{-0.9cm}
    \end{minipage}
    \begin{minipage}{0.95\textwidth}
      \includegraphics[width=2.9cm, angle=0, trim=0 -50 0 0]{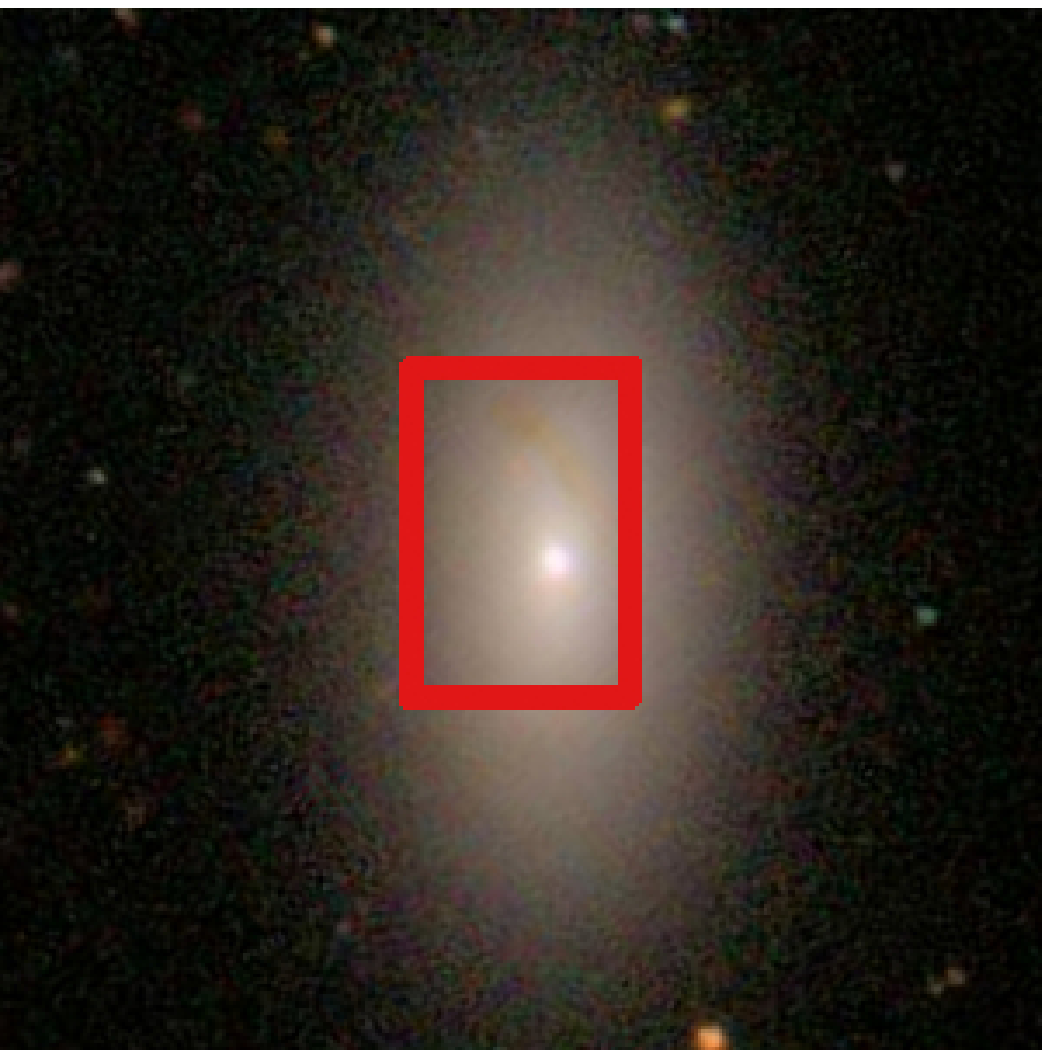}
      \includegraphics[height=4.0cm, angle=0, trim=70 0 0 0]{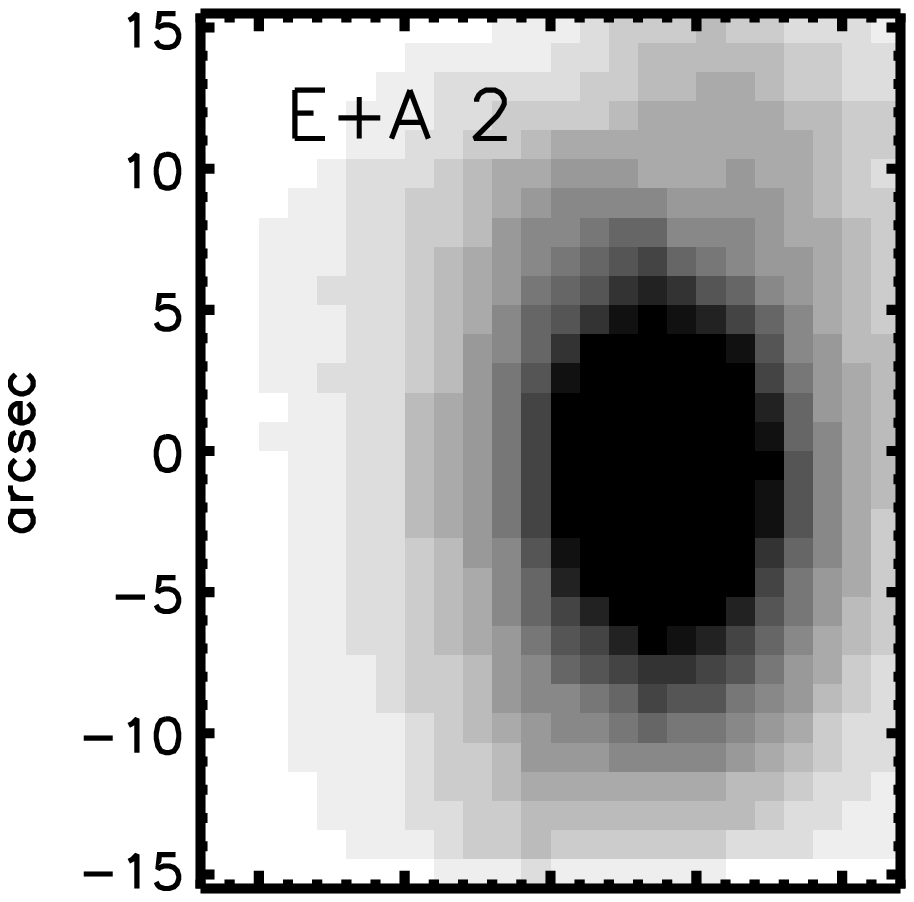}
      \includegraphics[height=3.6cm, angle=0, trim=210 -20 0 0]{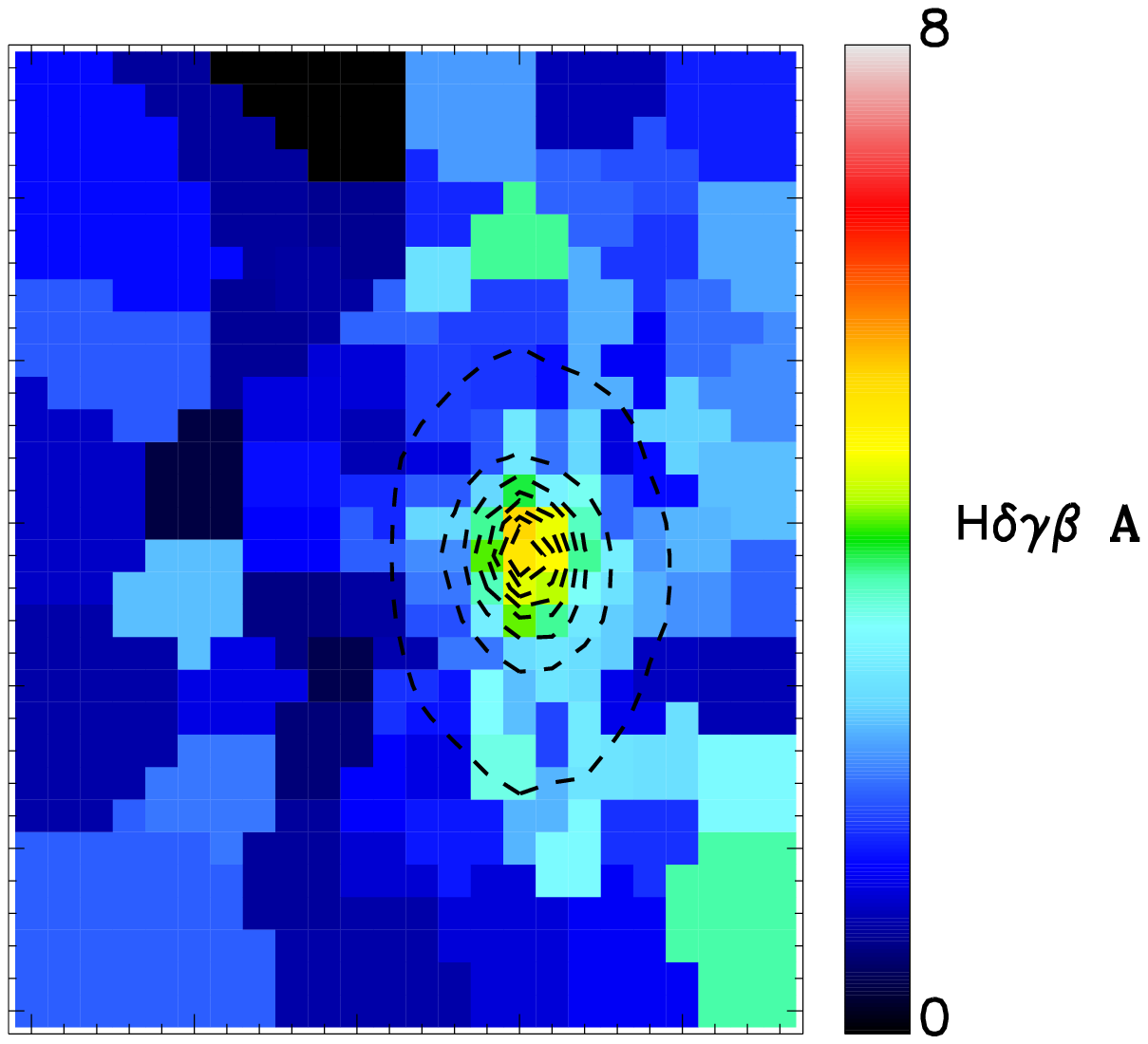}
      \includegraphics[height=4.0cm, angle=90, trim=-60 -30 0 100]{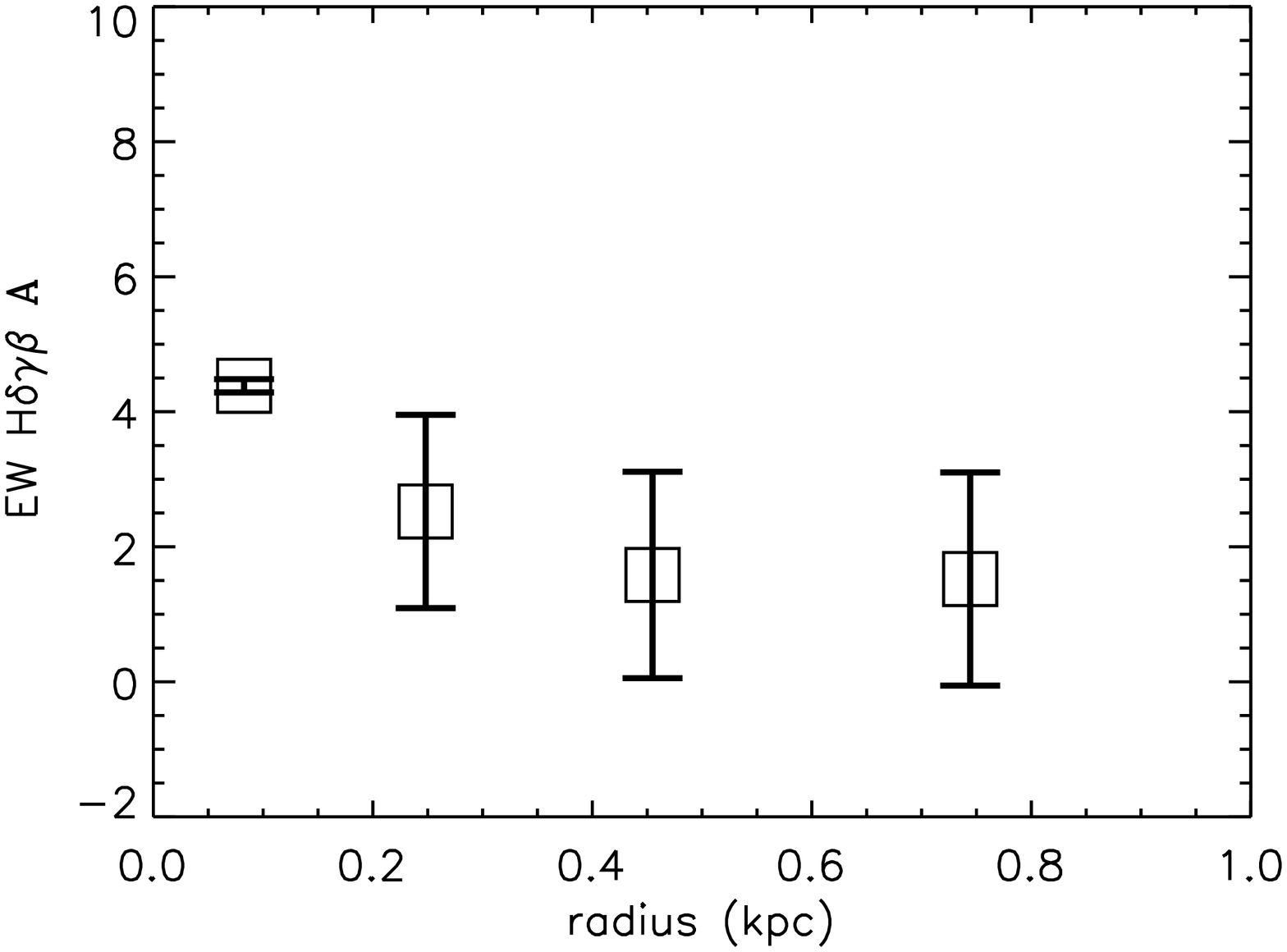}
      \vspace{-0.9cm}
    \end{minipage}
    \begin{minipage}{0.95\textwidth}
      \includegraphics[width=2.9cm, angle=0, trim=0 -50 0 0]{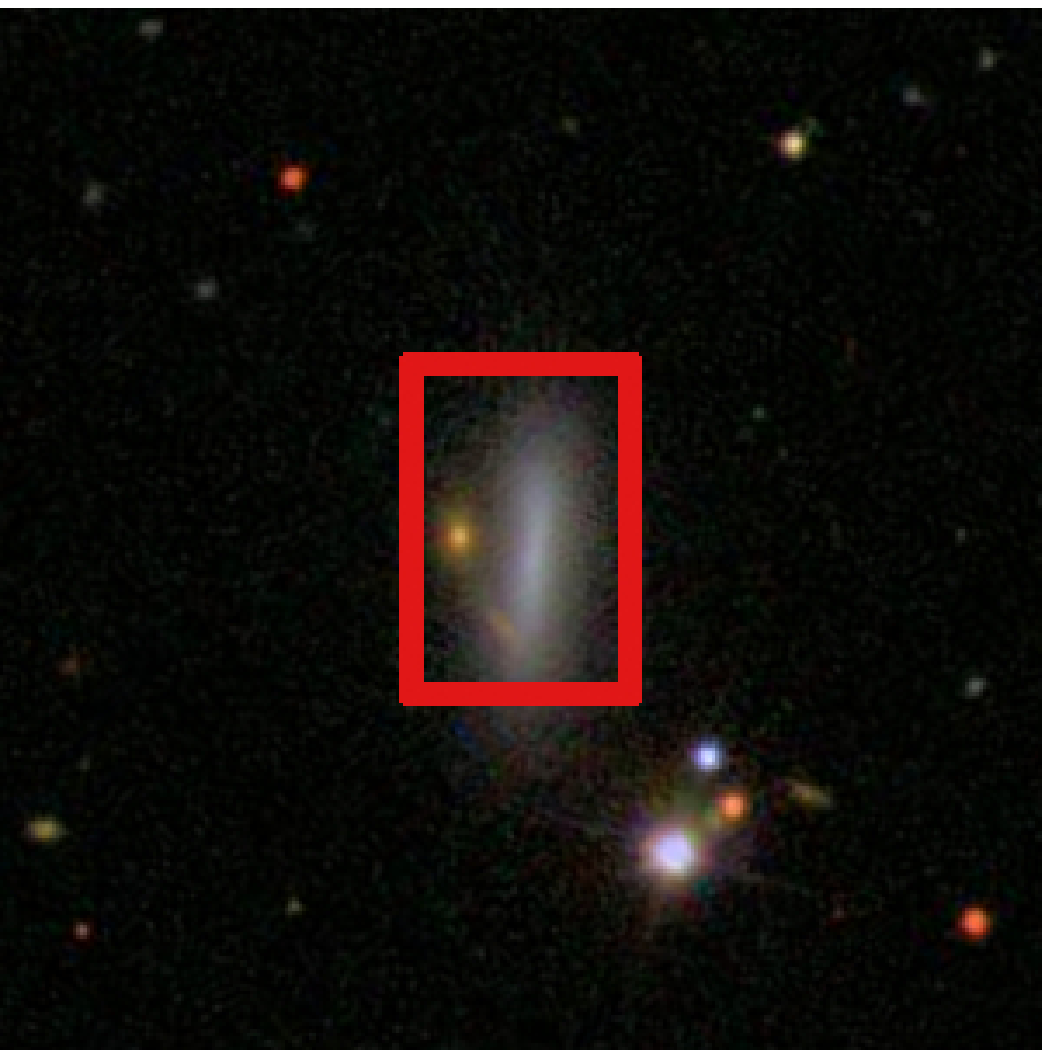}
      \includegraphics[height=4.0cm, angle=0, trim=70 0 0 0]{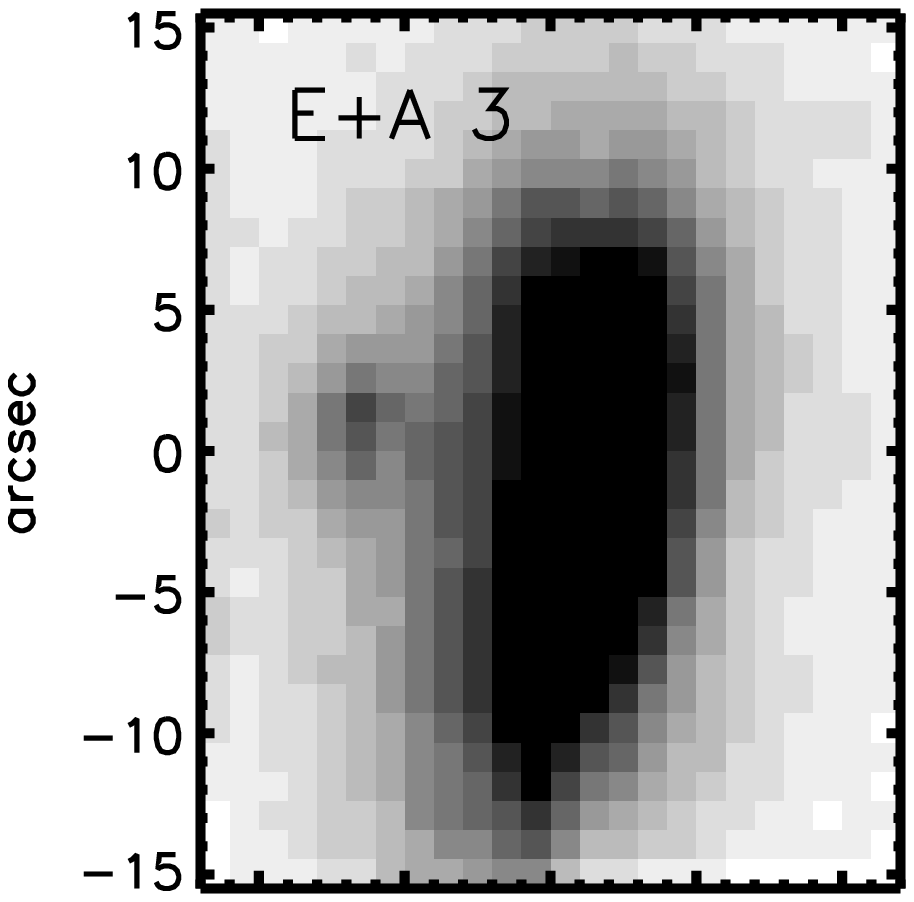}
      \includegraphics[height=3.6cm, angle=0, trim=210 -20 0 0]{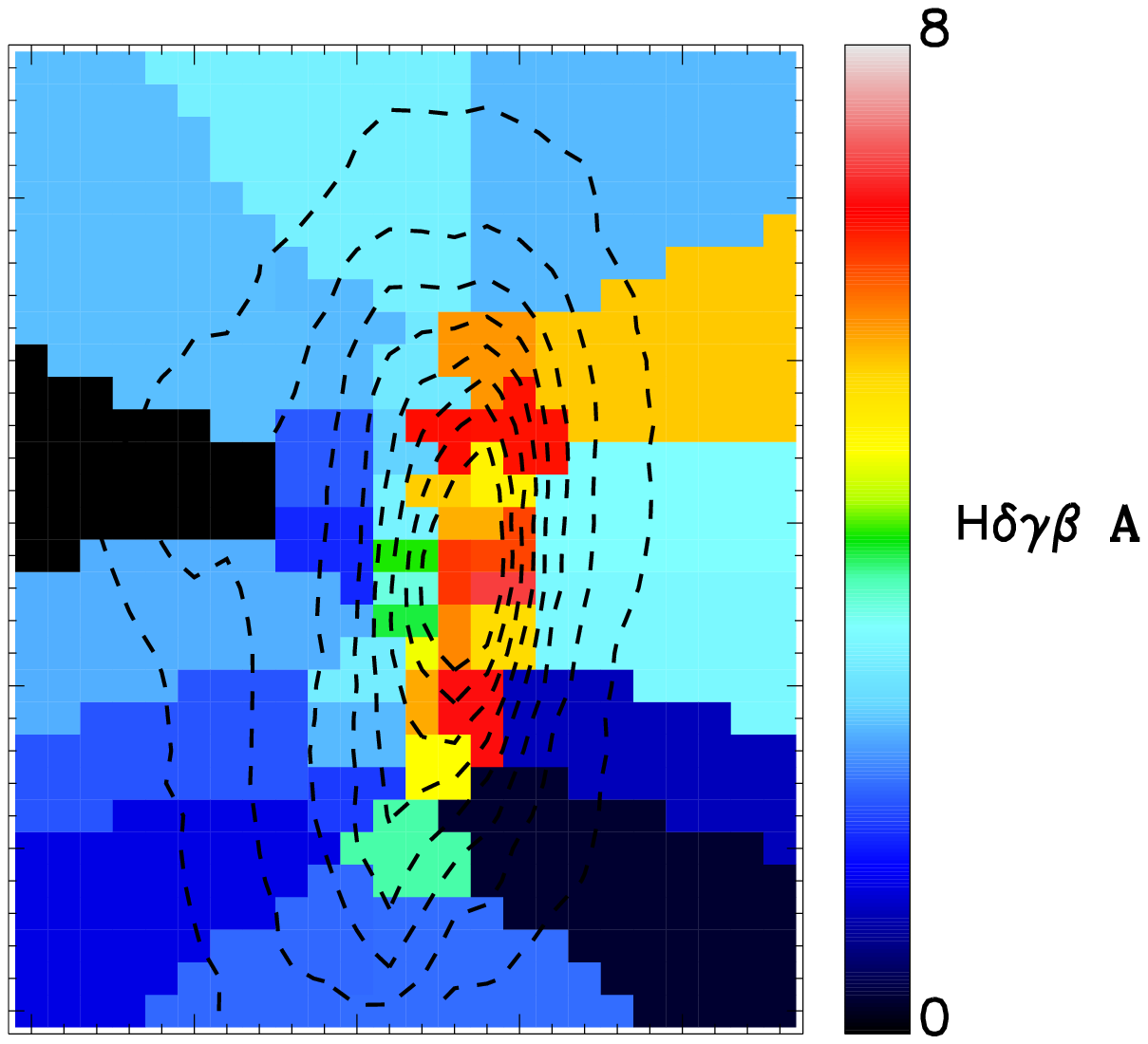}
      \includegraphics[height=4.0cm, angle=90, trim=-60 -30 0 100]{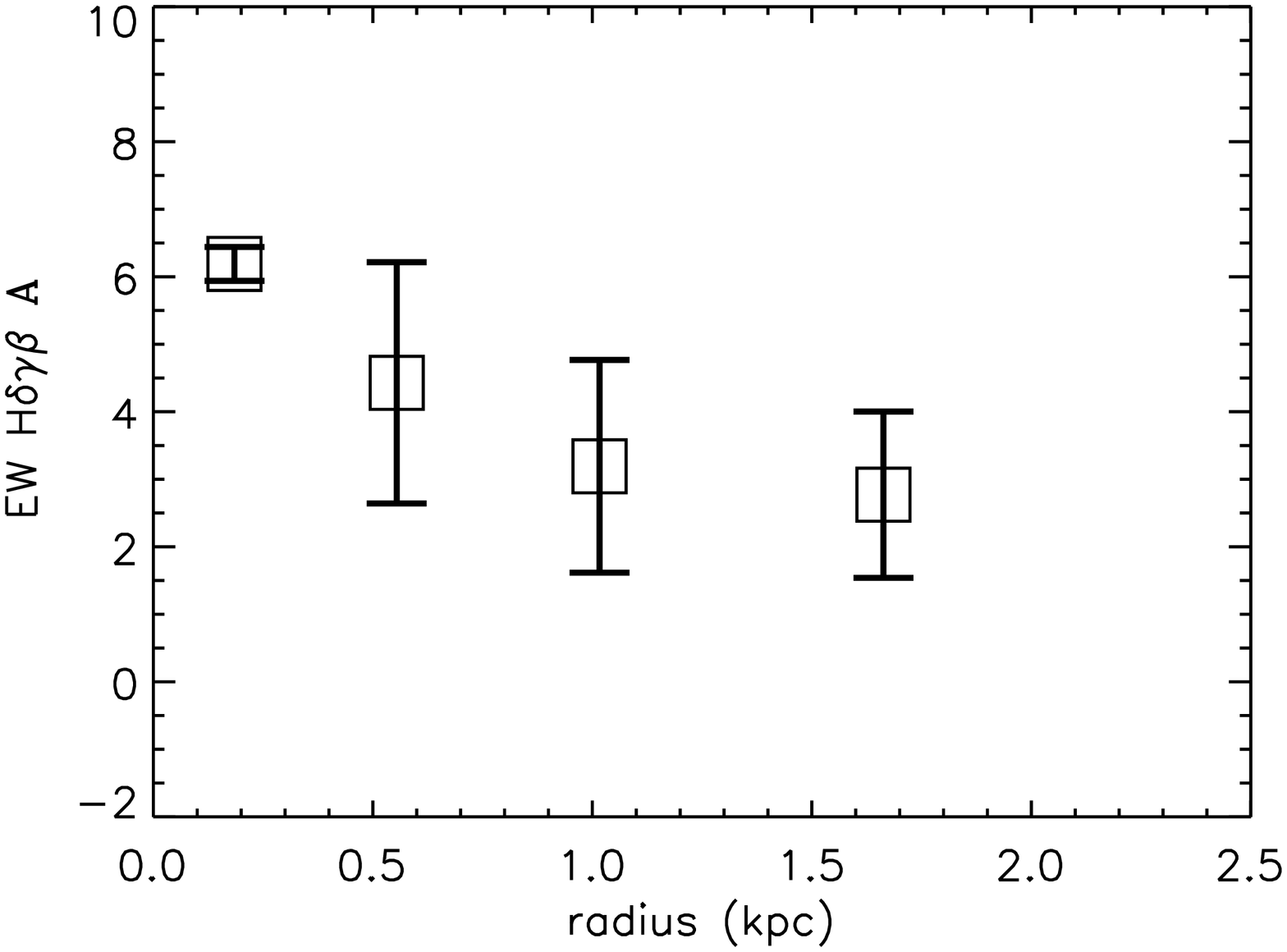}
      \vspace{-0.9cm}
    \end{minipage}
    \begin{minipage}{0.95\textwidth}
      \includegraphics[width=2.9cm, angle=0, trim=0 -50 0 0]{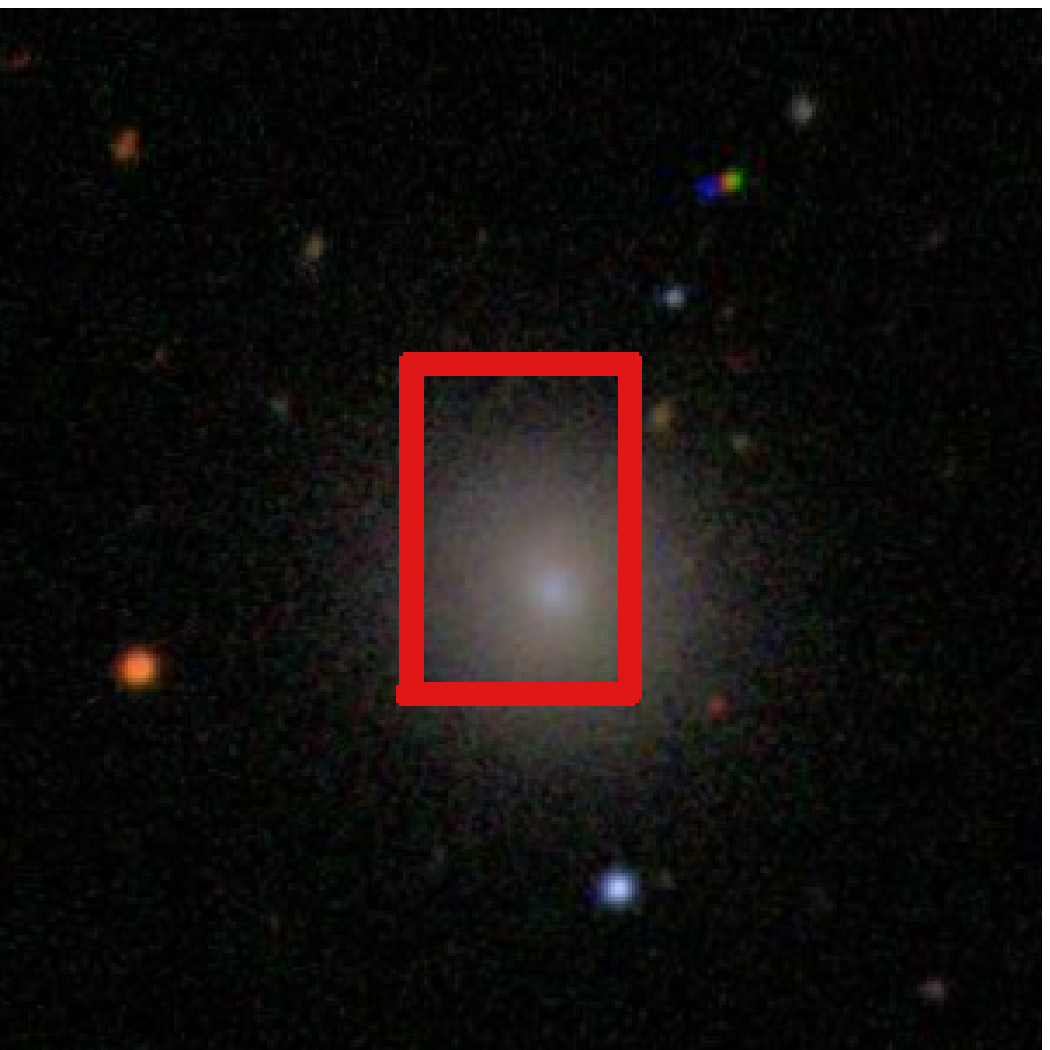}
      \includegraphics[height=4.0cm, angle=0, trim=70 0 0 0]{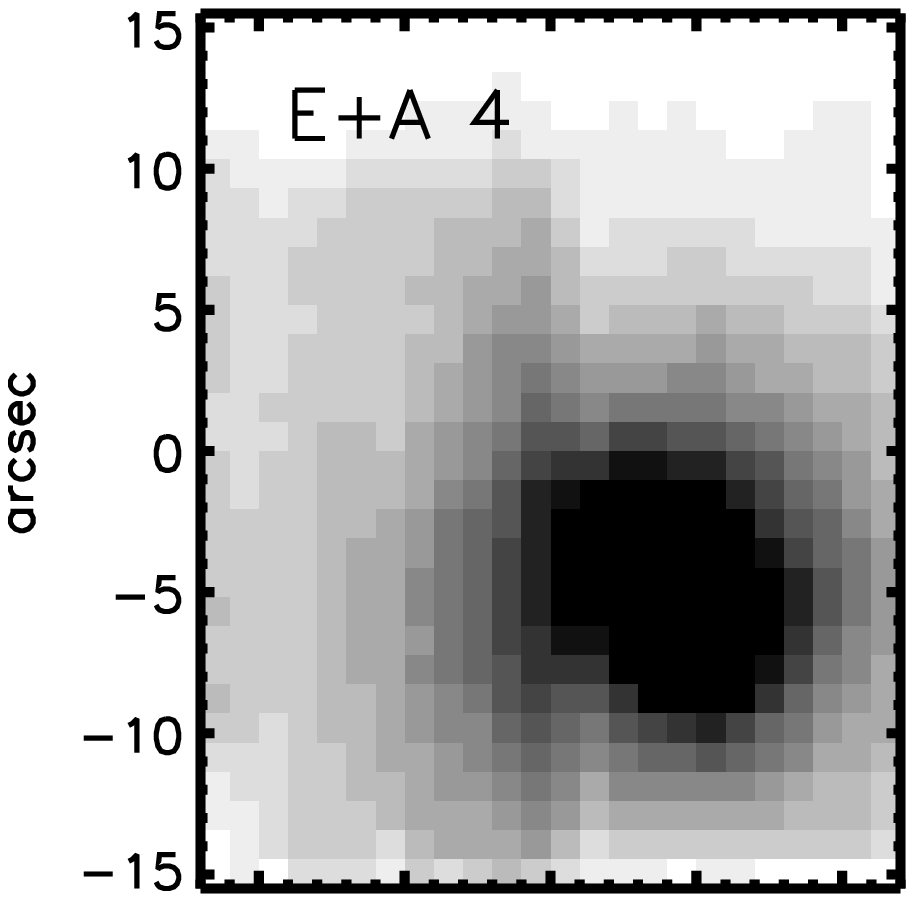}
      \includegraphics[height=3.6cm, angle=0, trim=210 -20 0 0]{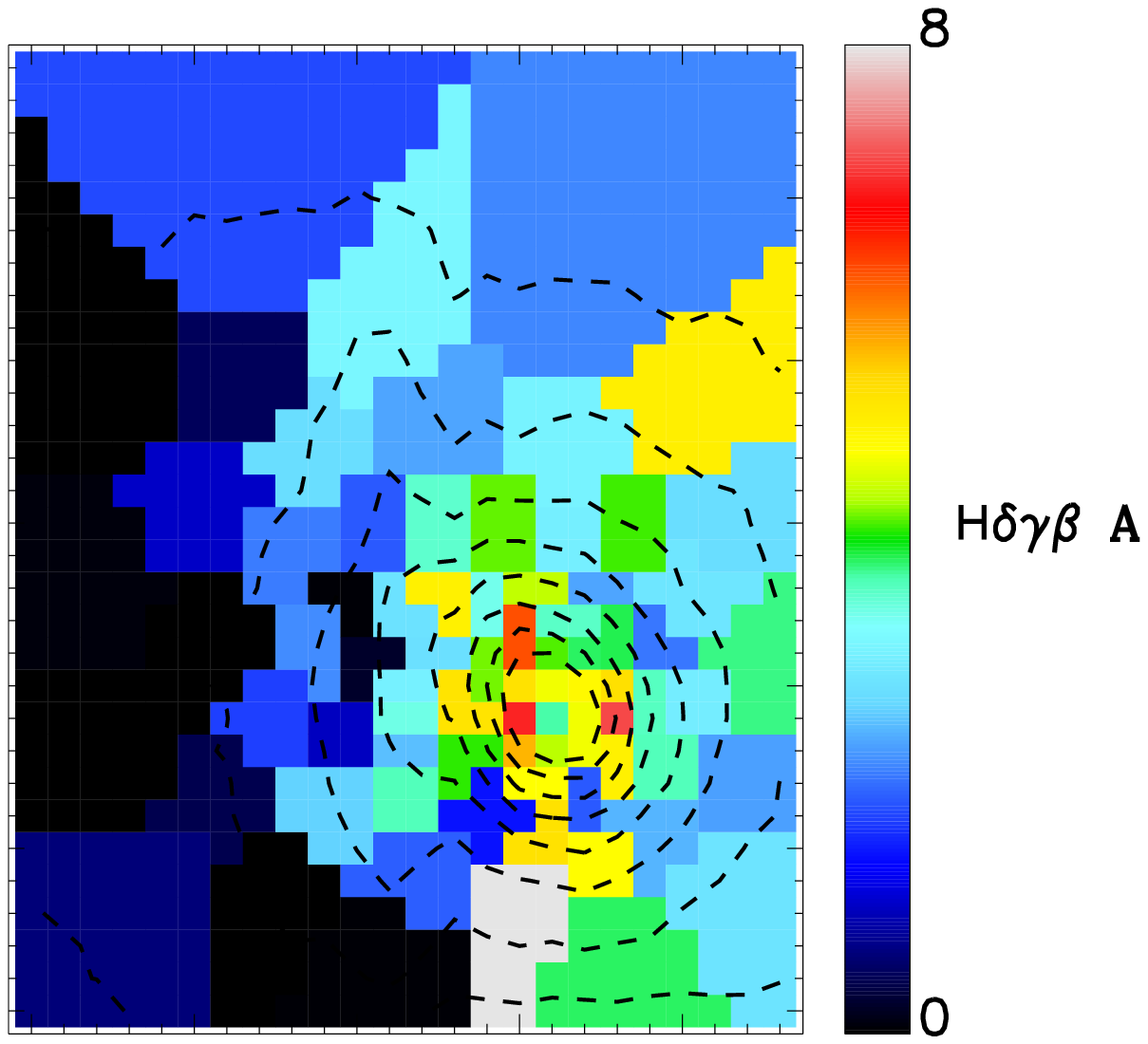}
      \includegraphics[height=4.0cm, angle=90, trim=-60 -30 0 100]{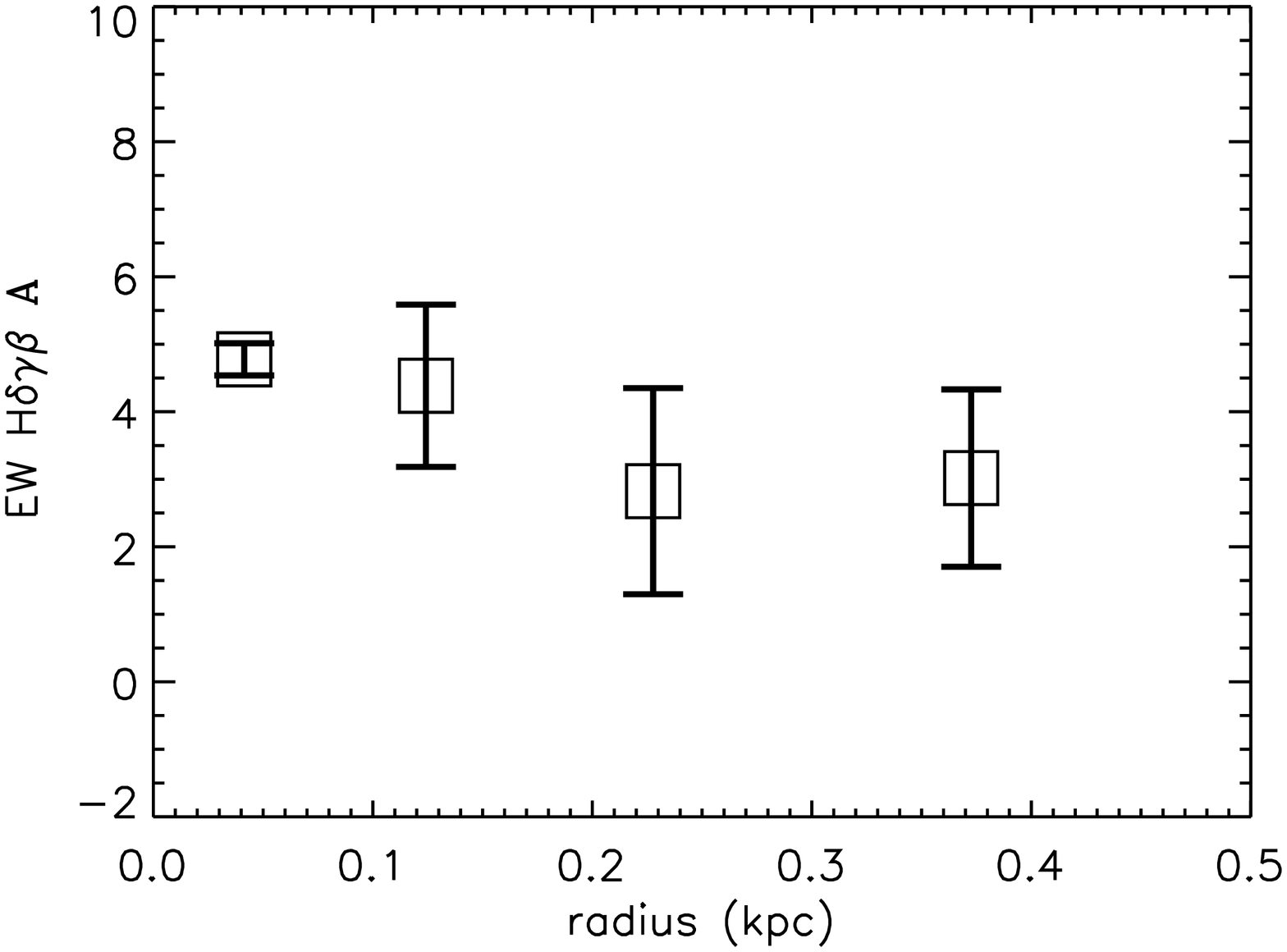}
      \vspace{-0.9cm}
    \end{minipage}
    \begin{minipage}{0.95\textwidth}
      \includegraphics[width=2.9cm, angle=0, trim=0 -50 0 0]{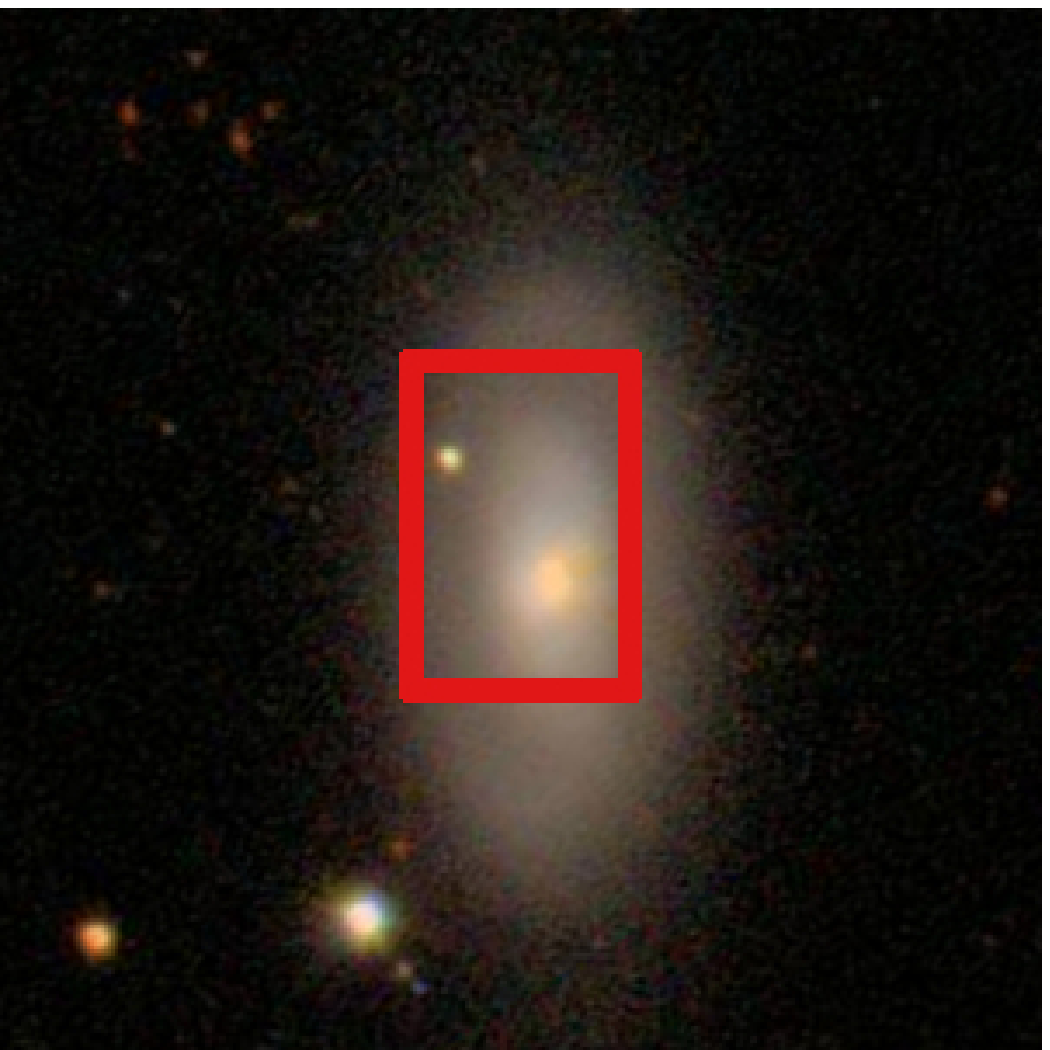}
      \includegraphics[height=4.0cm, angle=0, trim=70 0 0 0]{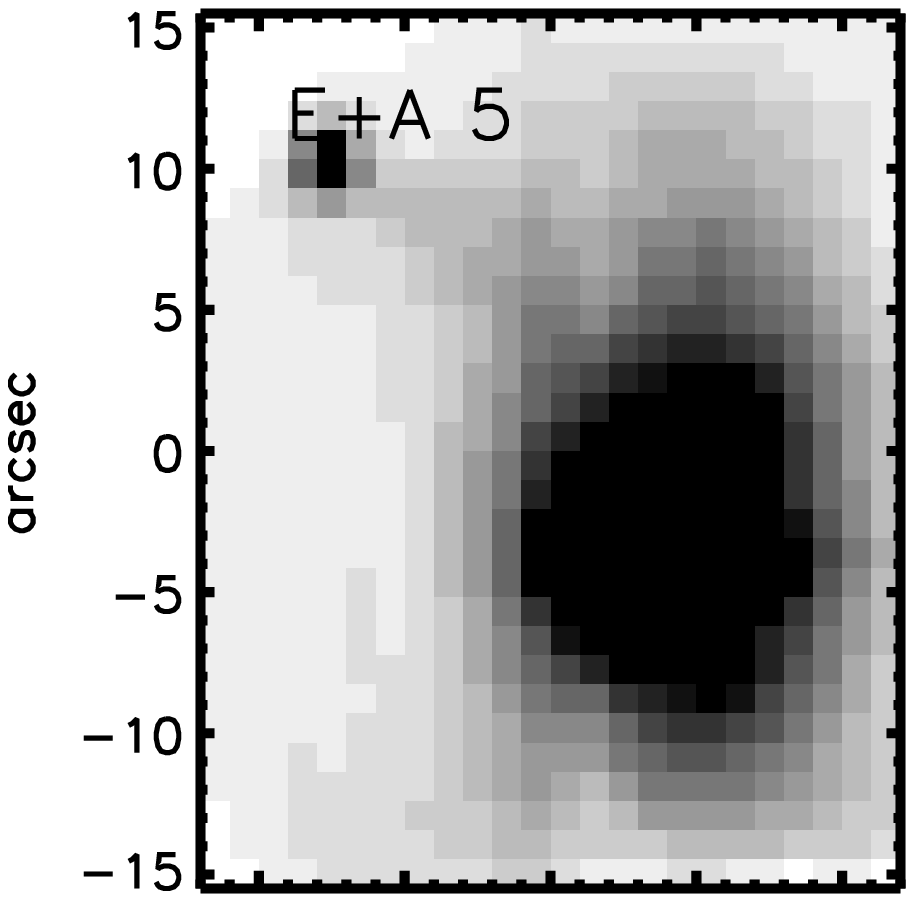}
      \includegraphics[height=3.6cm, angle=0, trim=210 -20 0 0]{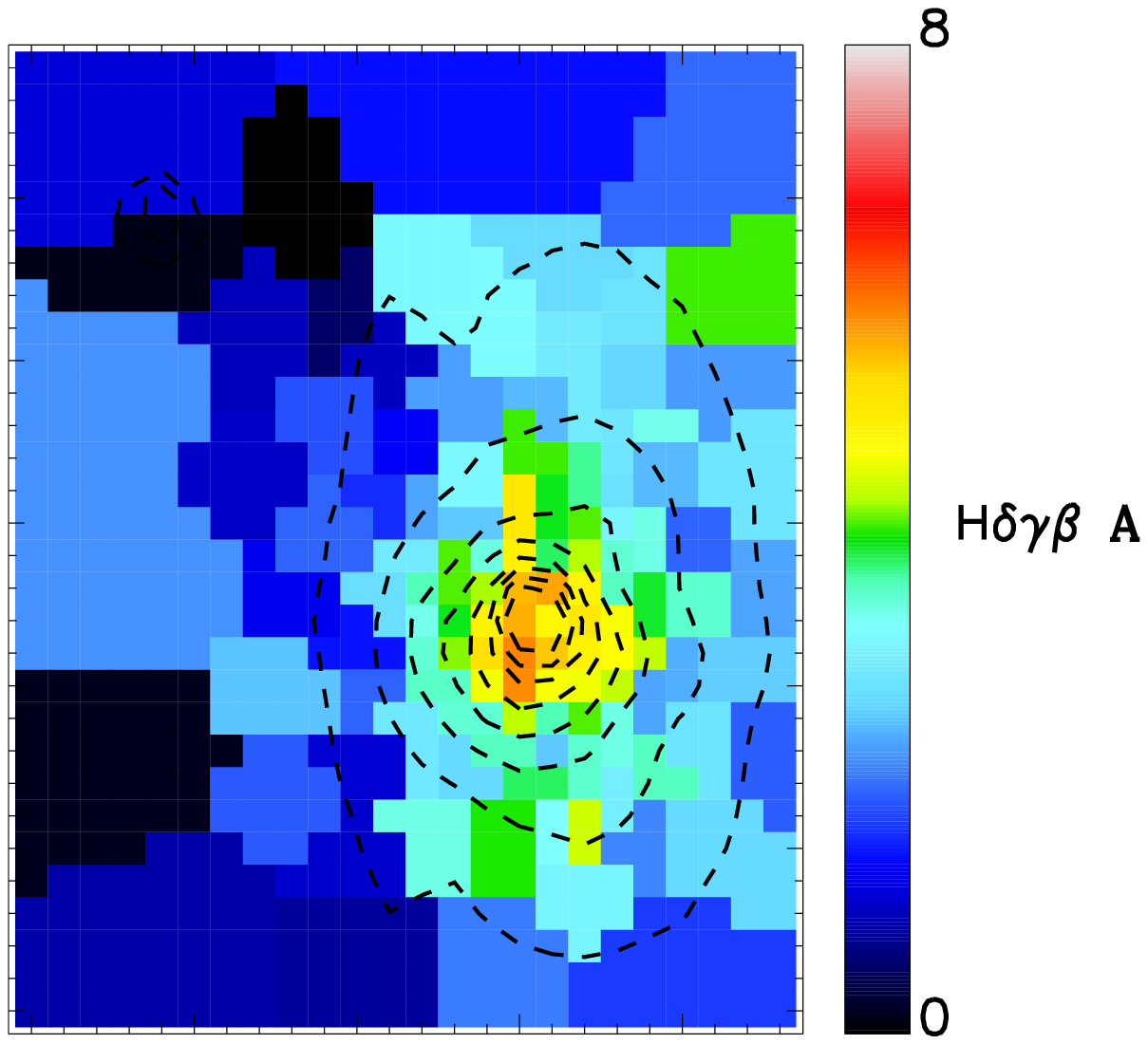}
      \includegraphics[height=4.0cm, angle=90, trim=-60 -30 0 100]{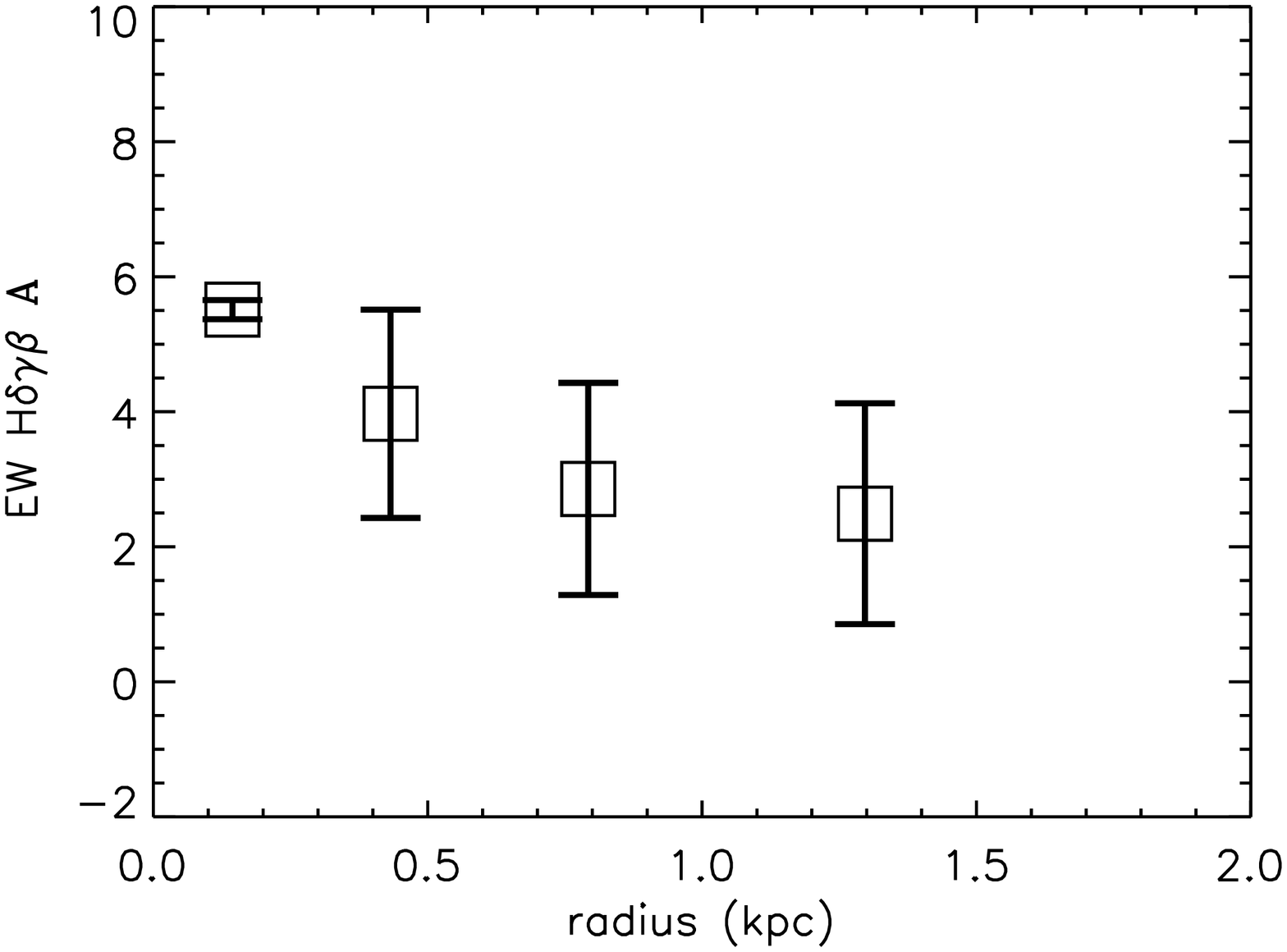}
      \vspace{-0.9cm}
    \end{minipage}
    \begin{minipage}{0.95\textwidth}
      \includegraphics[width=2.9cm, angle=0, trim=0 -50 0 0]{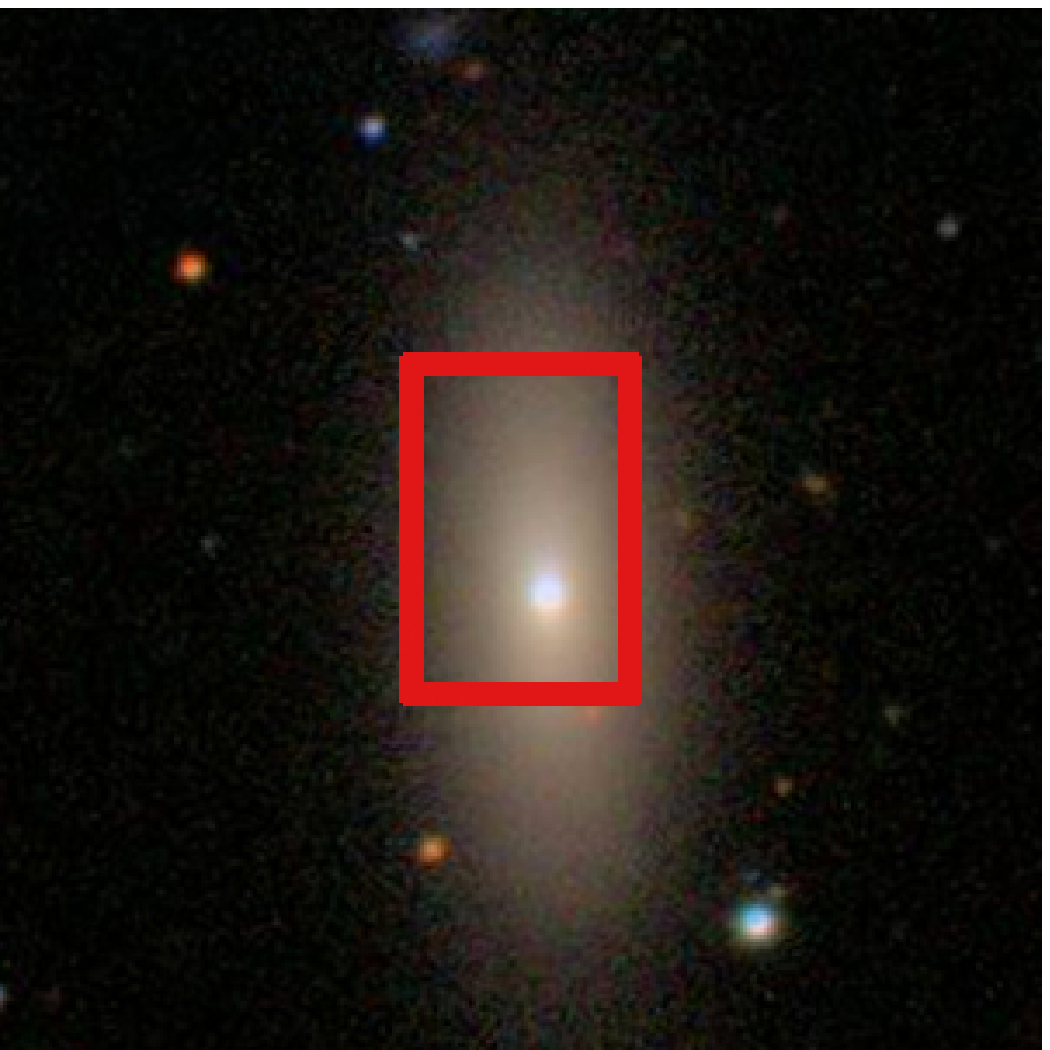}
      \includegraphics[height=4.0cm, angle=0, trim=70 0 0 0]{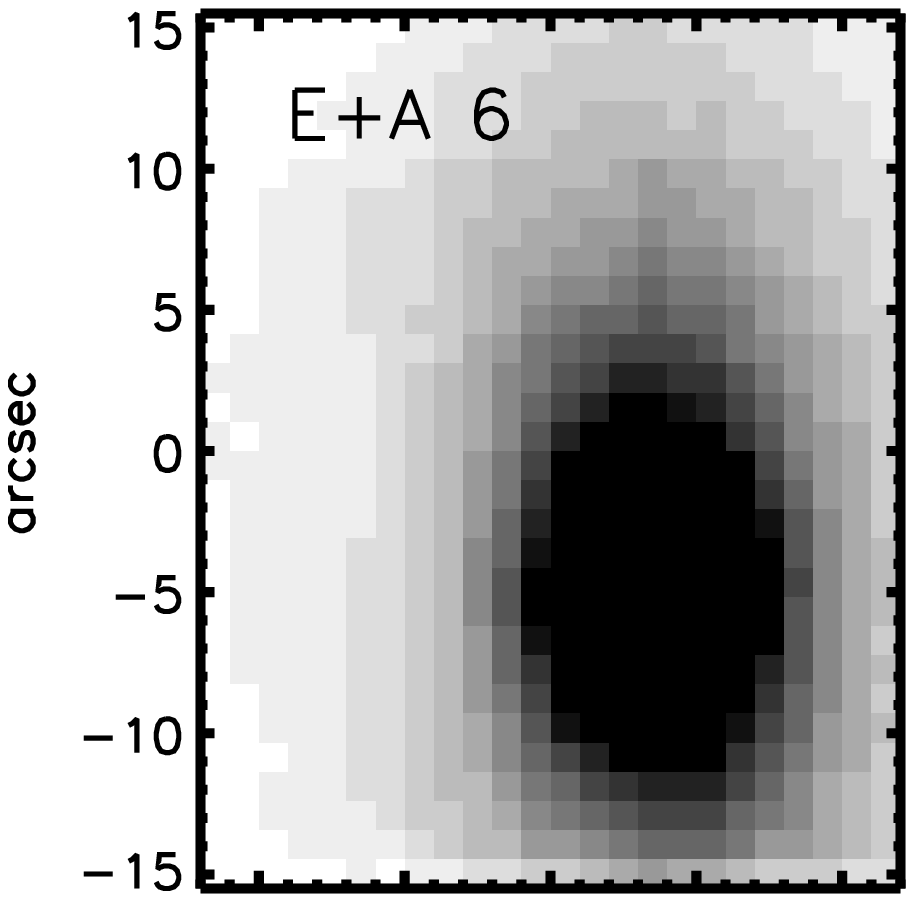}
      \includegraphics[height=3.6cm, angle=0, trim=210 -20 0 0]{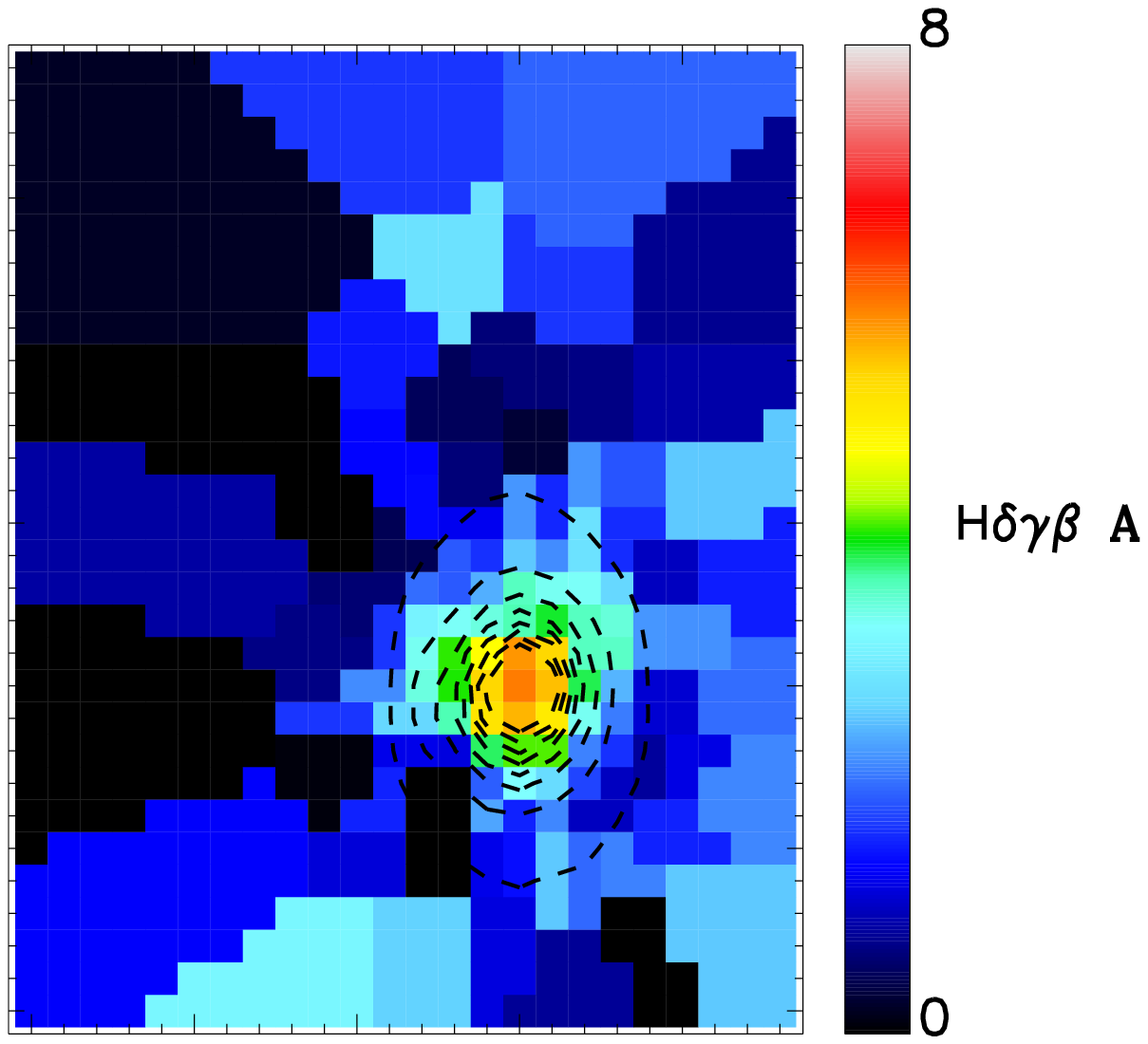}
      \includegraphics[height=4.0cm, angle=90, trim=-60 -30 0 100]{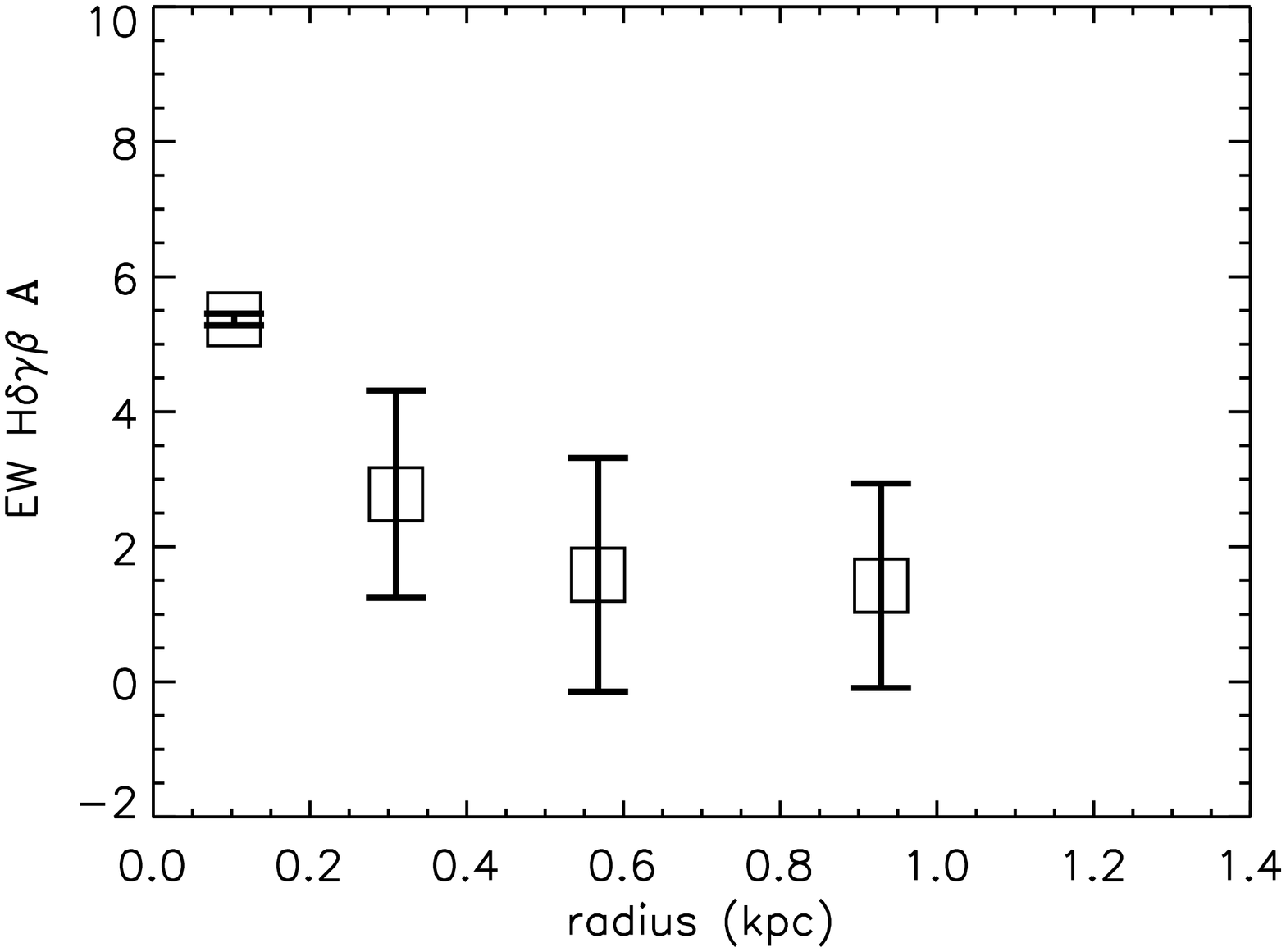}
      \vspace{-0.9cm}
    \end{minipage}
    \begin{minipage}{0.95\textwidth}
      \includegraphics[width=2.9cm, angle=0, trim=0 -50 0 0]{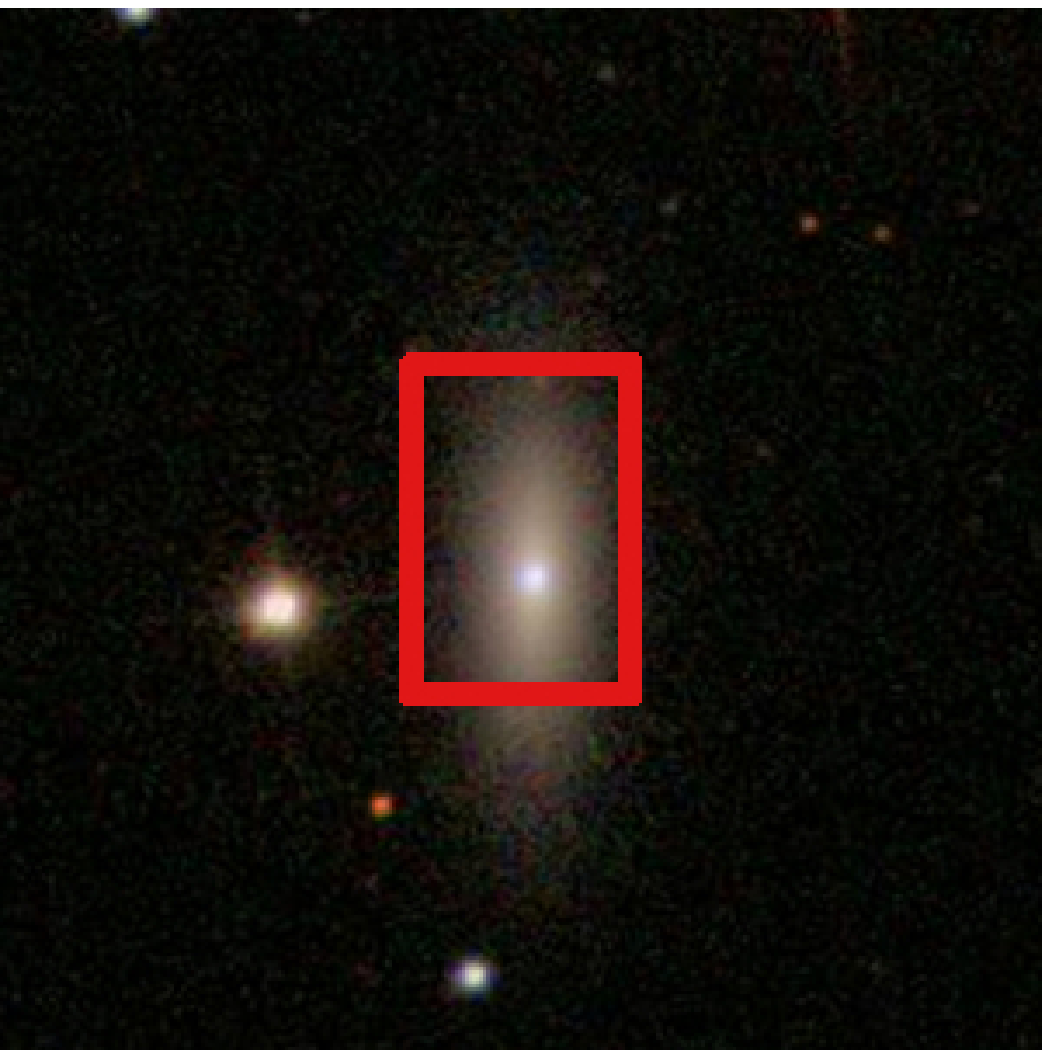}
      \includegraphics[height=4.0cm, angle=0, trim=70 0 0 0]{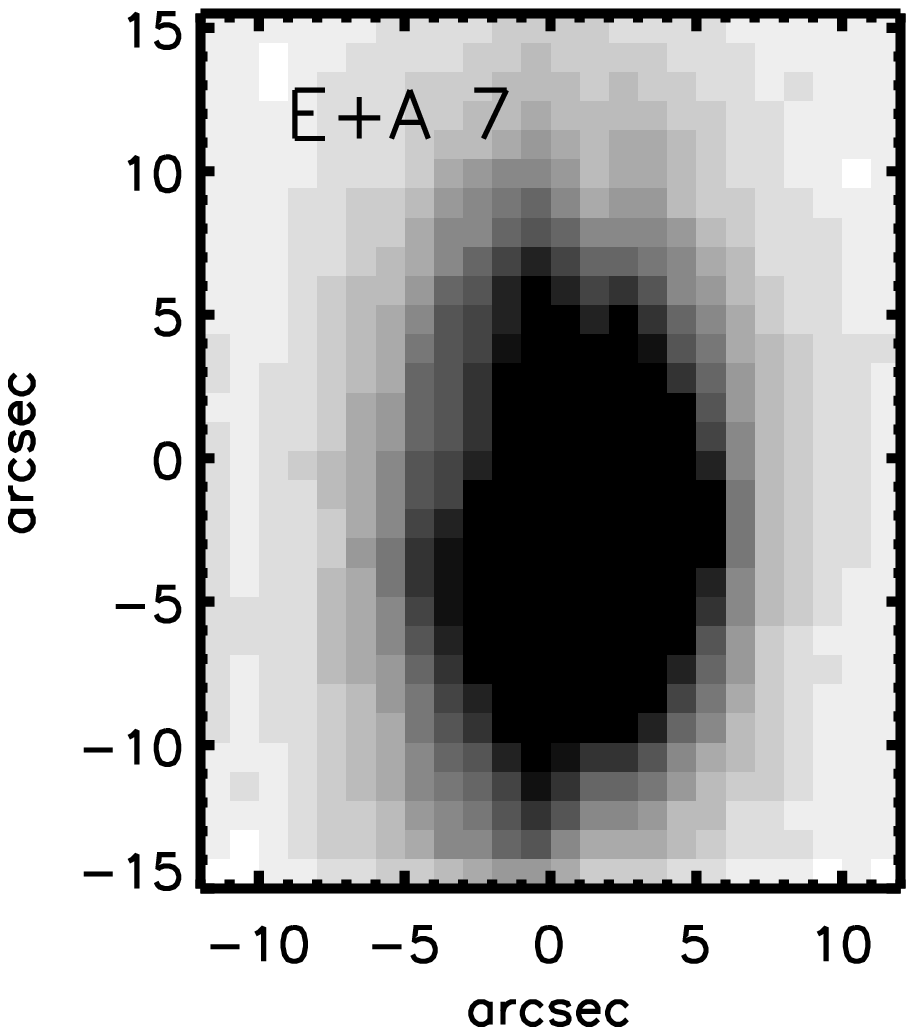}
      \includegraphics[height=3.6cm, angle=0, trim=210 -20 0 0]{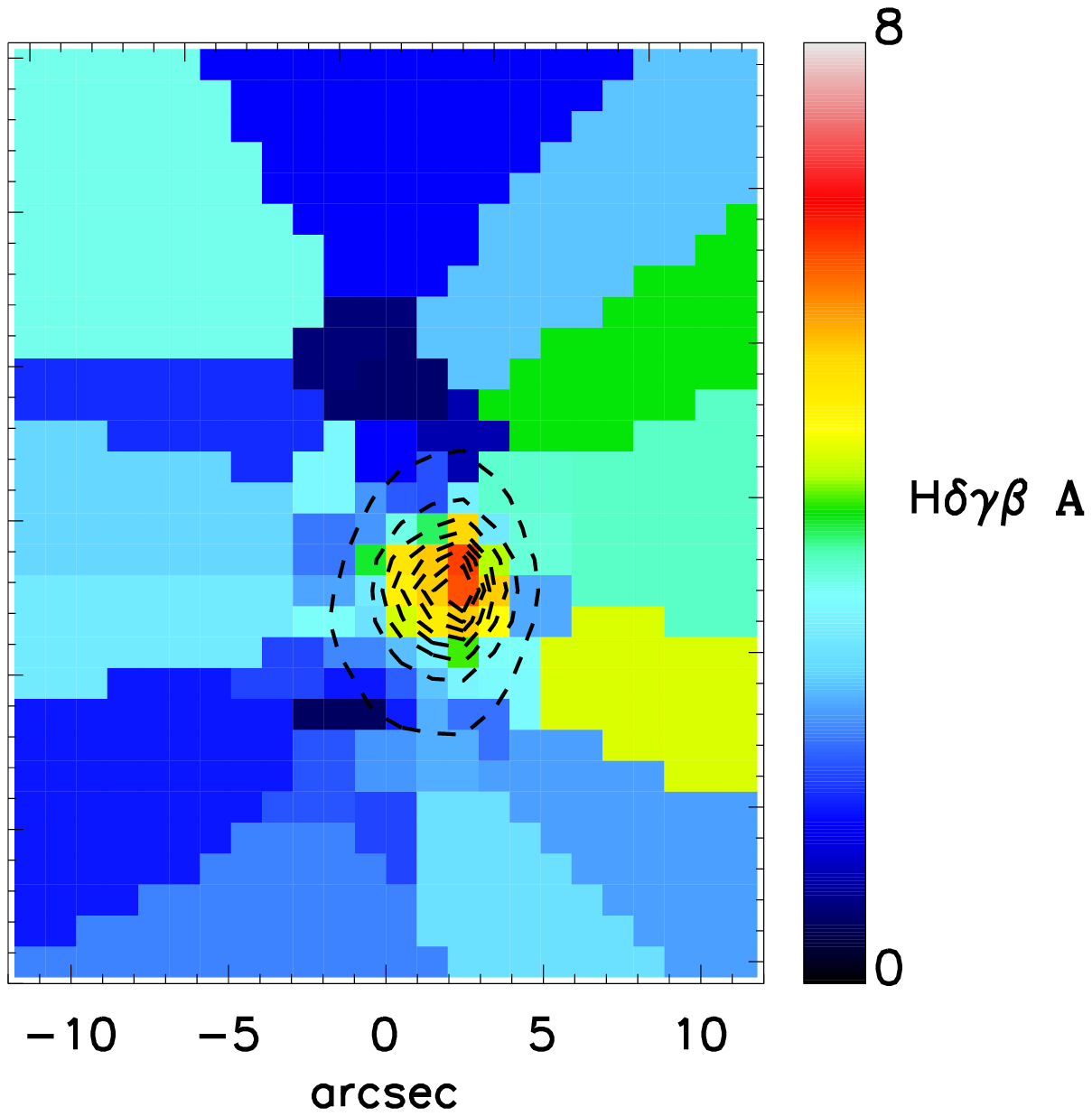}
      \includegraphics[height=4.0cm, angle=90, trim=-60 -30 0 100]{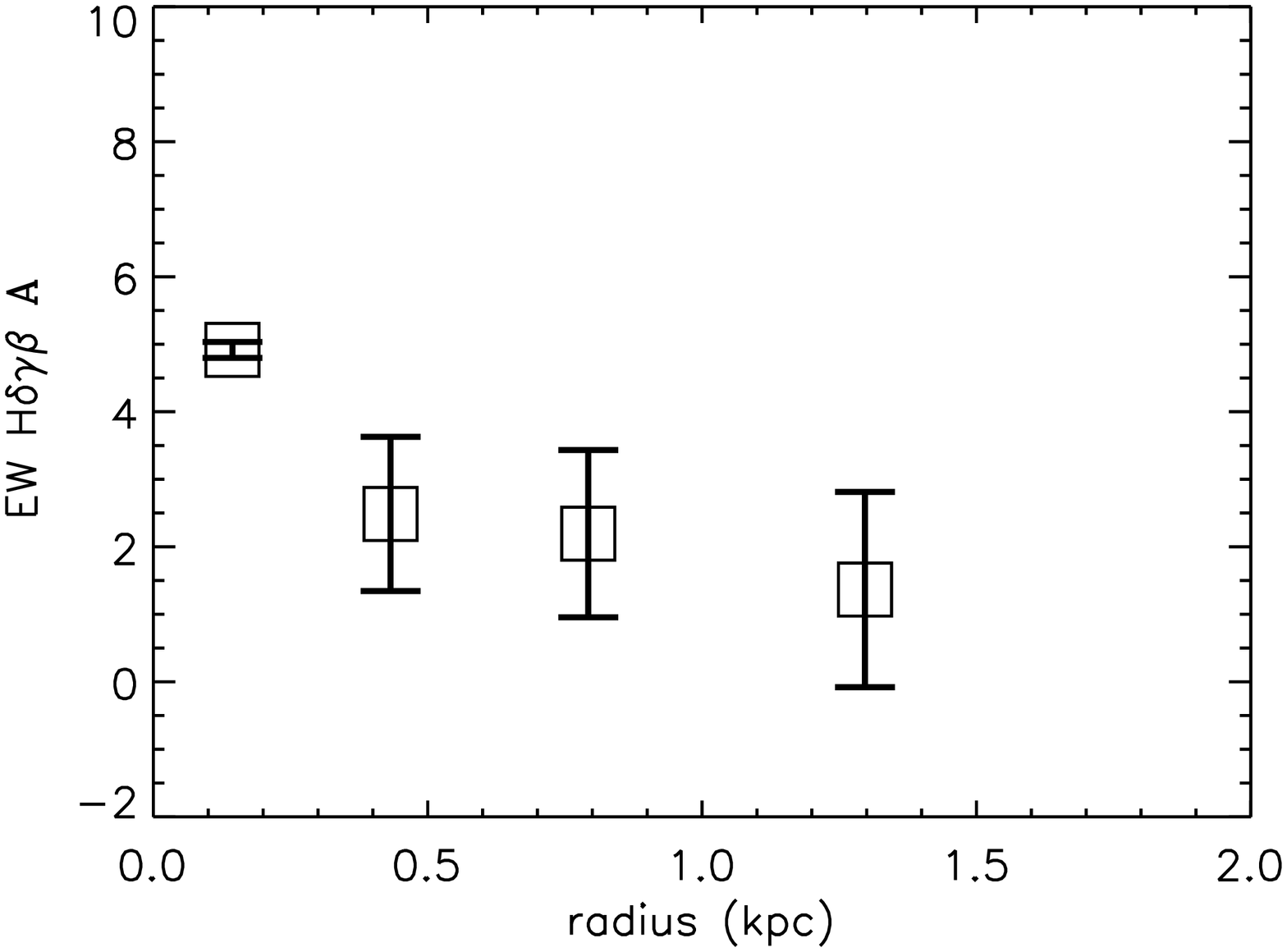}
\vspace{-0.3cm}
    \end{minipage}
\end{center}
    \caption{One object per row -- from left to right: SDSS image with the size, orientation and position of
the WiFeS field-of-view superimposed; reconstructed image by collapsing the WiFeS data in the spectral direction;
 Balmer line equivalent width maps for ${\rm H}\delta + {\rm H}\gamma + {\rm H}\beta$. The overlaid contours are the continuum light profiles; radial Balmer line gradients
based on azimuthal binning of the WiFeS spectra. Note that the horizontal axis scale in the final column changes. }
    \label{fig:images}
\end{figure*}

In the third column of Fig. \ref{fig:images} we show the equivalent width maps for the
H$\delta$, H$\gamma$ and H$\beta$ lines combined as a straight average. The trends in these maps are also
present in each of the individual line maps but combining the lines results in greater signal-to-noise. The equivalent width maps
can be compared with the distribution of overall galaxy light shown in column 2 by collapsing
the light through the original unbinned IFU elements along the wavelength direction and column 1
which shows an SDSS colour image with the IFU field-of-view superimposed. {\it In every case except 
E+A 1 there is strong enhancement in the Balmer line absorption equivalent width concentrated
in the galaxy core on scales $<1$\,kpc}. E+A 1 has a uniform very strong H$\delta$ signature
spread across the entire extent of the galaxy. It also differs to the rest of the sample
in other respects: it is the only irregular system and easily the bluest 
galaxy in the sample. As mentioned in Table \ref{tab:targets} E+A 2 (NGC 3156) is a member of the  Spectroscopic Areal Unit
For Research on Optical Nebulae (SAURON) sample \citep{dezeeuw02} enabling us to compare our Balmer absorption line map with the 
H$\beta$ map of \citet{kuntschner10} and we note that the maps are in good agreement.
In the last column of Fig. \ref{fig:images} we have azimuthally binned
the data before measuring line indices to construct radial Balmer line gradients. The
Balmer line enhancement seen in the two dimensional maps are seen as negative Balmer line
radial gradients. These radial profiles of the Balmer line equivalent width are 
similar to those predicted from merger models c.f. Fig. 9 of \citet{pracy05}. The observed
gradients are confined to a smaller radius than the models (which show enhancements with a 
negative gradient over $\sim$2\,kpc) but this is expected given the models are for significantly
brighter galaxies which should have a larger central starburst region.

\begin{figure*}
  \begin{center}
    \begin{minipage}{0.95\textwidth}
      \includegraphics[width=6.5cm, angle=0, trim=0 0 0 0]{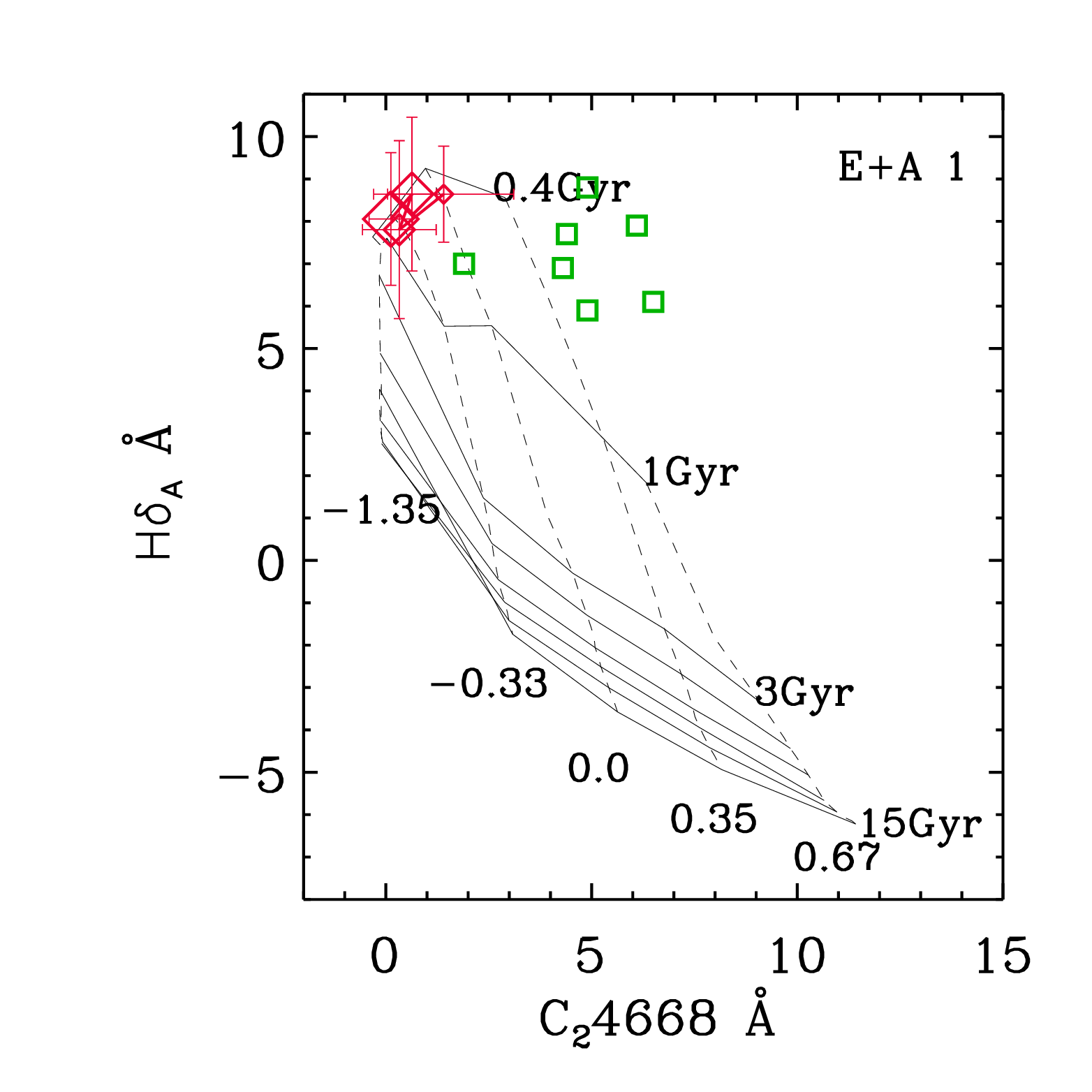}
\hspace{-1.0cm}
      \includegraphics[width=6.5cm, angle=0, trim=0 0 0 0]{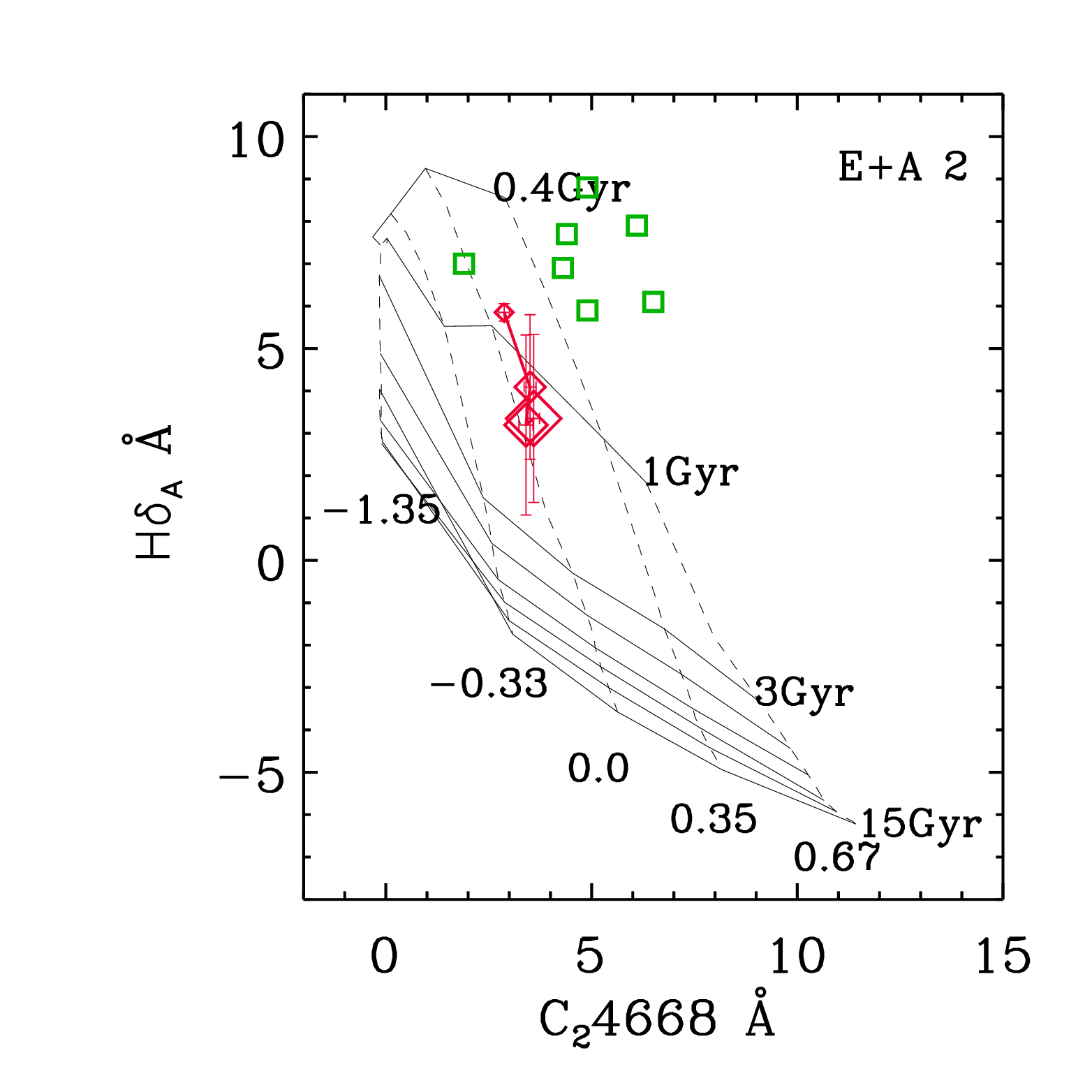}
\hspace{-1.0cm}
      \includegraphics[width=6.5cm, angle=0, trim=0 0 0 0]{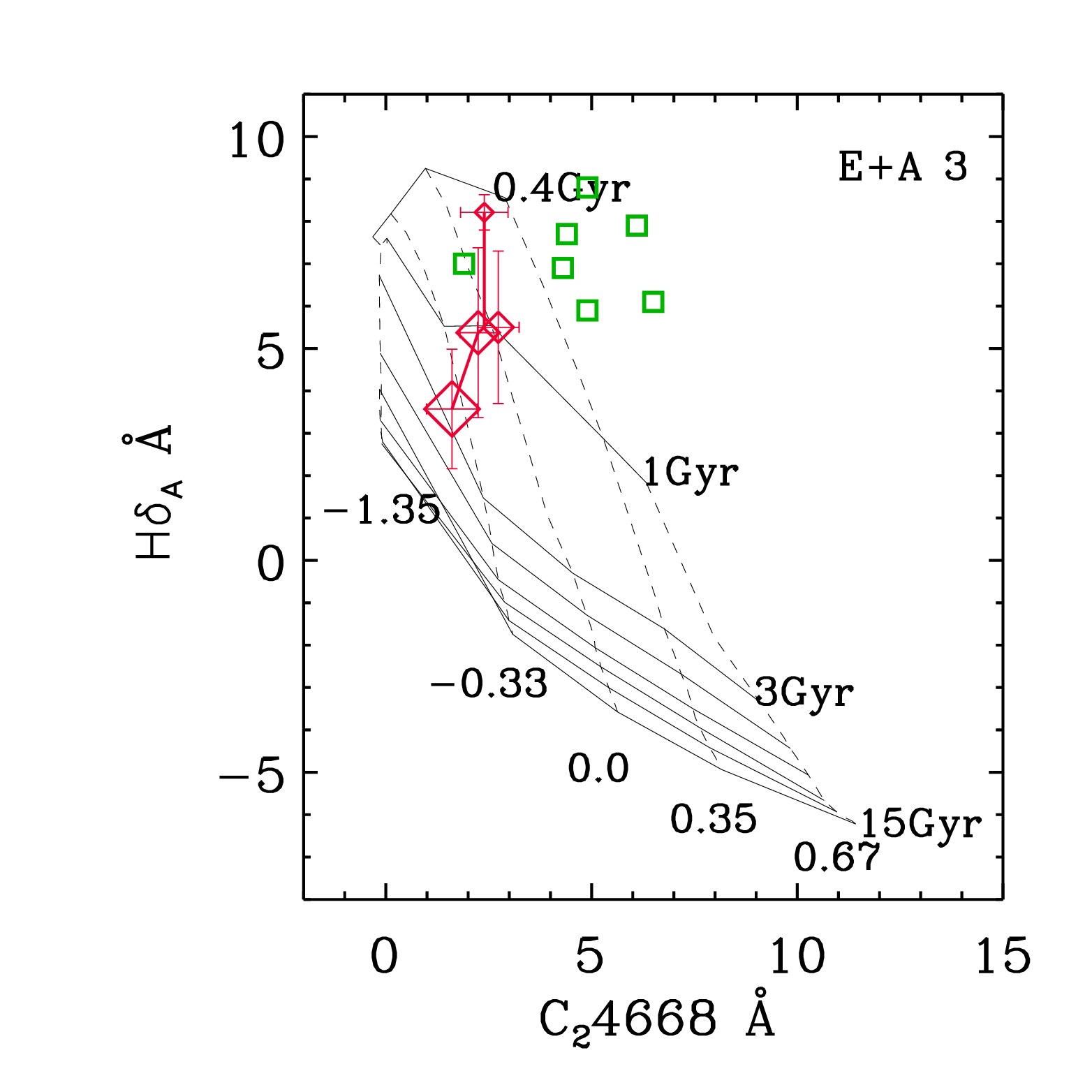}
      \vspace{-0.2cm}
    \end{minipage}

    \begin{minipage}{0.95\textwidth}
     \includegraphics[width=6.5cm, angle=0, trim=0 0 0 0]{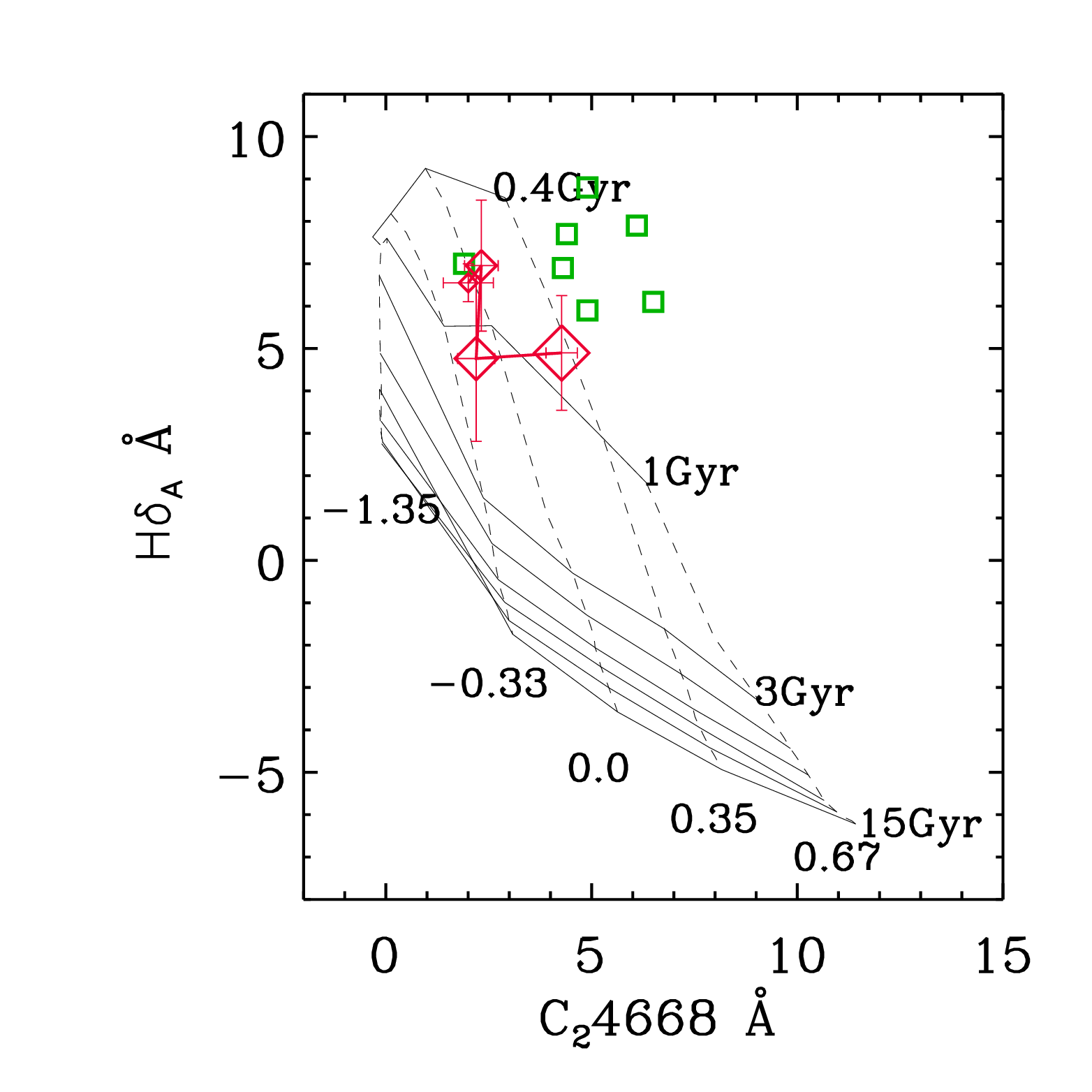}
\hspace{-1.0cm}
      \includegraphics[width=6.5cm, angle=0, trim=0 0 0 0]{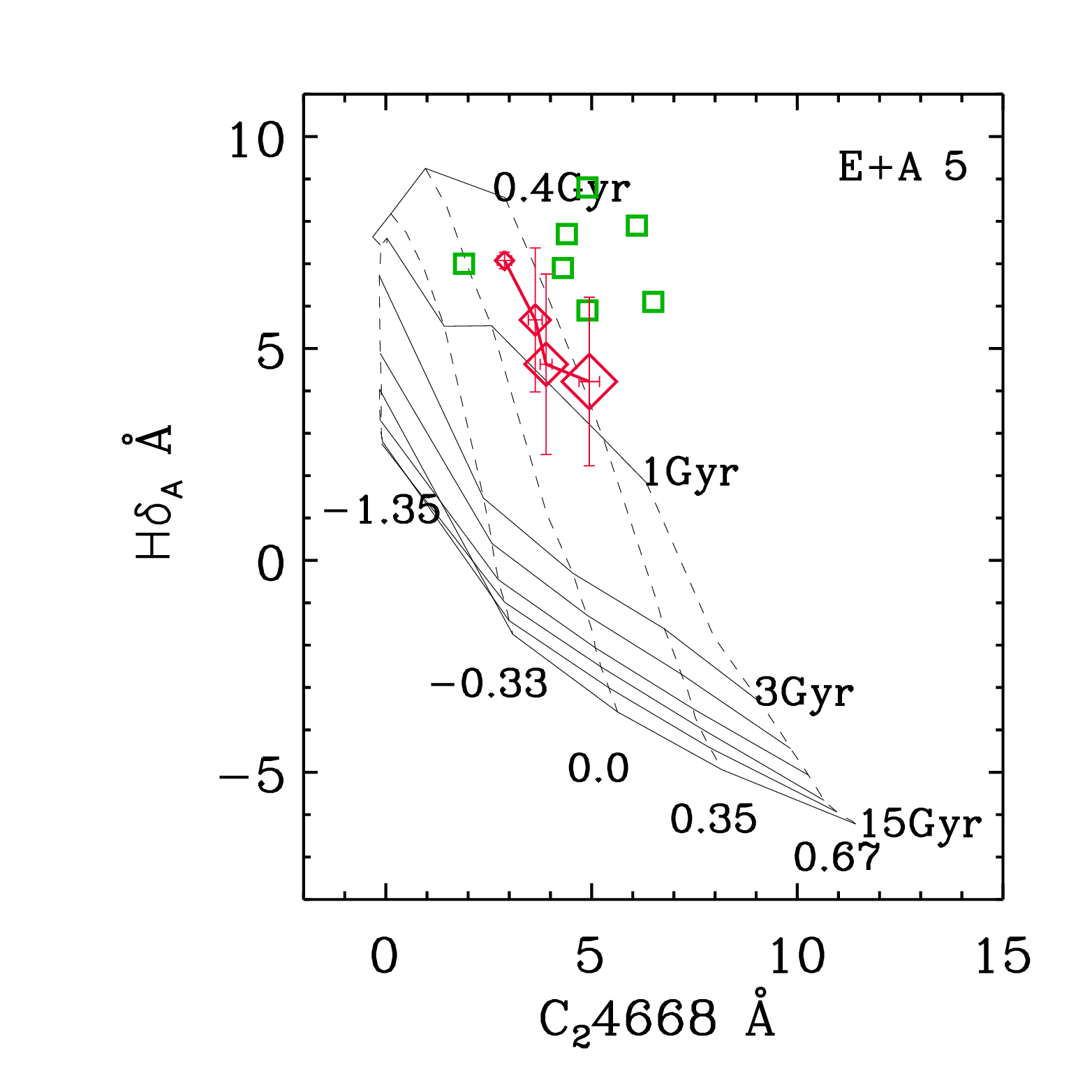}
\hspace{-1.0cm}
      \includegraphics[width=6.5cm, angle=0, trim=0 0 0 0]{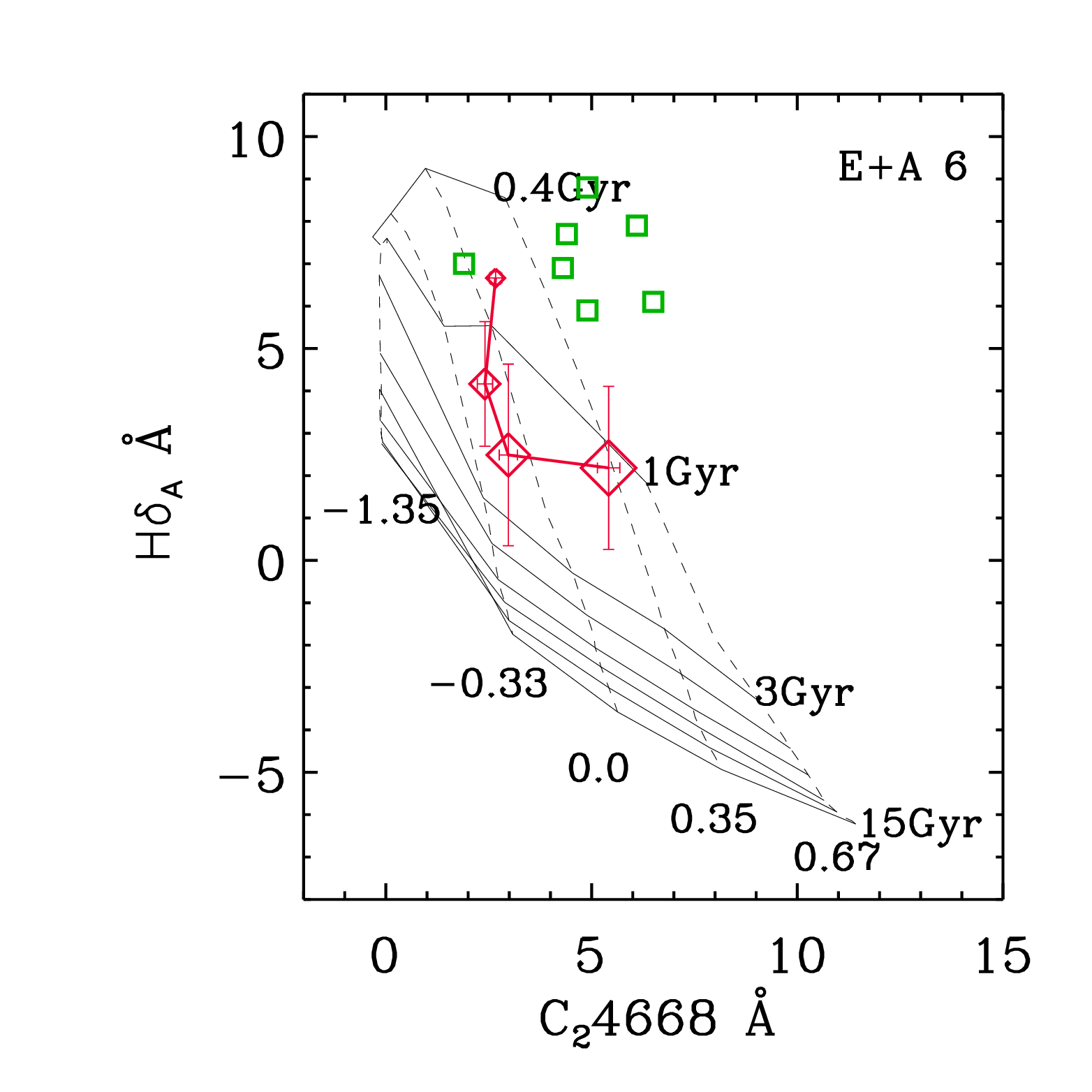}
\vspace{-0.3cm}
    \end{minipage}

    \begin{minipage}{0.95\textwidth}
      \includegraphics[width=6.5cm, angle=0, trim=0 0 0 0]{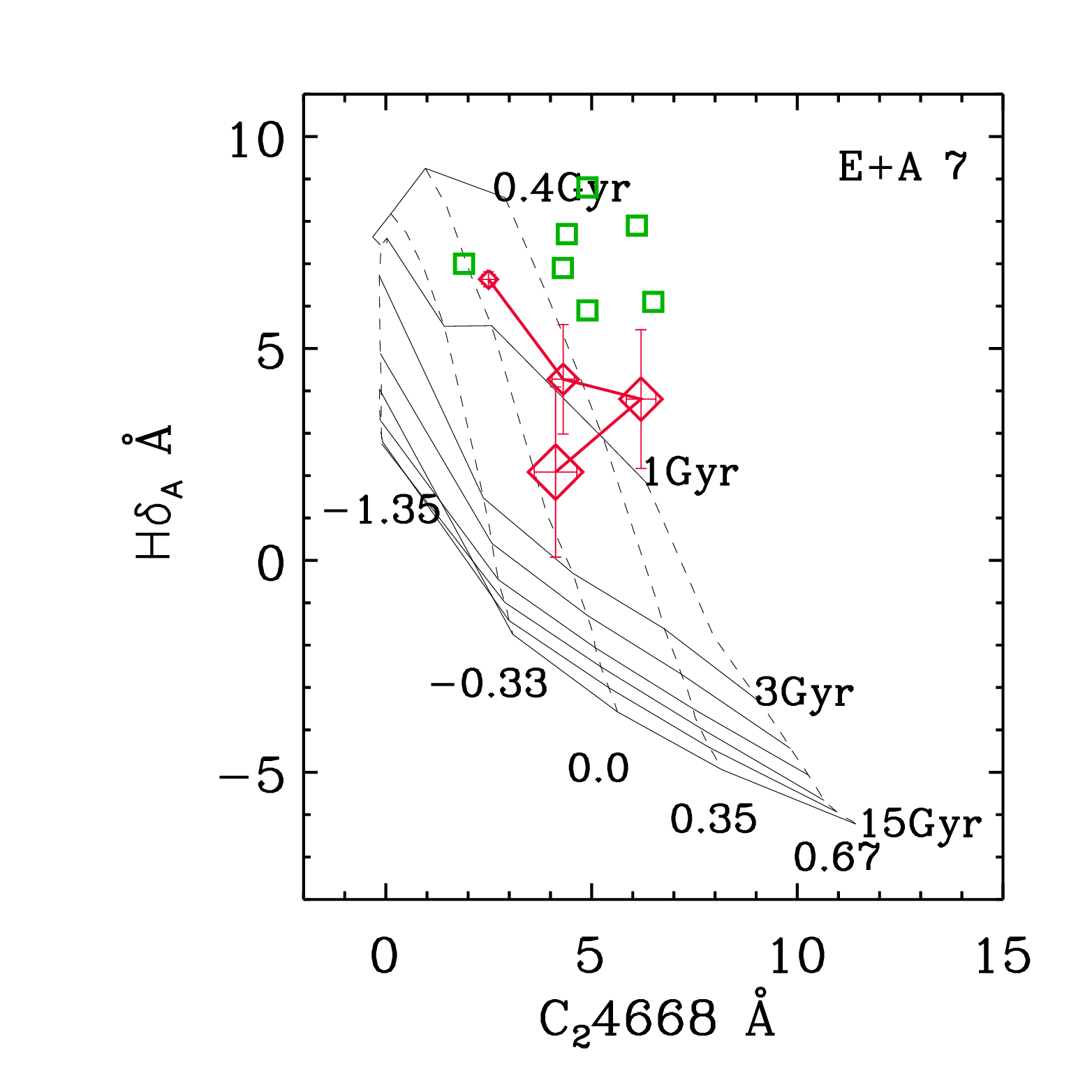}
\vspace{-0.3cm}
    \end{minipage}
\end{center}
    \caption{Age metallicity diagnostic diagrams using the H$\delta$ and C$_{2}4668$ lines. The radial annular equivalent width measurements are displayed as
{\it red diamonds} with symbol sizes increasing with galacto-centric radius. The {\it green squares} are the global values from the \citet{pracy09} sample.}
    \label{fig:lindia}
\end{figure*}

In Figure \ref{fig:lindia} we show an age-metallicity diagnostic diagram for our galaxies using the H$\delta$ and
C$_{2}4668$ lines. The line ratios are shown for the radial data with the smallest {\it red diamond} representing the innermost radial point and increasing radii
represented by increasing larger symbol sizes. The age-metallicity grids are from the single stellar population models of \citet{thomas03,thomas04} and assume solar
abundance ratios. The Balmer line radial gradients can be seen as luminosity weighted age gradients in this plot. In general, the smaller inner radial data points have 
younger ages than those at larger radii. The inner radial regions have ages $\lesssim 1$\,Gyr consistent with what is expected for E+A galaxies and similar to the global
ages of the \citet{pracy09} E+A sample which are overlaid as {\it green squares}.

\subsection{Mass fractions}
We estimated the mass in recently formed stars from the 
best fitting stellar population templates as described in Section 3.4. 
In this case we constructed a spatially integrated spectrum for each of 
our sample by summing over all spaxels within a circular aperture of 
radius 11 arcseconds. We fitted the spectra with a combination of a single 
old stellar population of age $\sim 13$\,Gyr and a set of young stellar 
population models with ages ranging from 100\,Myr to $\sim 1$\,Gyr (all 
model stellar populations assumed solar metallicity) and calculated the 
mass fraction of the young stellar population. This results in burst mass 
fraction estimates of between $\sim 6$ and 12\,per cent for all 
galaxies in our sample except E+A 1 which is best fitted with 100\,per cent young 
mass fraction i.e. we are unable to quantify the old stellar population 
in this galaxy. These mass fractions correspond to the young stellar population 
contributing between 51\,per cent and 63\,per cent of the light between 4500\AA\, and 5500\AA.
The calculated burst mass fractions are not sensitive to the 
precise age chosen for the old stellar population since the spectra of 
old stellar populations do not change rapidly with age. For the young 
population there are degeneracies between age and burst fraction in 
the sense that decreasing the age of the stellar population has a similar
effect as increasing the burst fraction. To illustrate this we fitted 
again allowing the combination of only two models. 
First we fitted an old $\sim 13$\,Gyr model and a young stellar population 
of age 890\,Myrs which resulted in the mass fraction estimates increasing to between
 11\,per cent and 17\,per cent (E+A 1 again returns a 100\,per cent young mass 
fraction and is poorly fitted by this combination). We also fitted 
the combination of a $\sim 13$\,Gyr model and a younger 500\,Myr template 
which resulted in the range of mass fractions decreasing to $\sim$3--4\,per 
cent (E+A 1 has a young mass fraction of $\sim$25\,per cent under these assumptions). 
These lower burst fractions by mass still contribute significantly to the overall light
(between 40 and 50\,per cent) since the younger stellar populations have a smaller mass-to-light
ratio. Reducing the age of the young fraction reduces the burst mass fractions further but results 
in progressively poorer fits (again the exception is E+A 1 which is better fitted by a younger
population). Overall, the recently formed stars account for of order 10\,per cent of the total 
stellar mass but contribute more than half the total light at $\sim$5000\AA.

\subsection{Aperture effects}
\begin{figure}
\hspace{-0.6cm}
      \includegraphics[width=6.8cm, angle=90, trim=0 0 0 0]{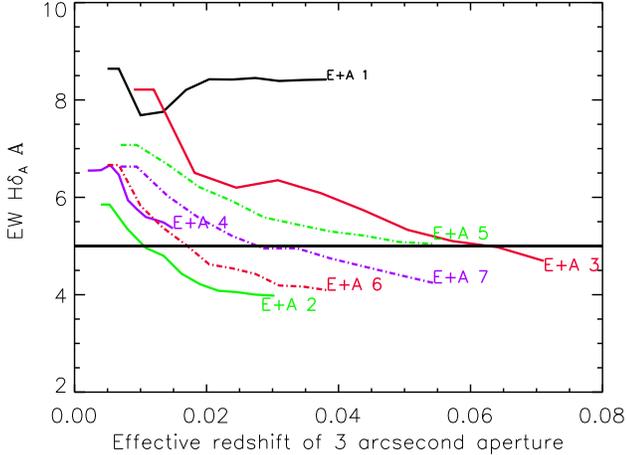}
\caption{\label{fig:aperture} Shows the effect of changing the redshift each galaxy is observed at on the H$\delta$ equivalent width measured by an SDSS fibre. We plot
H$\delta$ equivalent width measured in different apertures versus the redshift that galaxy would be at to make the aperture 3 arcseconds -- the diameter of the
SDSS fibres.}
\end{figure}

The presence of such strong radial gradients in the Balmer lines means the classification of
these objects are strongly dependent on aperture effects. The measured equivalent width of the Balmer
lines will depend on the physical scale probed which in turn depends on both the
distance of the object and the angular size subtended by the fibre. At a given distance a smaller fibre will measure a
larger absorption line strength since the light is dominated by the central region with a large equivalent width, as
the fibre size is increased the light becomes more contaminated by the outer regions where the absorption equivalent
width is small. Likewise, for a fixed aperture the Balmer line equivalent width measured will decrease with distance
as the angular size of the galaxy decreases and more of the outer regions of the galaxy are within the fibre aperture.
This means, independent of apparent magnitude constraints, galaxies that fit the equivalent width selection 
criteria at low redshift may not if they were observed at higher redshift. This is illustrated for our sample in Fig. \ref{fig:aperture}  
where we have measured the H$\delta$ equivalent width for different sized apertures. The horizontal axis shows the effective redshift the object would be at
to subtend a three arcsecond aperture (which is the fibre aperture of the SDSS) i.e. we have measured the H$\delta$ equivalent width in our IFU data with an
aperture that is the same physical size as a 3 arcsecond aperture would be at that redshift. This demonstrates how the H$\delta$ equivalent width measured
would decrease for our objects if they were observed at higher redshift. The horizontal {\it solid black line} indicates the H$\delta$ equivalent width selection
limit for our sample. All of our sample (except E+A 1 which does not display a radial gradient) would have H$\delta$ equivalent widths below the selection limit
if observed at $z\gtrsim 0.06$. 

\subsection{Emission lines}
As mentioned in Section 2.2, three of the galaxies have integrated spectra that have H$\alpha$ in emission. This emission
is weak, by selection, since we required EW(H$\alpha$)$>$-2.5\AA. In all three cases the ratio [NII]$\lambda 6583$/{\rm H}$\alpha \lambda 6563 \gtrsim$1
which is indicative of AGN activity rather than star formation \citep[e.g][]{baldwin81,veilleux87,kewley01}. However, as a result of our
selection of spectra with both weak or non--existent H$\alpha$ emission  and a stellar continuum with strong Balmer lines in absorption,
we therefore expect strong H$\alpha$ absorption in the stellar continuum underlying the emission line. Hence the  [NII]$\lambda 6583$/{\rm H}$\alpha \lambda 6563$ 
ratio is significantly overestimated. 

\begin{figure*}
  \begin{center}
    \begin{minipage}{0.95\textwidth}
\vspace{-0.2cm}
\hspace{1.2cm}
      \includegraphics[height=3.2cm, angle=0, trim=0 0 0 0]{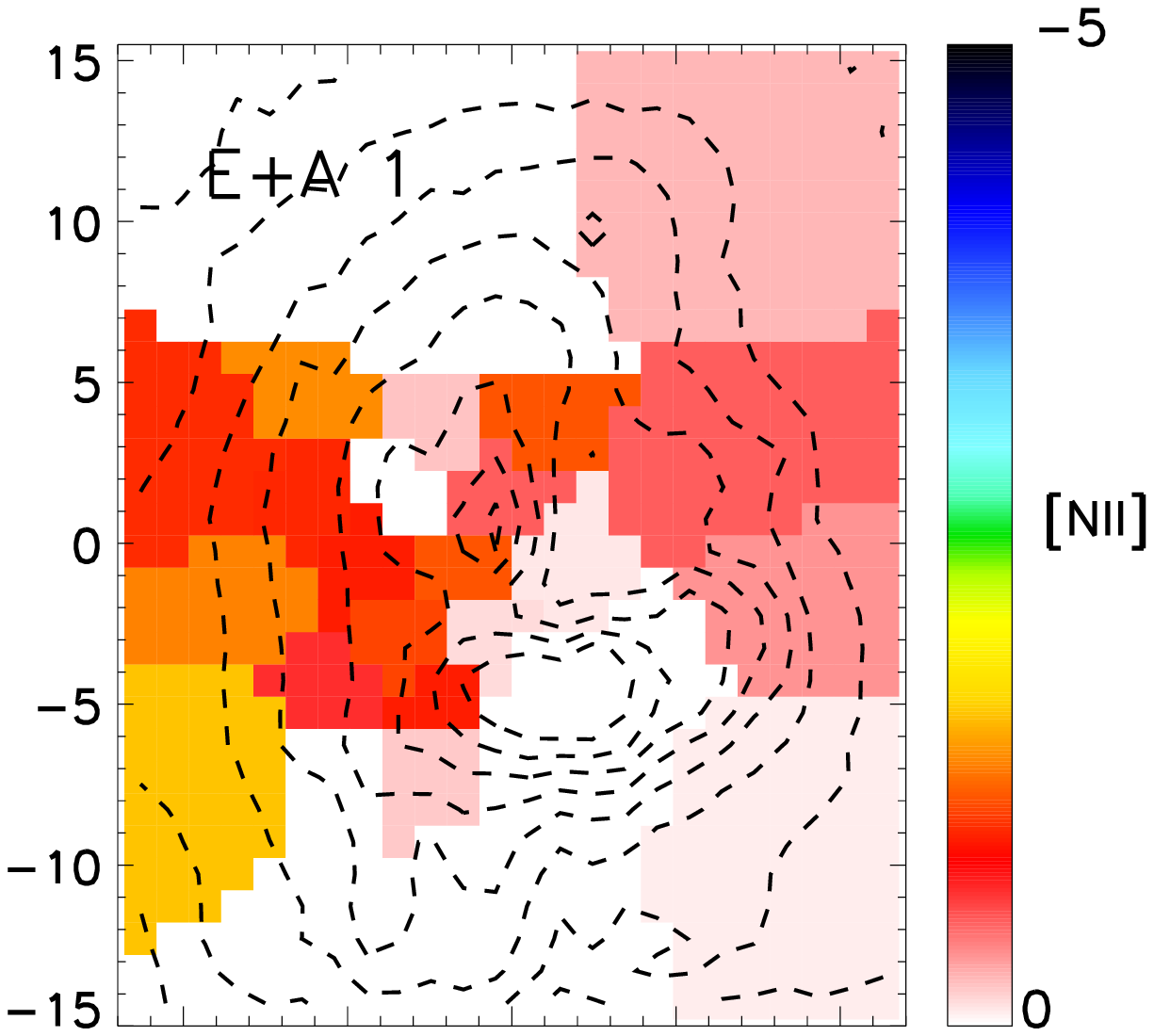}
\hspace{-1.8cm}
      \includegraphics[height=3.2cm, angle=0, trim=0 0 0 0]{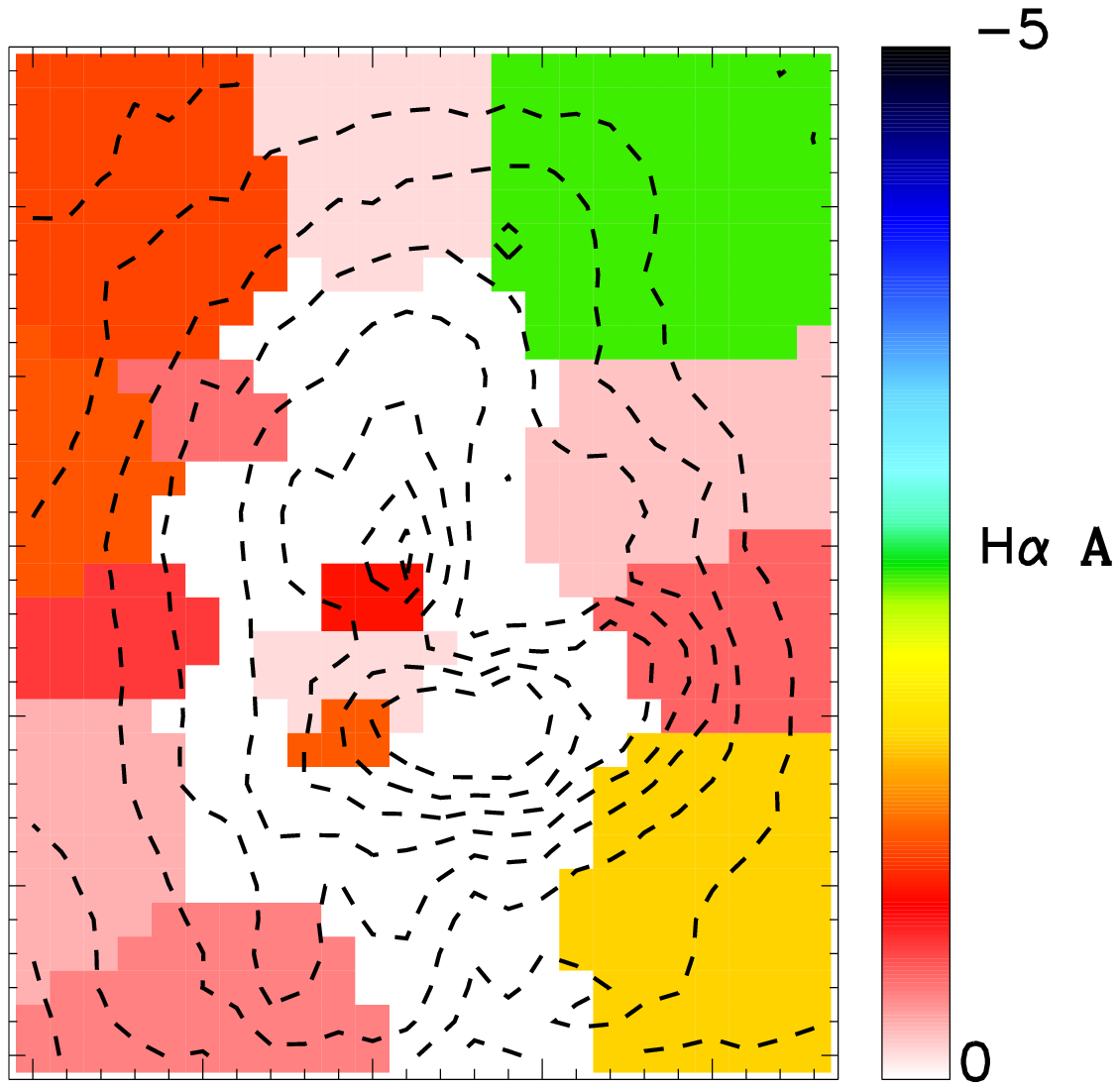}
\hspace{-1.8cm}
      \includegraphics[height=3.2cm, angle=0, trim=0 0 0 0]{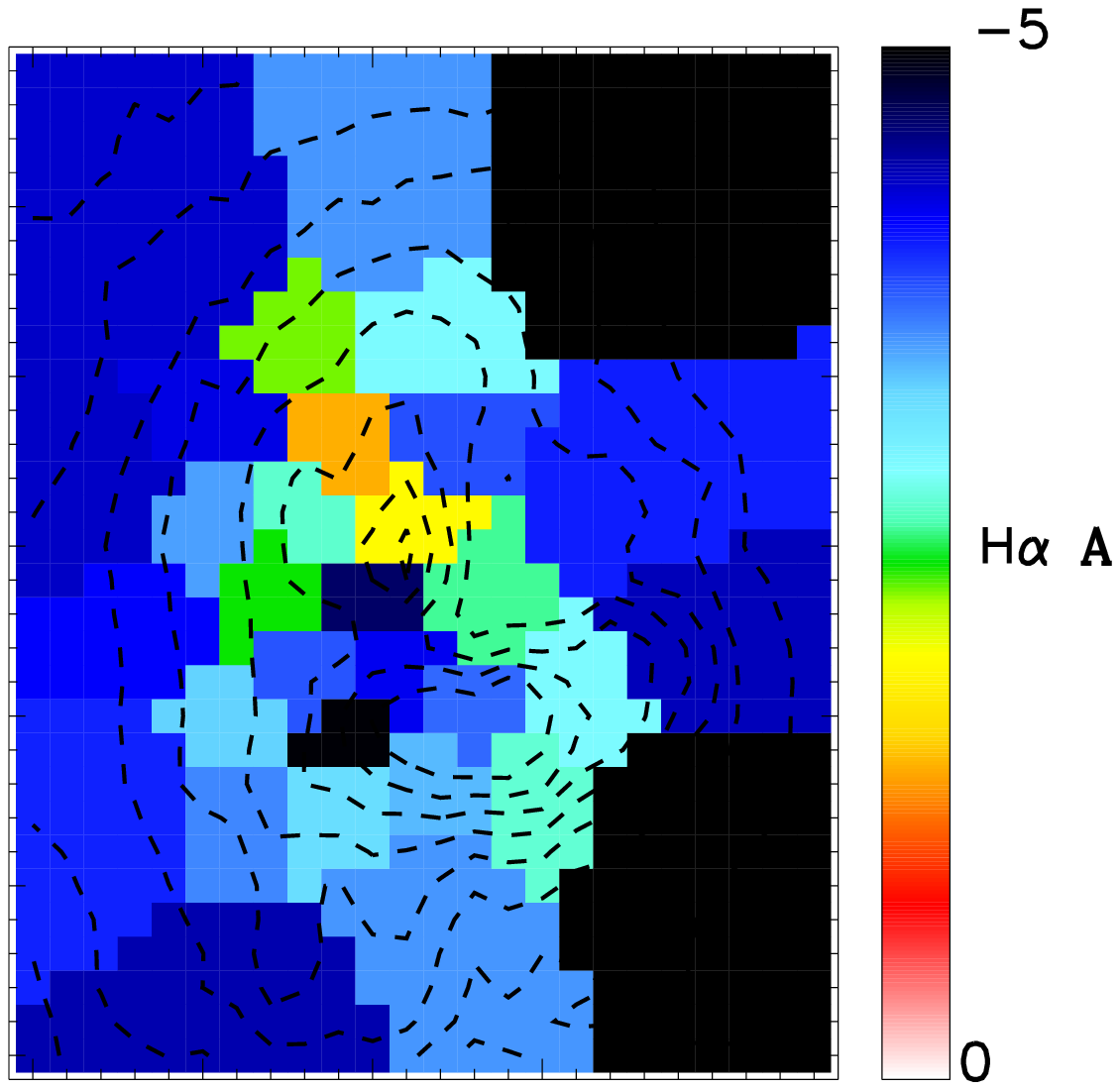}
\hspace{-1.8cm}
       \includegraphics[height=3.2cm, angle=0, trim=0 0 0 0]{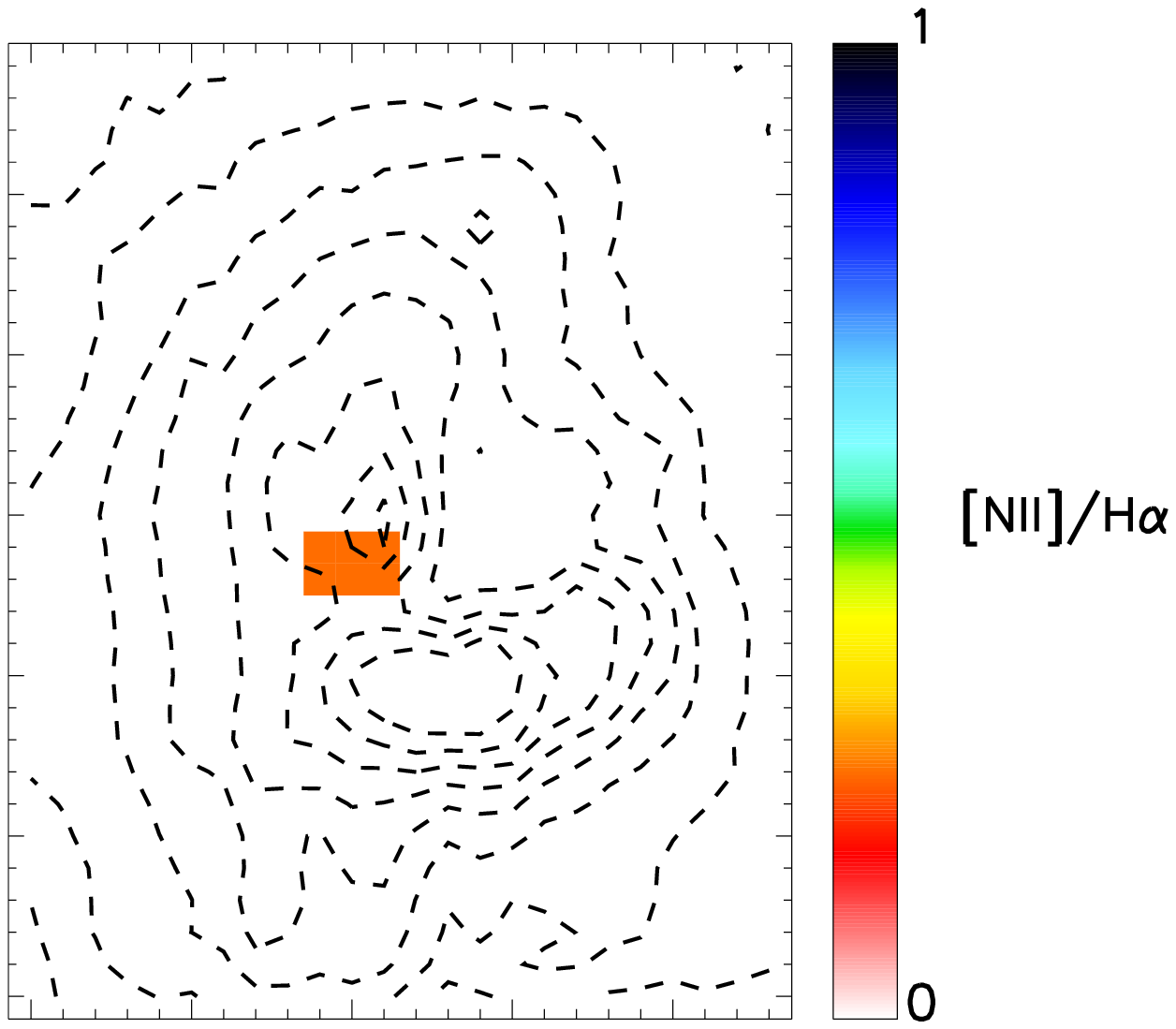}
    \end{minipage}
    \begin{minipage}{0.95\textwidth}
\hspace{1.2cm}
      \includegraphics[height=3.2cm, angle=0, trim=0 0 0 0]{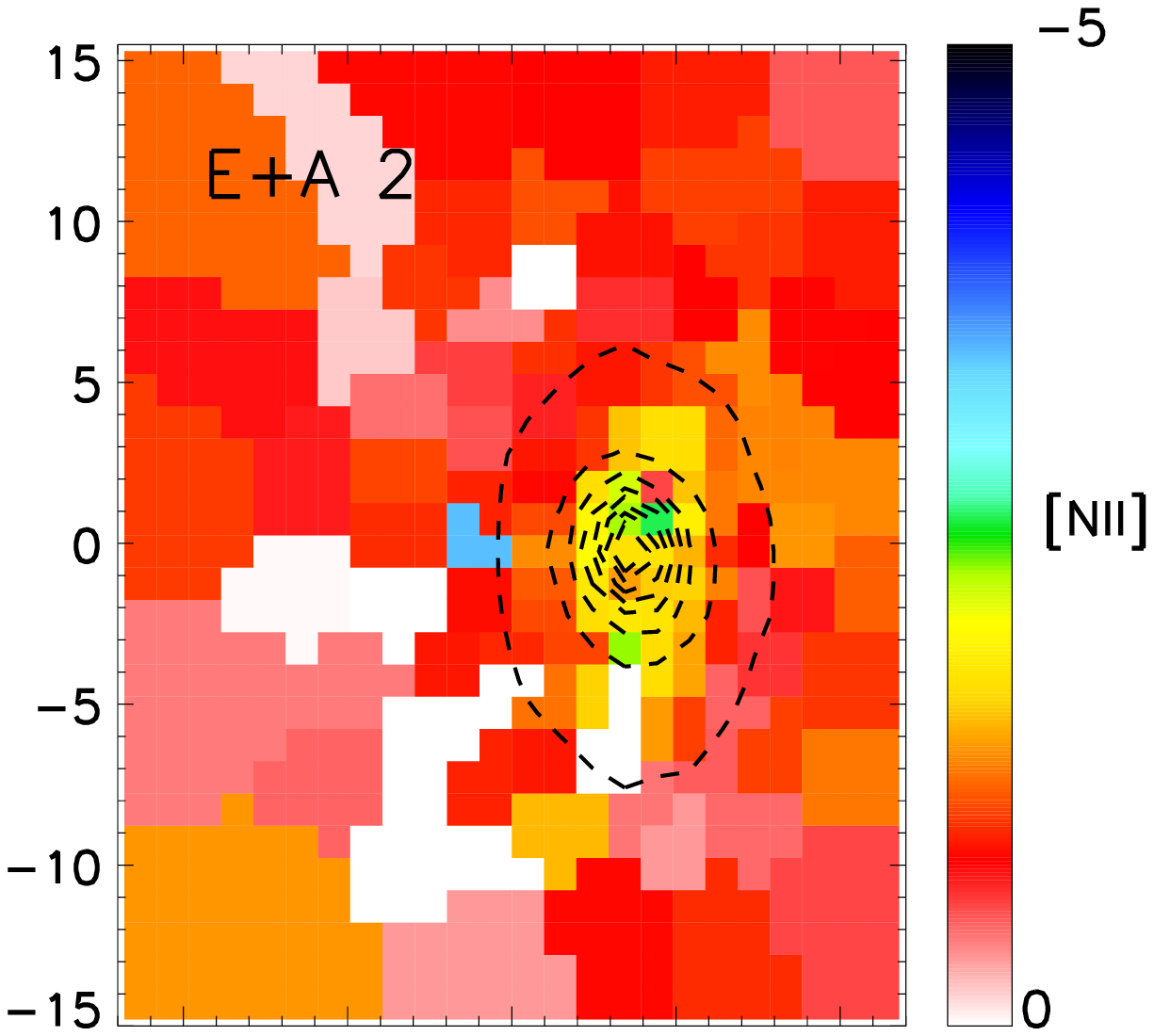}
\hspace{-1.8cm}
      \includegraphics[height=3.2cm, angle=0, trim=0 0 0 0]{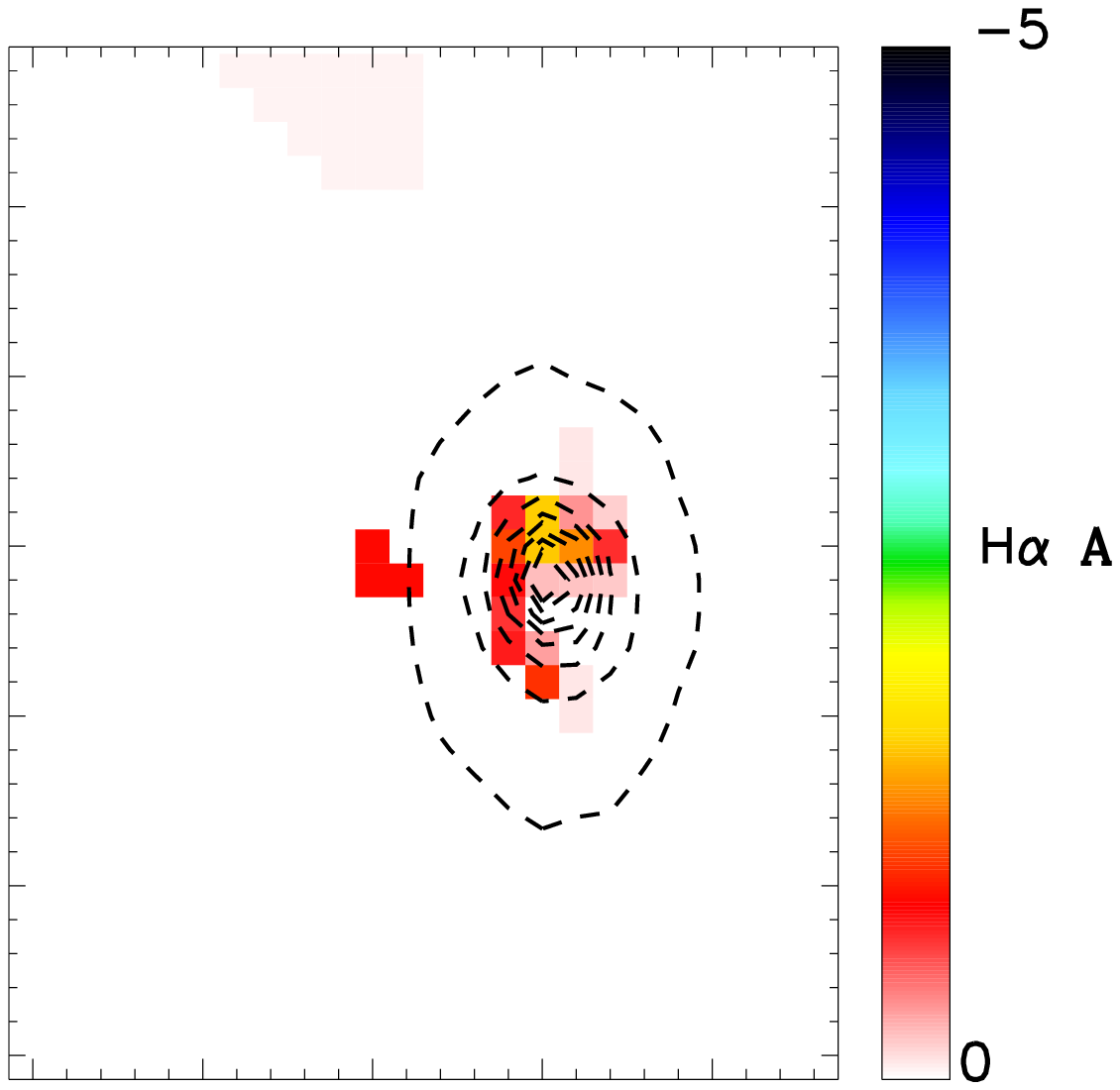}
\hspace{-1.8cm}
      \includegraphics[height=3.2cm, angle=0, trim=0 0 0 0]{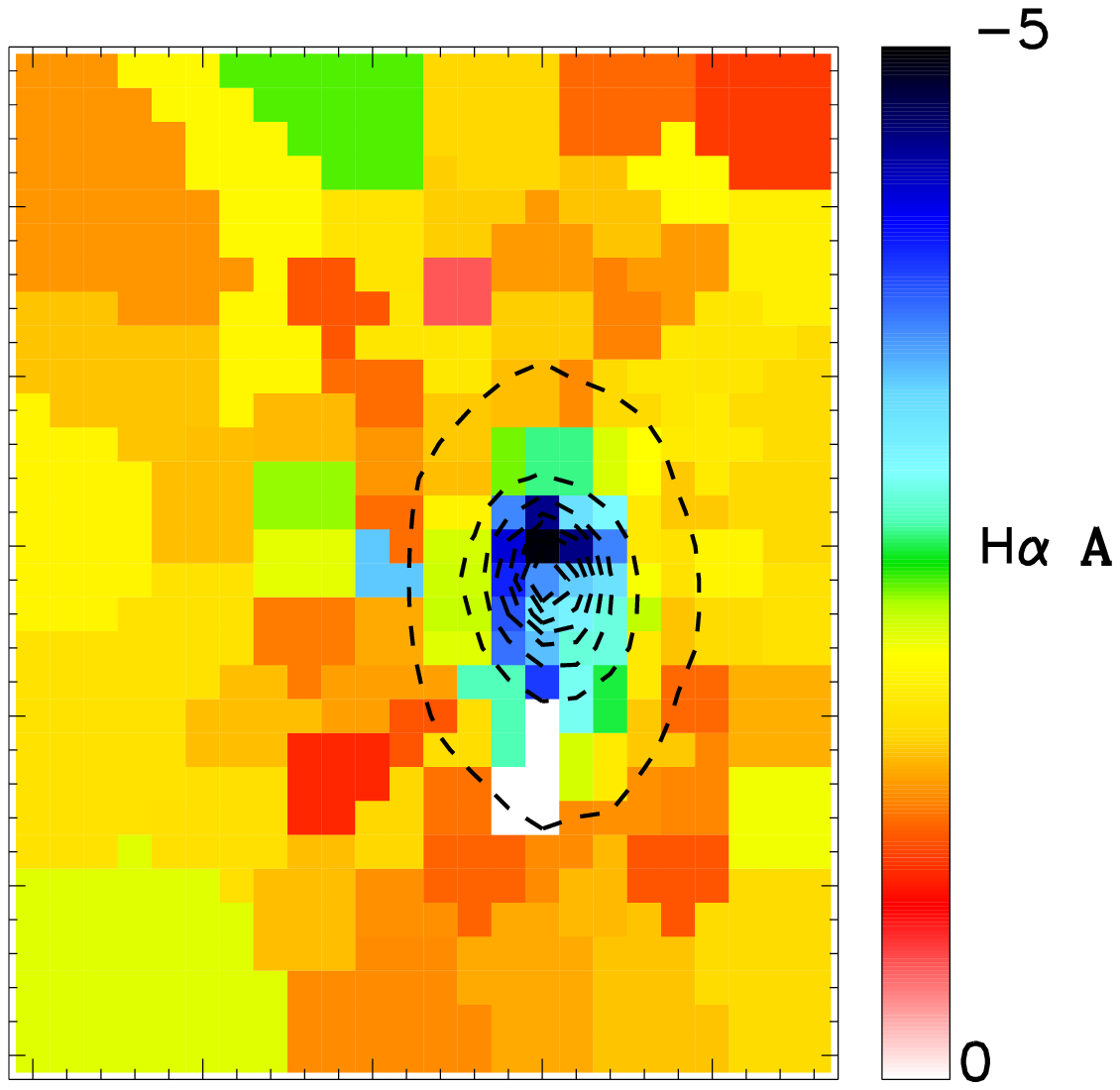}
\hspace{-1.8cm}
       \includegraphics[height=3.2cm, angle=0, trim=0 0 0 0]{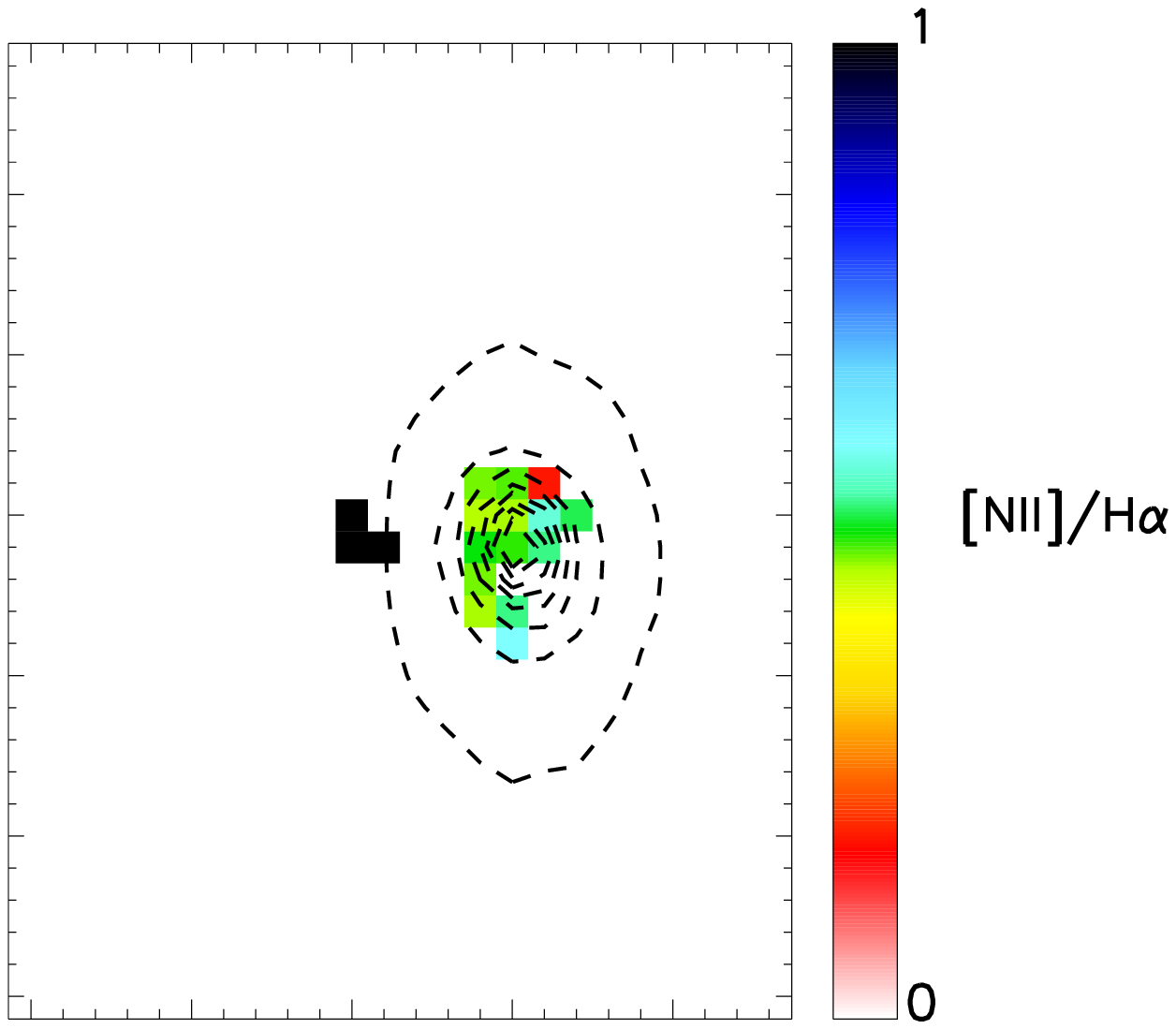}
    \end{minipage}
     \begin{minipage}{0.95\textwidth}
\hspace{1.2cm}
      \includegraphics[height=3.2cm, angle=0, trim=0 0 0 0]{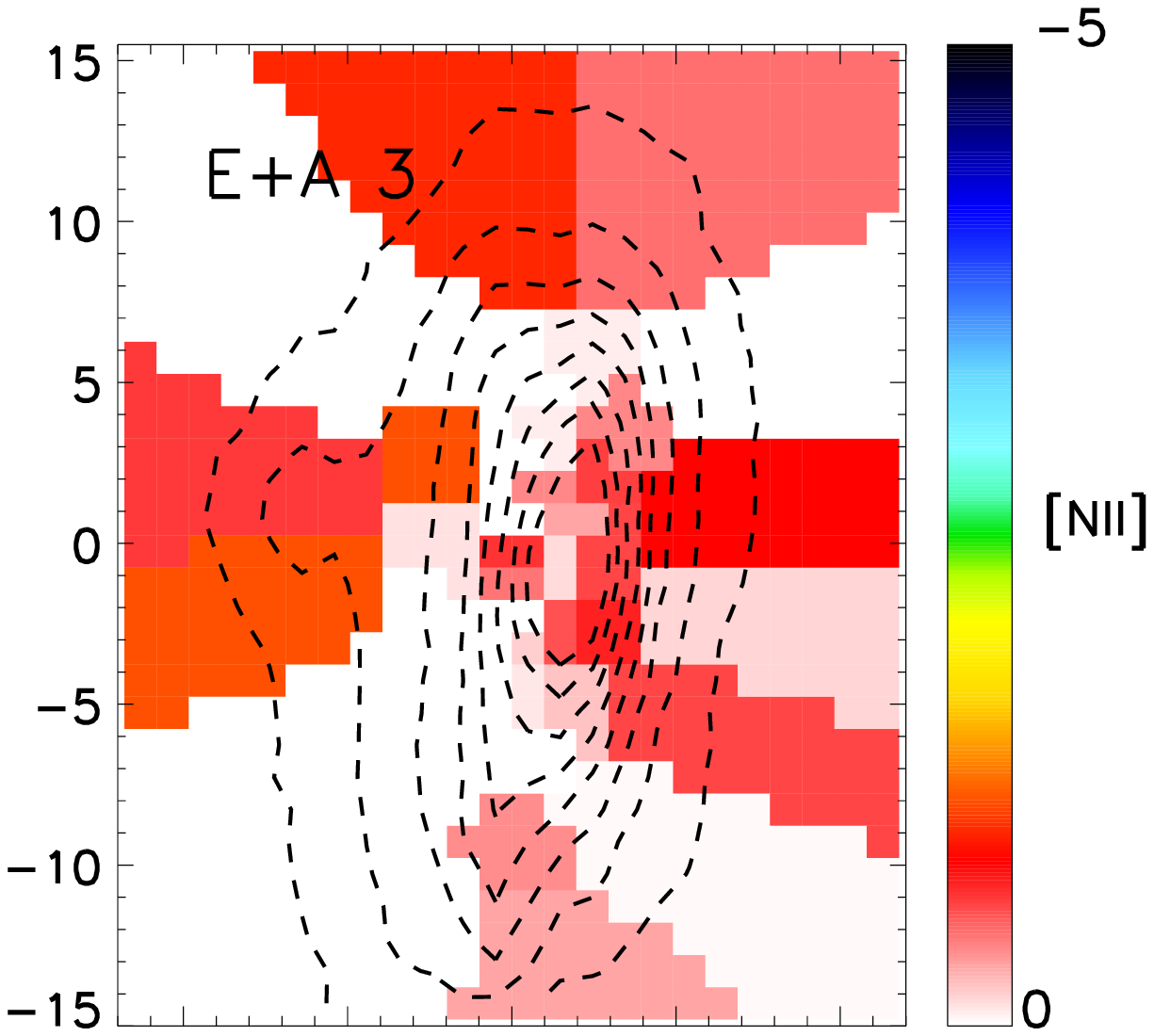}
\hspace{-1.8cm}
      \includegraphics[height=3.2cm, angle=0, trim=0 0 0 0]{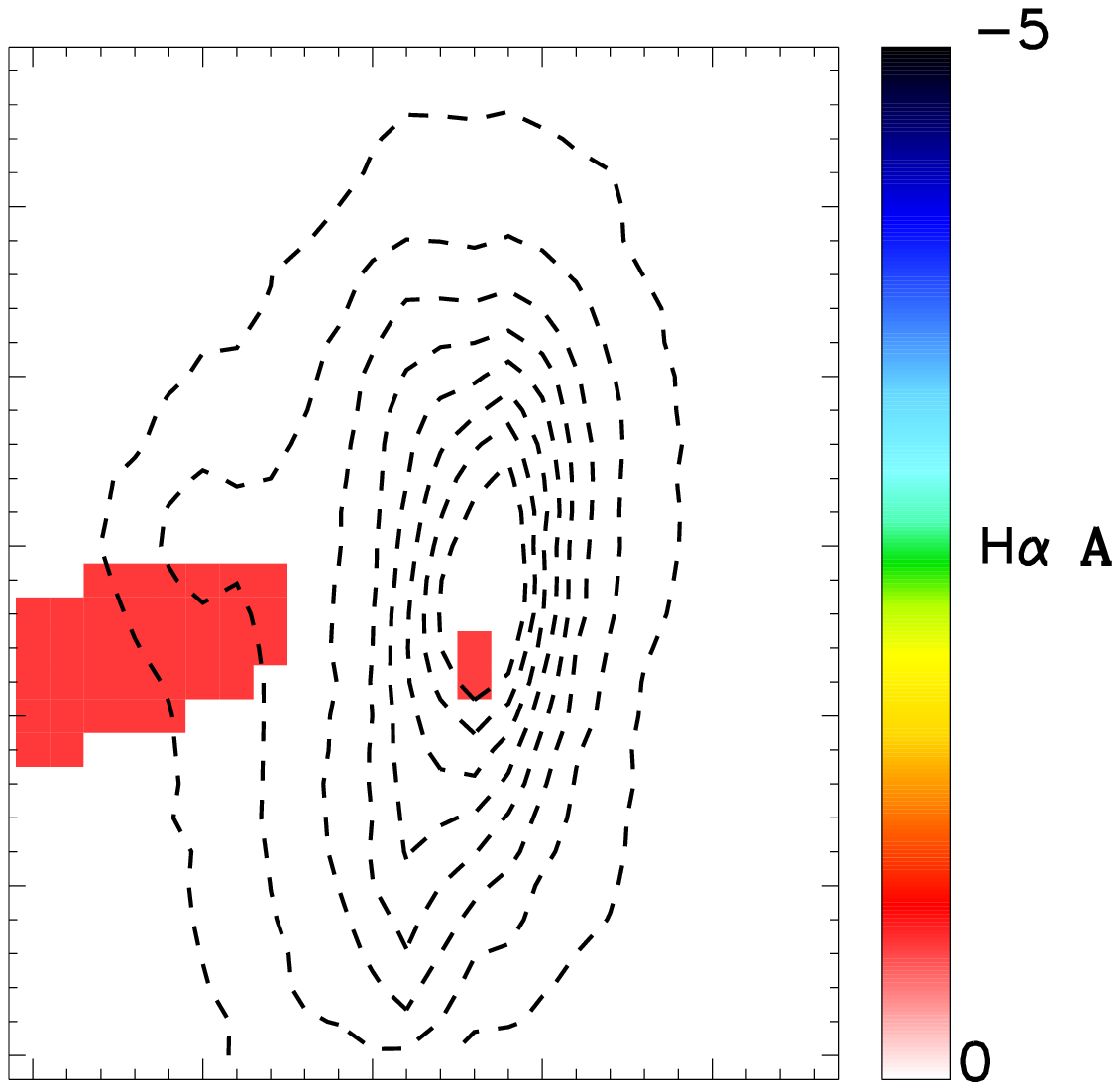}
\hspace{-1.8cm}
      \includegraphics[height=3.2cm, angle=0, trim=0 0 0 0]{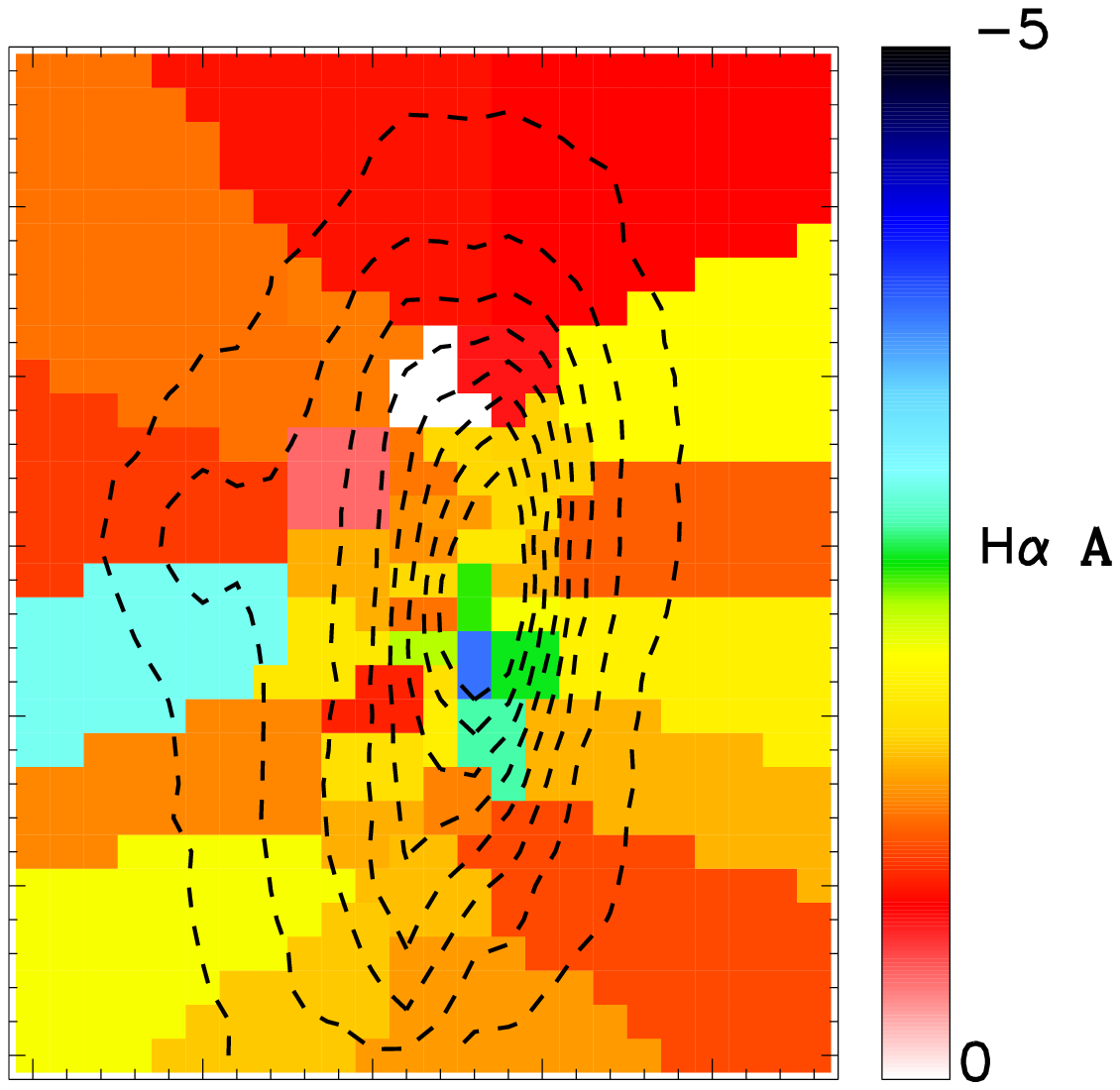}
\hspace{-1.8cm}
       \includegraphics[height=3.2cm, angle=0, trim=0 0 0 0]{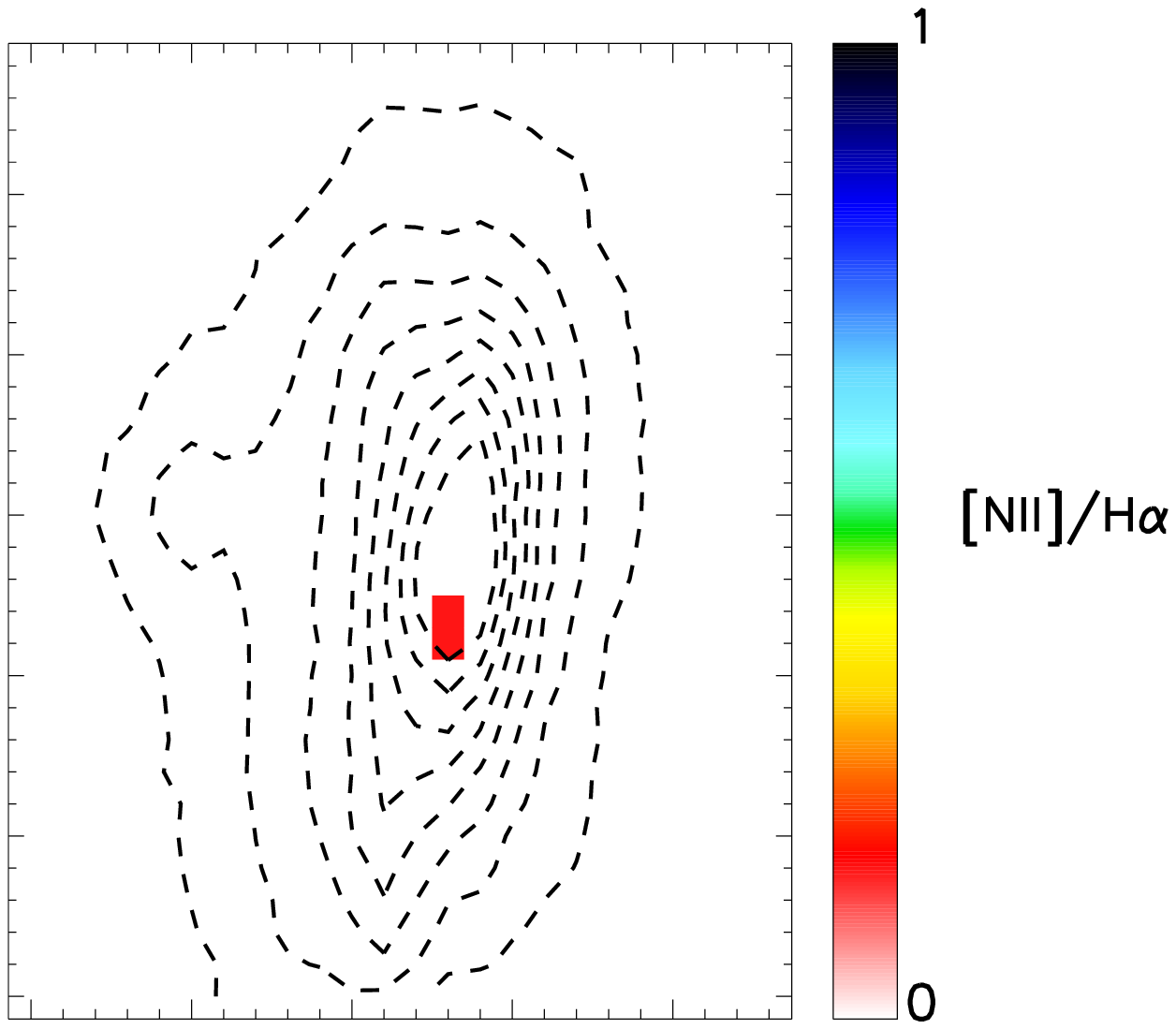}
    \end{minipage}
    \begin{minipage}{0.95\textwidth}
\hspace{1.2cm}
      \includegraphics[height=3.2cm, angle=0, trim=0 0 0 0]{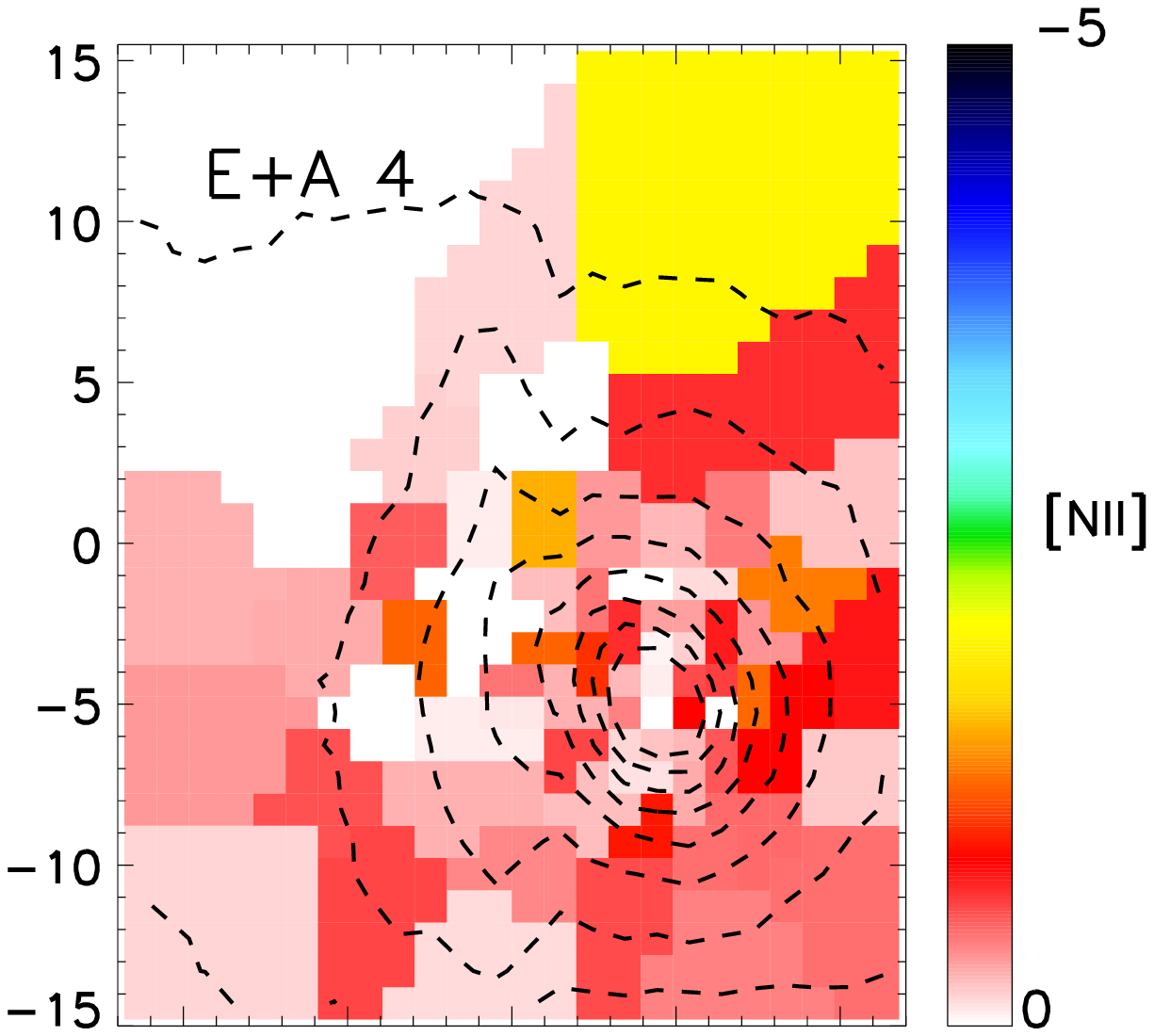}
\hspace{-1.8cm}
      \includegraphics[height=3.2cm, angle=0, trim=0 0 0 0]{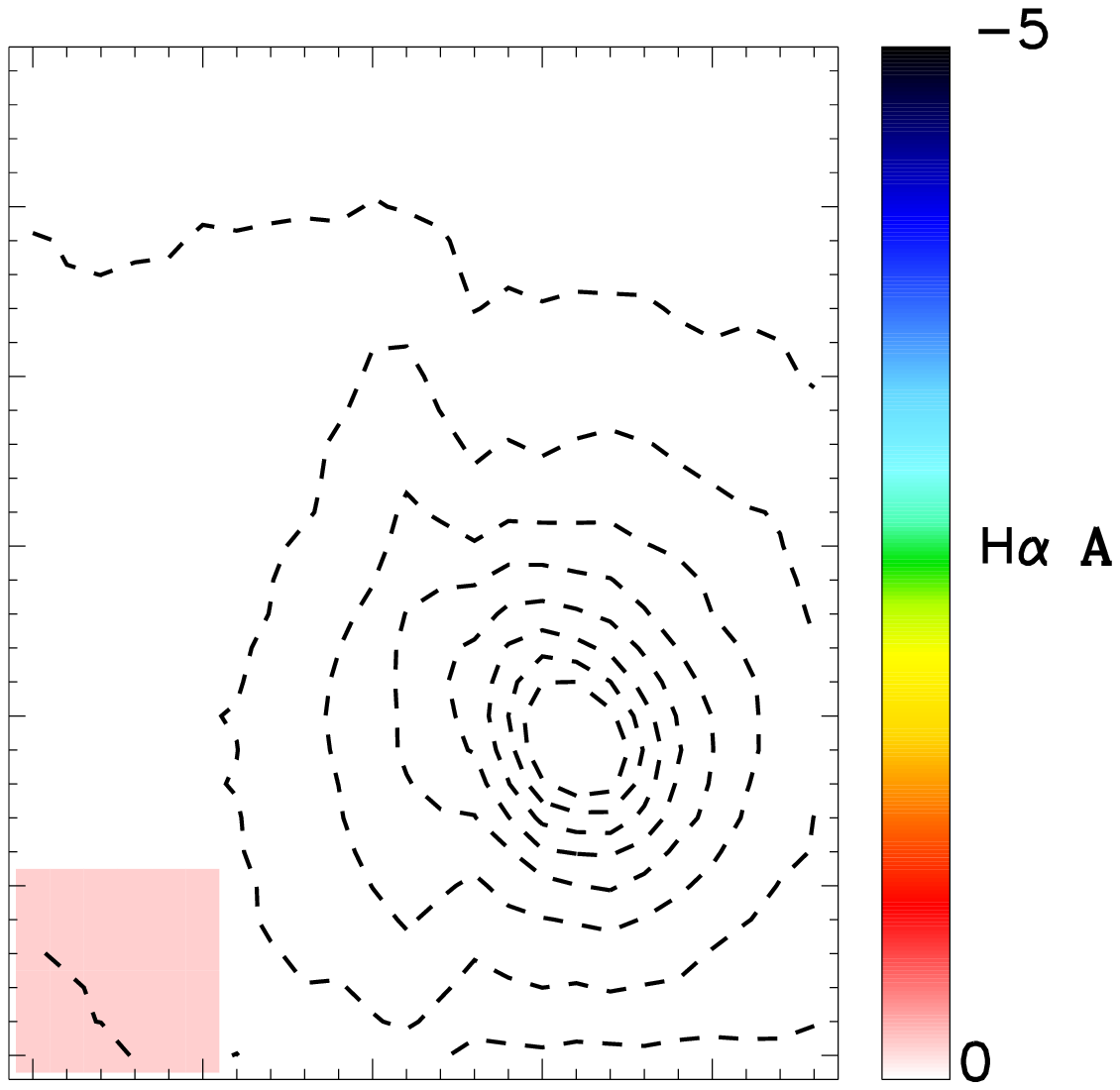}
\hspace{-1.8cm}
      \includegraphics[height=3.2cm, angle=0, trim=0 0 0 0]{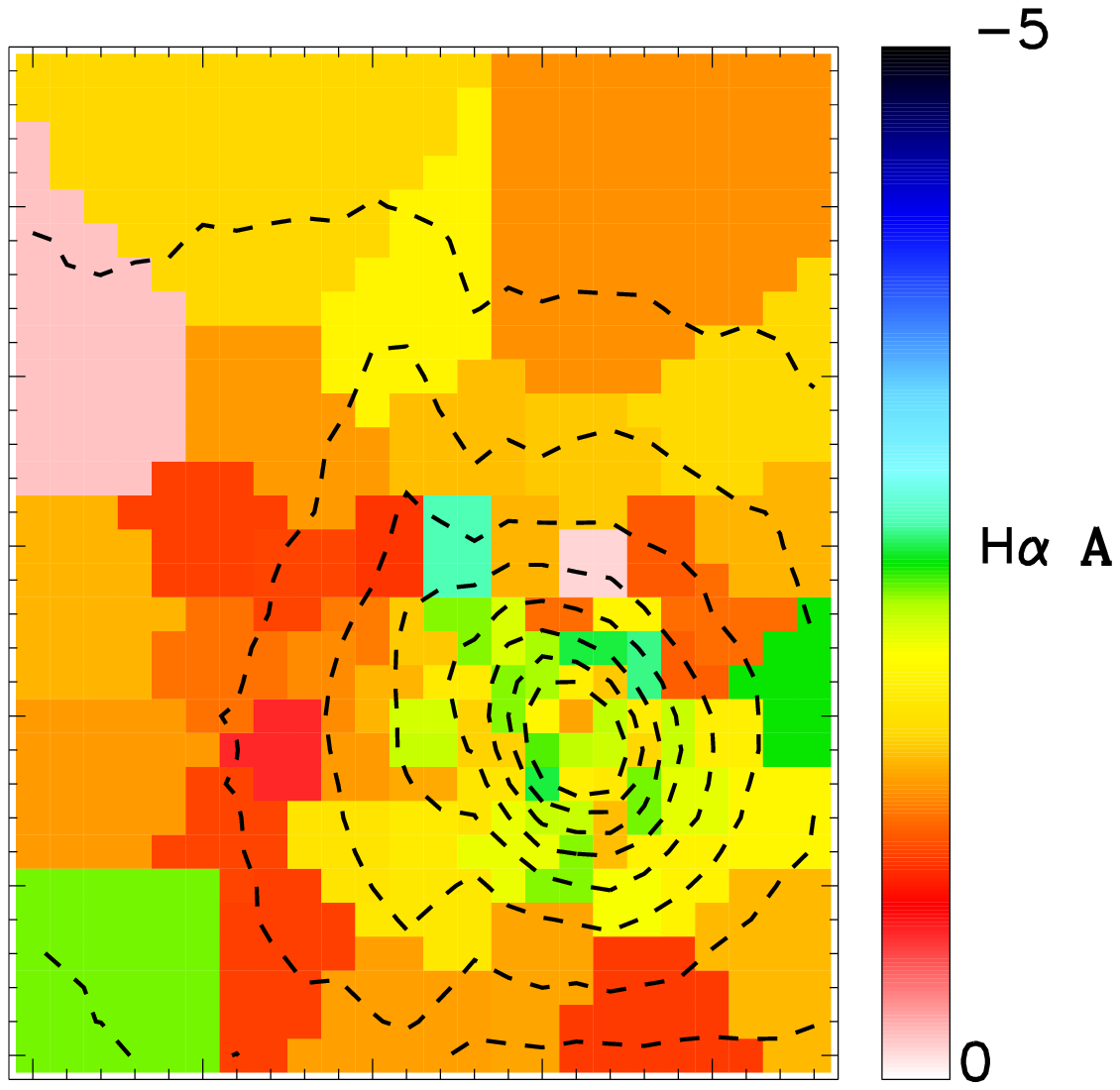}
\hspace{-1.8cm}
       \includegraphics[height=3.2cm, angle=0, trim=0 0 0 0]{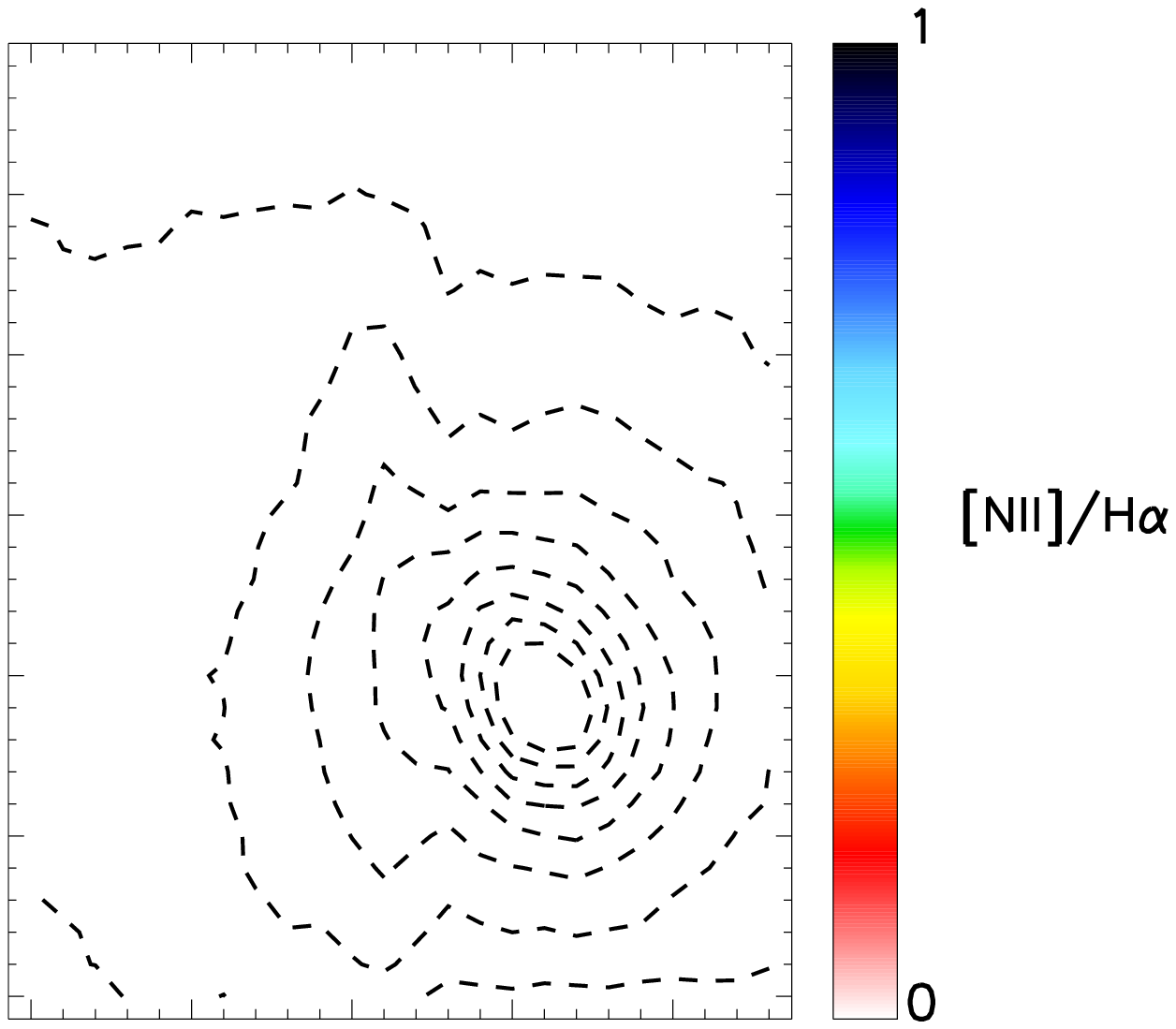}
    \end{minipage}
    \begin{minipage}{0.95\textwidth}
\hspace{1.2cm}
      \includegraphics[height=3.2cm, angle=0, trim=0 0 0 0]{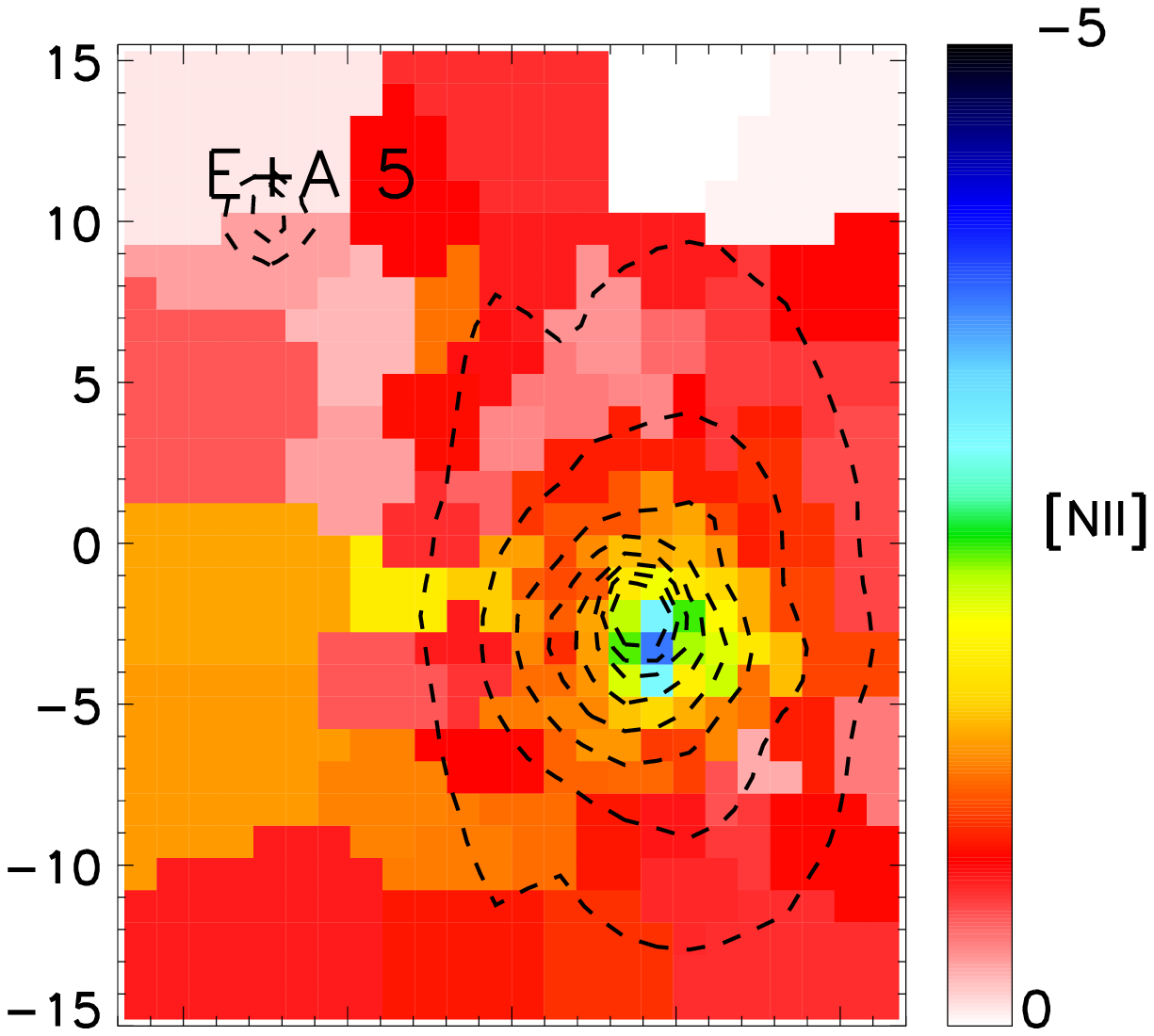}
\hspace{-1.8cm}
      \includegraphics[height=3.2cm, angle=0, trim=0 0 0 0]{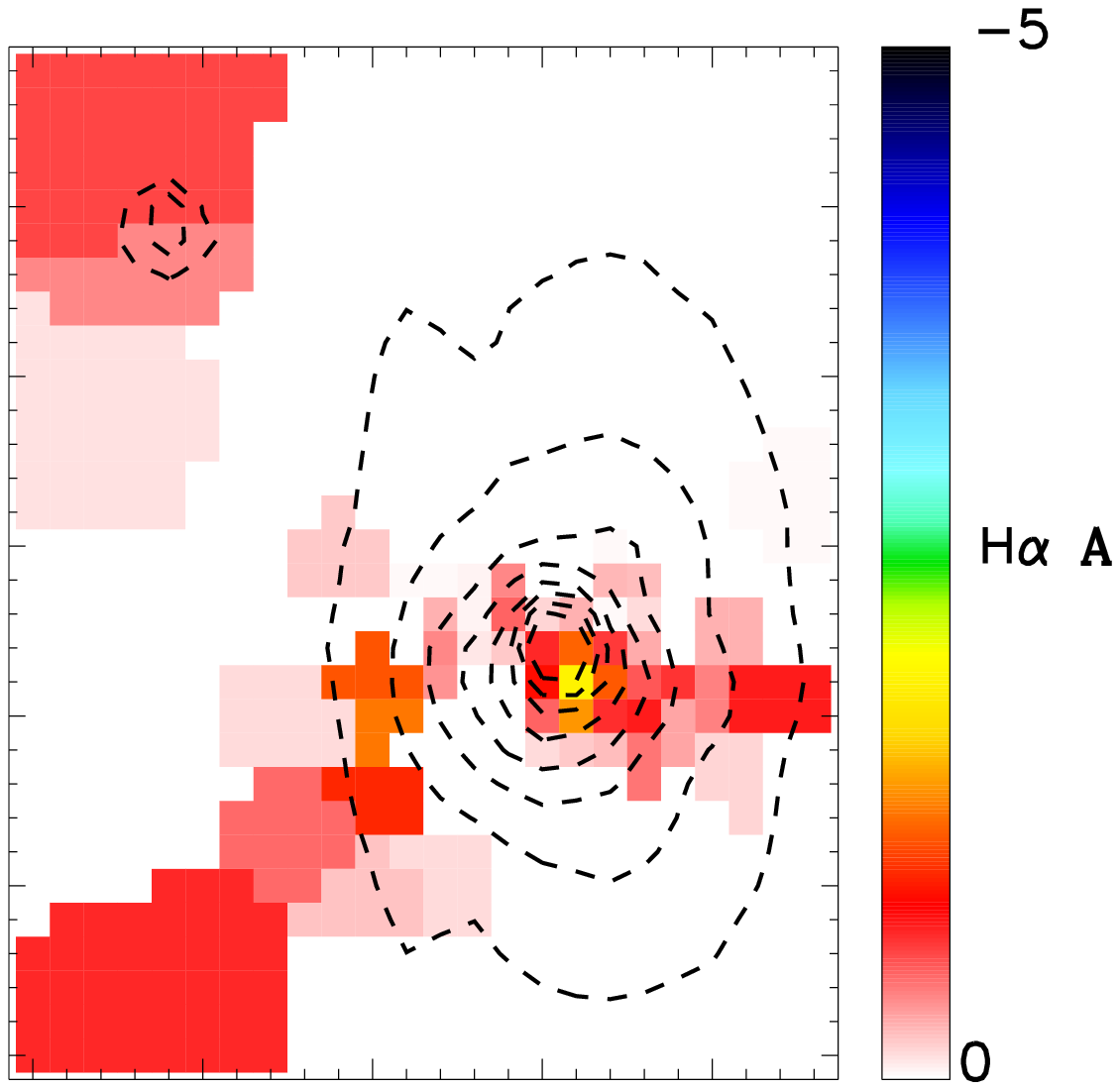}
\hspace{-1.8cm}
      \includegraphics[height=3.2cm, angle=0, trim=0 0 0 0]{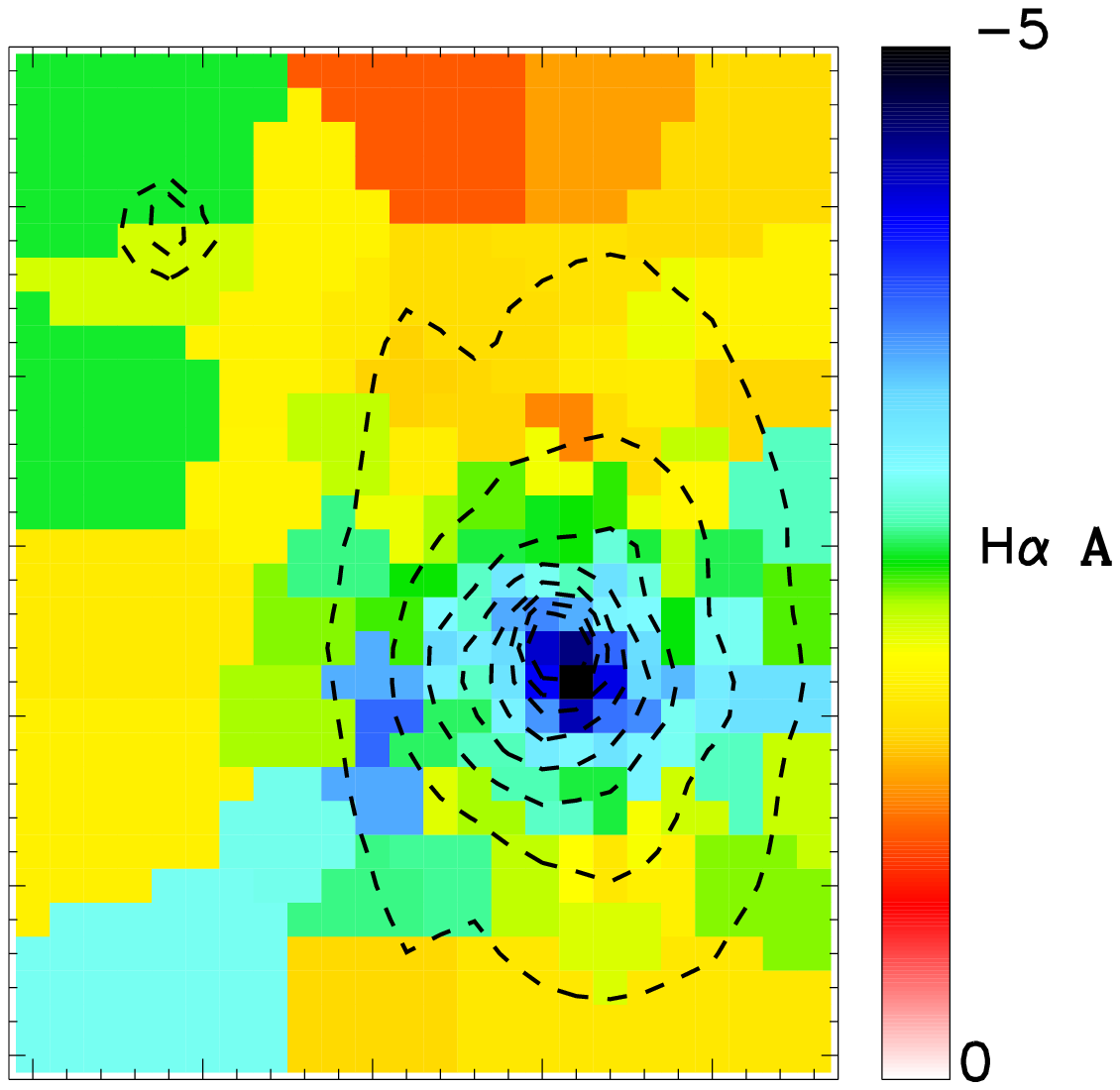}
\hspace{-1.8cm}
       \includegraphics[height=3.2cm, angle=0, trim=0 0 0 0]{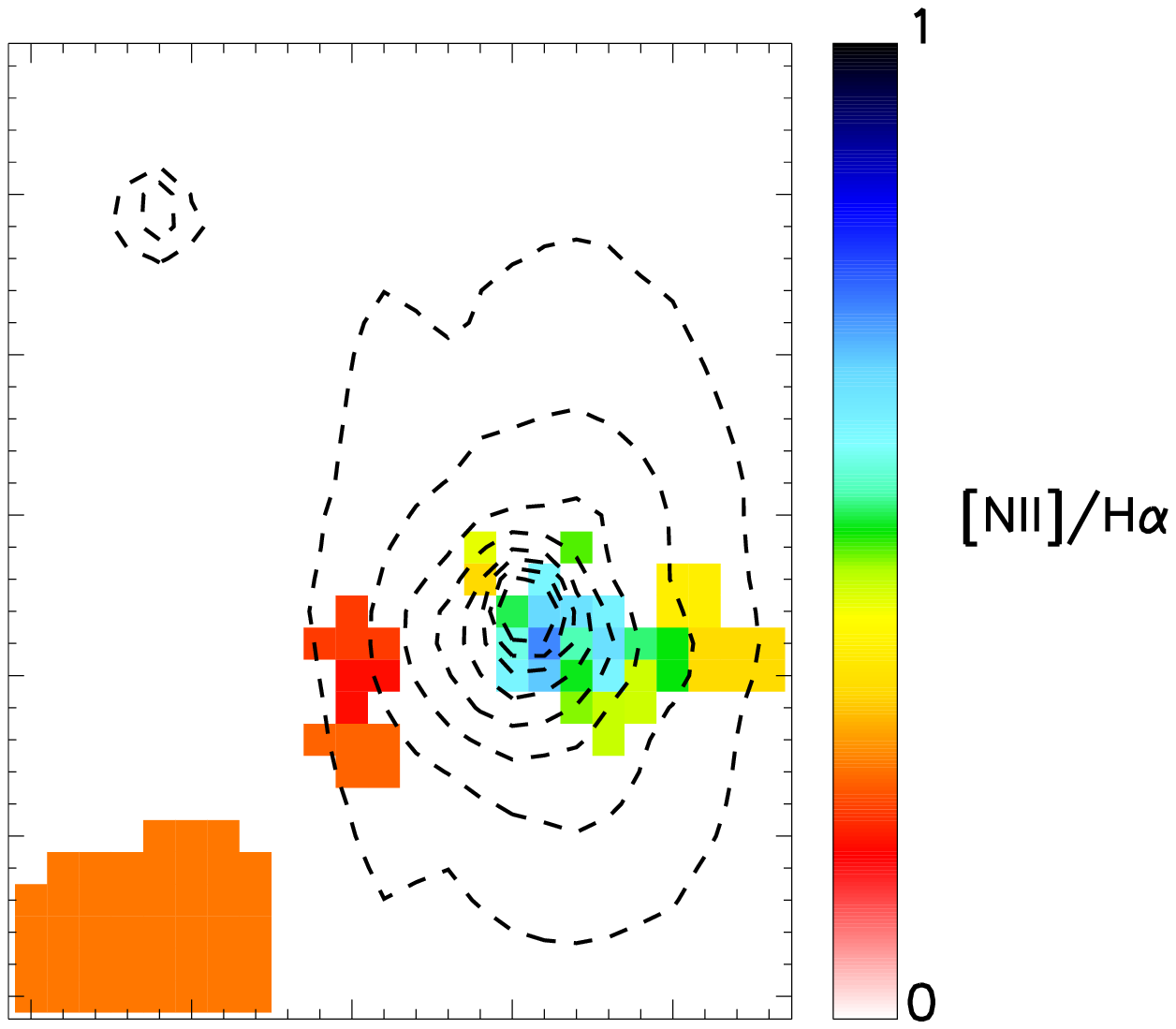}
    \end{minipage}
    \begin{minipage}{0.95\textwidth}
\hspace{1.2cm}
      \includegraphics[height=3.2cm, angle=0, trim=0 0 0 0]{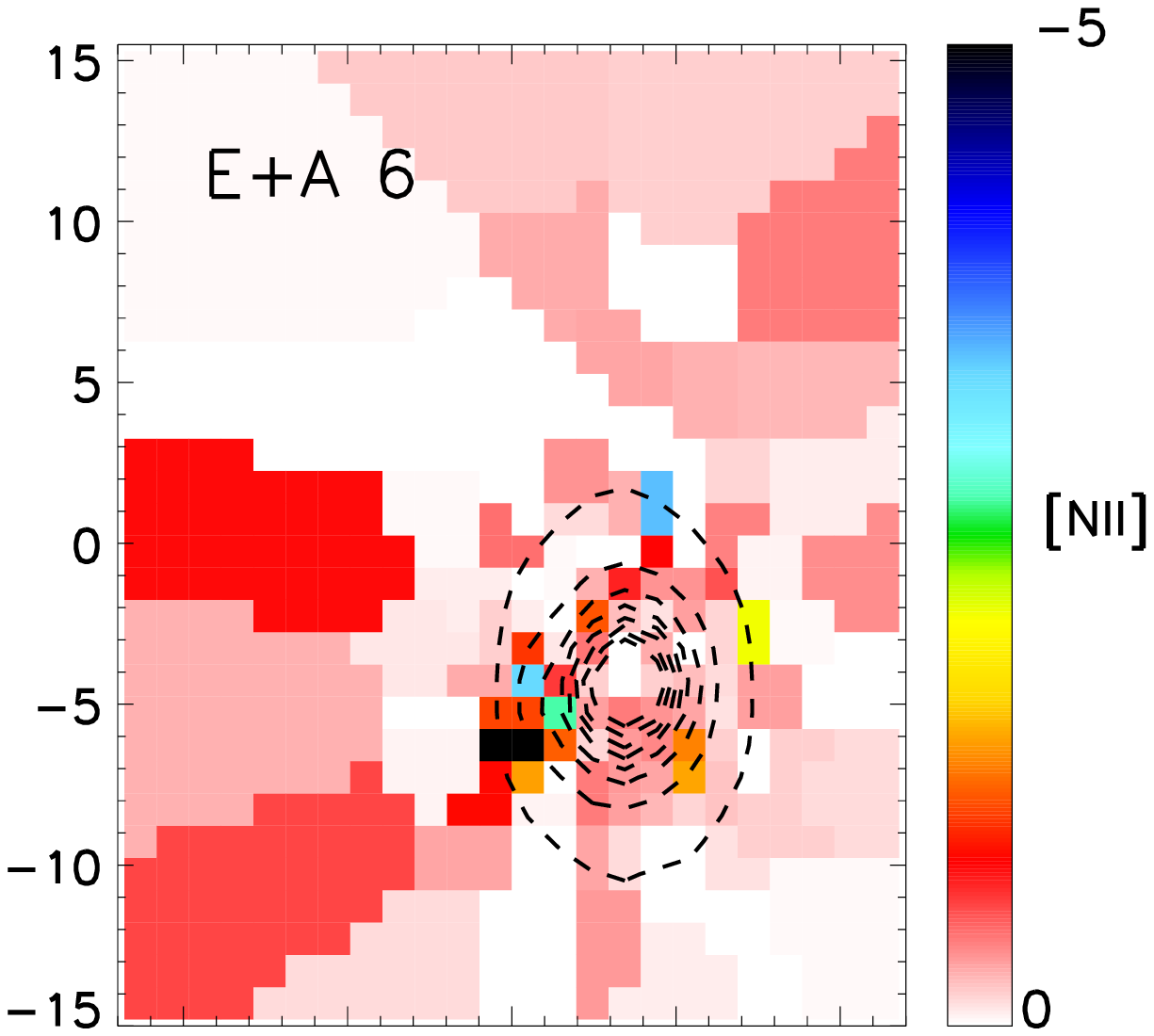}
\hspace{-1.8cm}
      \includegraphics[height=3.2cm, angle=0, trim=0 0 0 0]{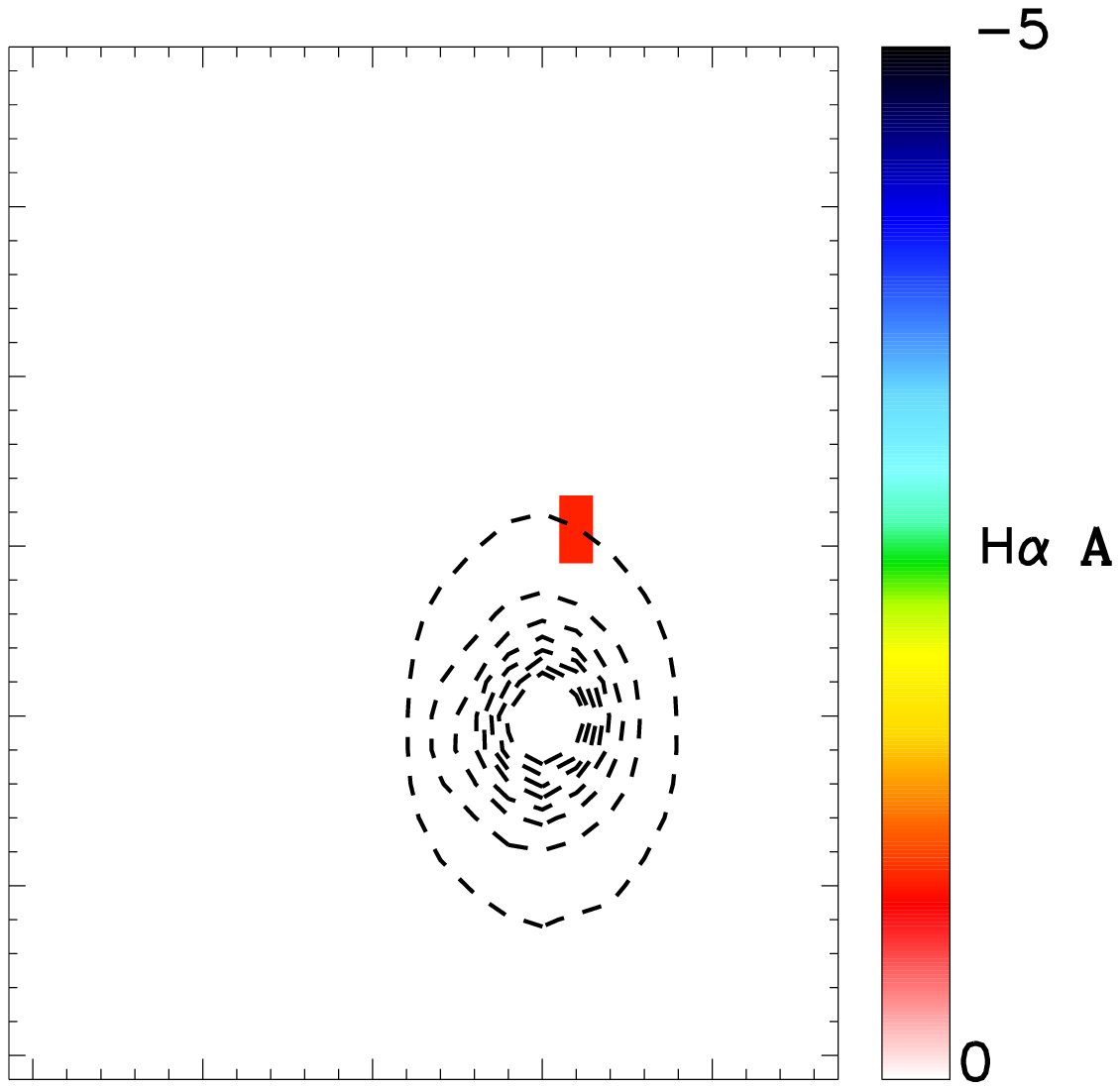}
\hspace{-1.8cm}
      \includegraphics[height=3.2cm, angle=0, trim=0 0 0 0]{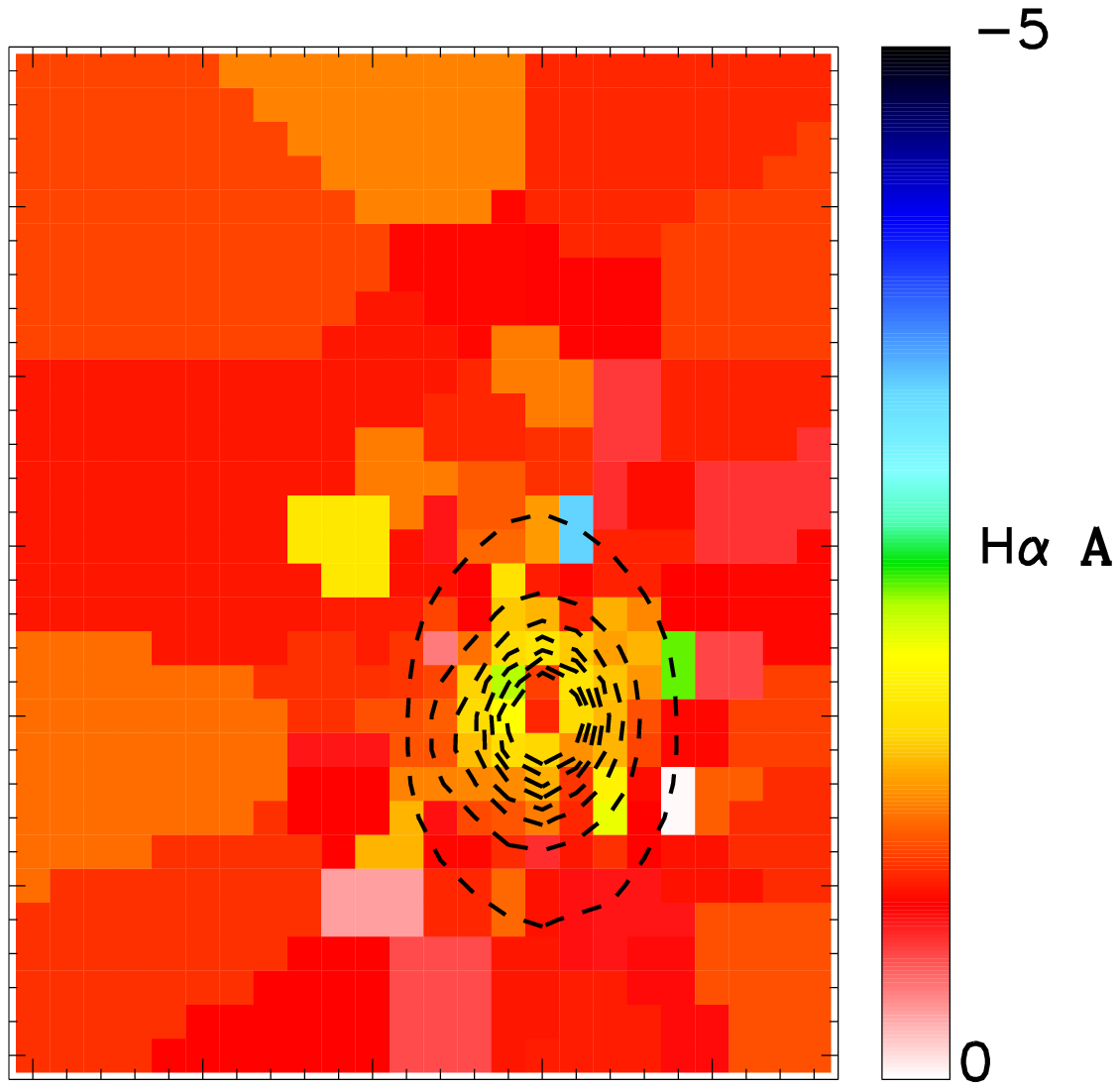}
\hspace{-1.8cm}
       \includegraphics[height=3.2cm, angle=0, trim=0 0 0 0]{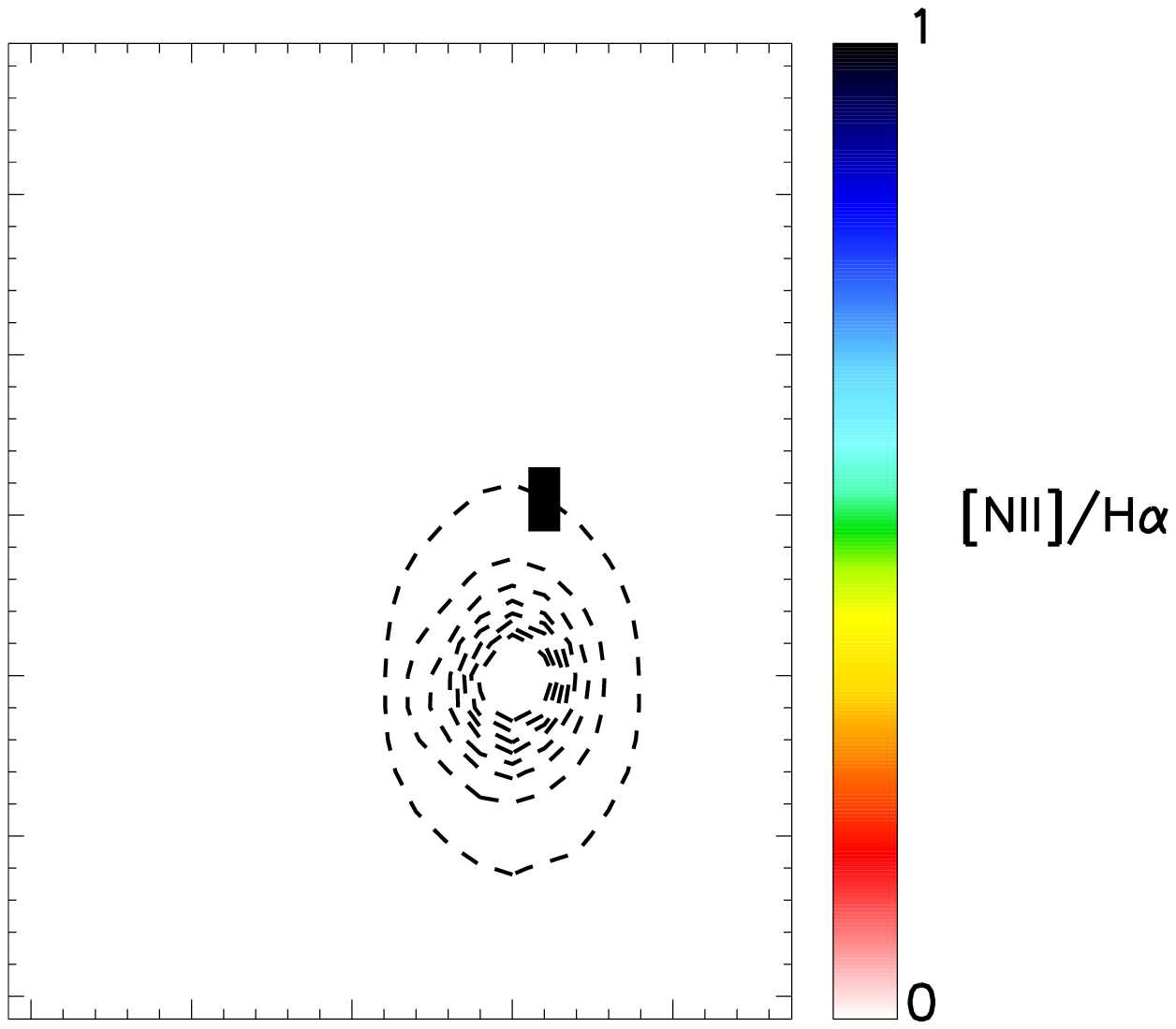}
    \end{minipage}
    \begin{minipage}{0.95\textwidth}
\hspace{1.2cm}
      \includegraphics[height=3.2cm, angle=0, trim=0 0 0 0]{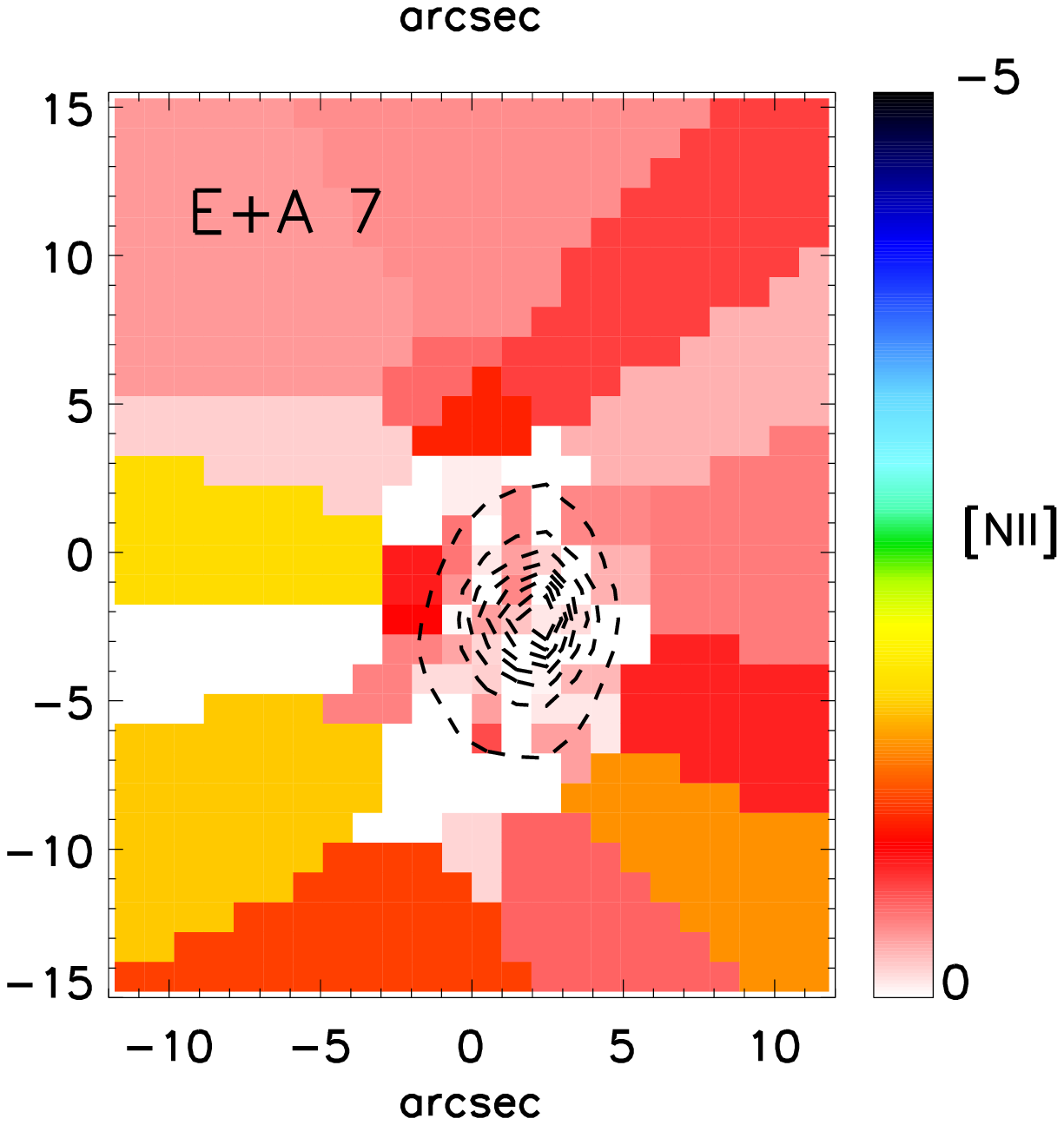}
\hspace{-1.8cm}
      \includegraphics[height=3.2cm, angle=0, trim=0 0 0 0]{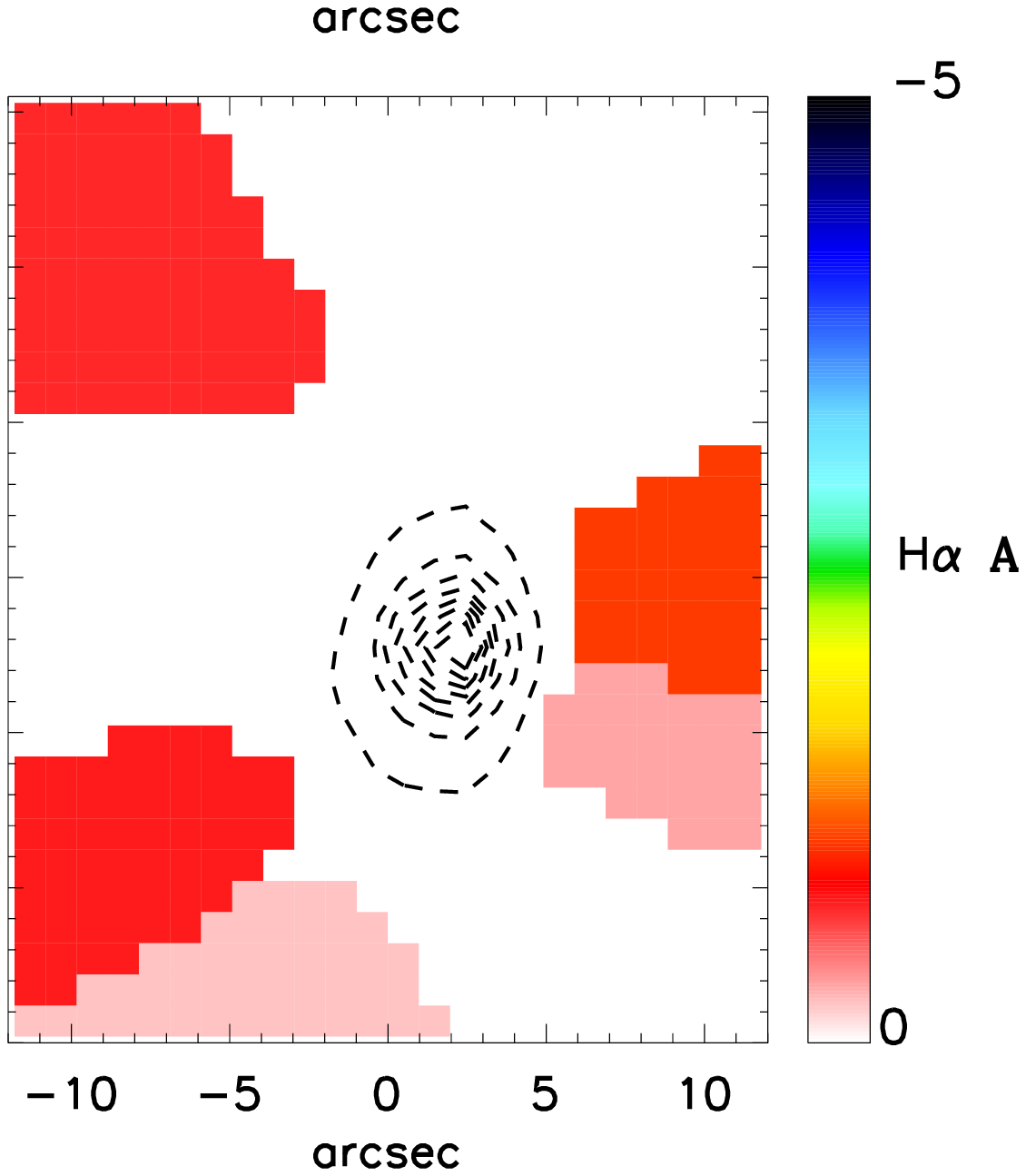}
\hspace{-1.8cm}
      \includegraphics[height=3.2cm, angle=0, trim=0 0 0 0]{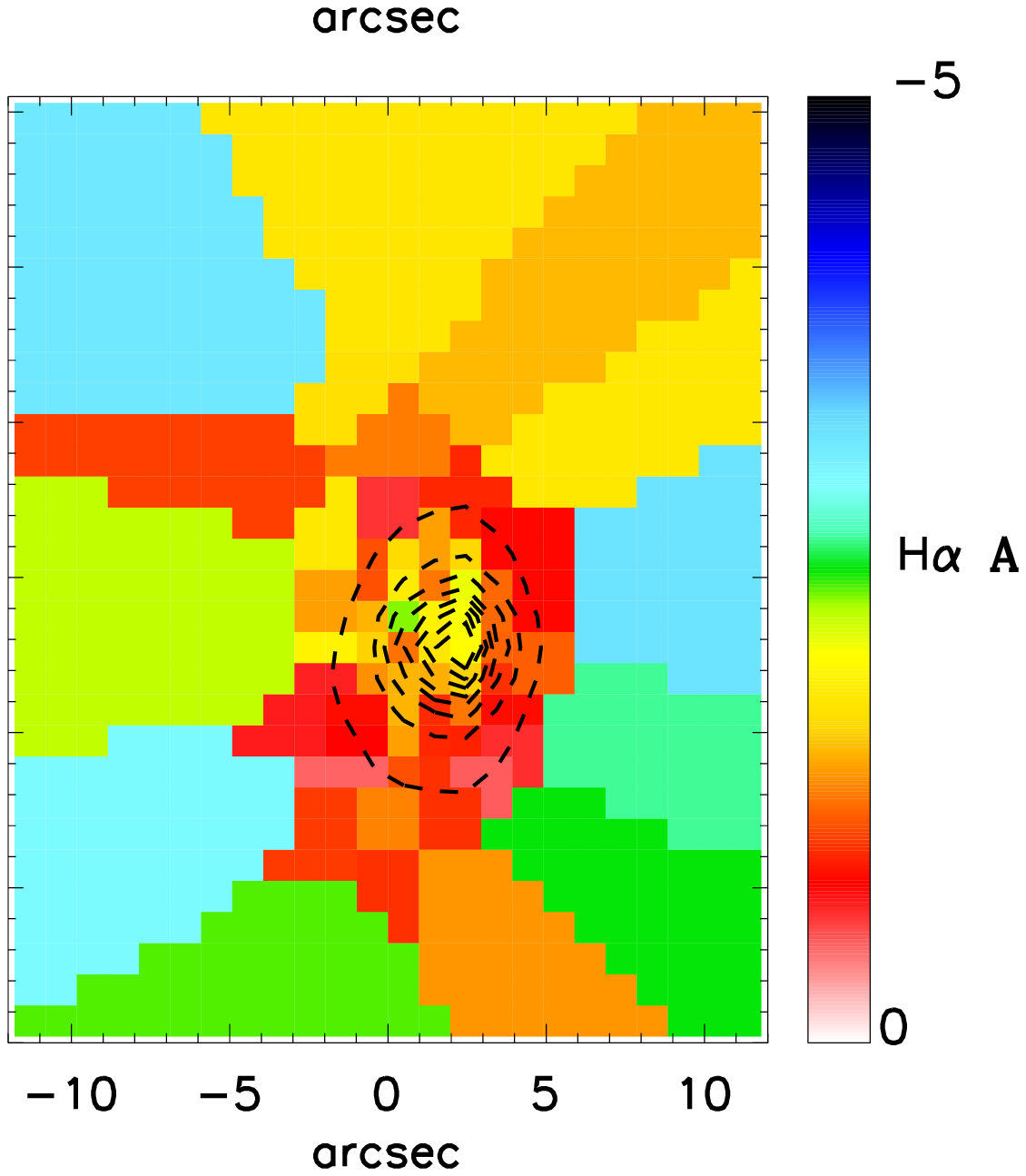}
\hspace{-1.8cm}
       \includegraphics[height=3.2cm, angle=0, trim=0 0 0 0]{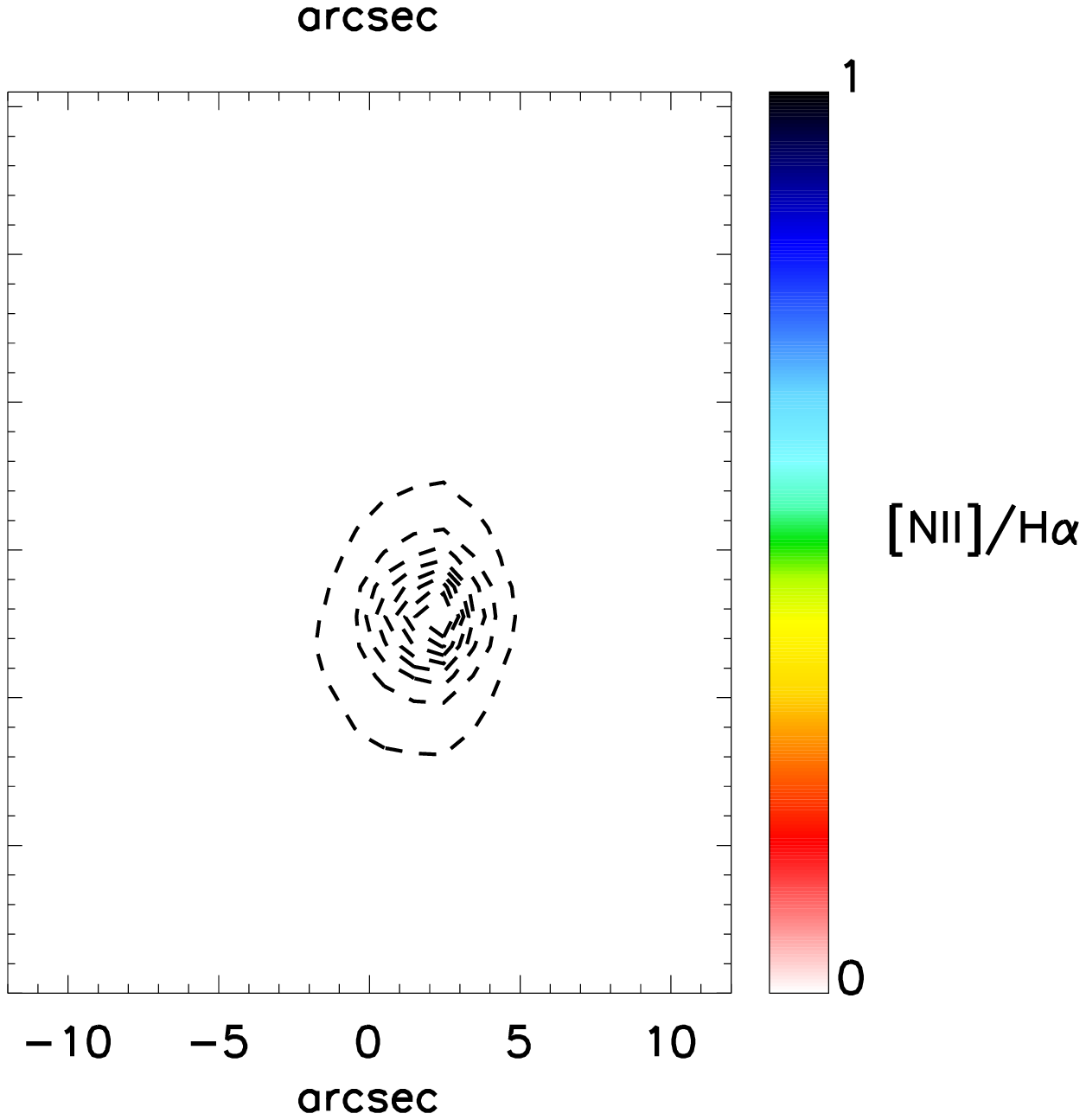}
      \vspace{-0.2cm}
    \end{minipage}
\end{center}
    \caption{Emission line strength and diagnostic maps. The first column is the [NII]$\lambda 6583$ equivalent width map; the second column is the 
H$\alpha$ emission equivalent width map; the third column is the H$\alpha$ equivalent width map after correction for underlying stellar absorption; and
the final column shows the [NII]$\lambda 6583$/H$\alpha$ ratio (corrected) which can be used to classify the source of the ionizing radiation.
The contours overlaid are the galaxy integrated light isophotes.}
    \label{fig:resid}
\end{figure*}

In Fig. \ref{fig:resid} we show spatially resolved emission line 
maps for the [NII] and H$\alpha$ lines in columns 1 and 2, respectively.
The three galaxies that show H$\alpha$ in emission in the integrated SDSS spectra (E+A 2, E+A 3 and E+A 5) in Fig. \ref{fig:spectra}, have their emission 
confined to the central region. In the case of E+A 3, there is convincing emission in only a single spaxel in the galaxy centre and most 
of the galaxy has H$\alpha$ in absorption. The rest of the sample shows no evidence for emission at larger galactic--centric radii 
that would not have been contained in the 3\,arcsecond\, SDSS aperture. The weak values of H$\alpha$ emission in E+A 1 are essentially 
due to noise, and inspection of the spectra (which are poor quality for this galaxy) shows little evidence for emission or absorption.  

In column 3 of Fig. \ref{fig:resid} we attempt to correct for the issue of stellar absorption on the H$\alpha$ emission line index
measurements. This is done by measuring the H$\alpha$ line equivalent width from the spectrum of the best fitting stellar population model to the 
blue arm data. This gives an estimate of the strength of the underlying H$\alpha$ absorption due to integrated light from the
stellar atmospheres alone. This can then be subtracted from the observed line equivalent width to obtain an estimate of 
equivalent width of the emission. Maps with this correction to the H$\alpha$ emission are shown in column 3 of Fig. \ref{fig:resid}.
The central H$\alpha$ emission is still clear in E+A 2, E+A 3 and E+A 5 with stronger absolute equivalent widths. In addition E+A 4 and E+A 6 also
show a weak enhancement of emission in the galaxy centre and E+A 7 shows greater emission in the galaxy outskirts. E+A 1 has uniform emission
after correction of $\sim$5\AA, which is the result of the observed spectra showing little emission or absorption at H$\alpha$ but strong absorption
in the bluer Balmer lines causing a large correction. The outer regions of the
rest of the sample show $\sim$1 to 2\AA\, of emission line filling indicating there may be small amounts of ongoing star formation or that the
fit to the higher order Balmer lines is slightly overestimating the H$\alpha$ equivalent width. 

In the final column of Fig. \ref{fig:resid} we plot the ratio of [NII]$\lambda 6583$/{\rm H}$\alpha \lambda 6563$ where we have used the corrected H$\alpha$
value. We only plot spaxels which have EW(H$\alpha$) and EW([NII]$\lambda 6583$) $>$ 0.2\AA\, and a signal-to-noise ratio in both these lines greater than
2.  The most common diagnostic plot for classifying emission line galaxies is the BPT diagram \citep{baldwin81} and uses the position 
of a galaxy in the two dimensional [OIII]/H$\beta$ versus [NII]/H$\alpha$ plane. However, this diagnostic cannot be used for many weak line galaxies
or low signal-to-noise spectra where the H$\beta$ and/or [OIII] lines cannot be satisfactorily measured -- which is the case here. For this reason
\citet{fernandes11} developed several more inclusive diagnostics for emission line taxonomy, the simplest of which uses the EW(H$\alpha$) versus [NII]/H$\alpha$
plane, which is available here. Optimal transpositions of the common division lines for classifying objects as star-forming, Seyfert or LINER in the 
traditional BPT diagram \citep{kewley01,kauffmann03,stasinska06,kewley06} are derived by \citet{fernandes11}. In this plane the 
standard division between star formation and AGN of \citet{kauffmann03} translates to [NII]$\lambda 6583$/{\rm H}$\alpha \lambda 6563 \sim$0.5 with values
lower than this being in the star formation regime. The hard upper limit for star forming galaxies of \citet{kewley01} translates to 
[NII]$\lambda 6583$/{\rm H}$\alpha \lambda 6563 \sim$0.8 with galaxies between these two values being composite systems and galaxies above
the \citet{kewley01} value being pure AGN \citep{fernandes11}. Comparing these values with the maps in the final column of Fig. 6 for the
galaxies with H$\alpha$ emission shows E+A 2 and E+A 5  have [NII]$\lambda 6583$/{\rm H}$\alpha \lambda 6563 \sim$0.5--0.7 corresponding to the
transition region whilst E+A 3, the only late type disk system in the sample, has a line ratio consistent with pure star--formation. A bright point-like
continuum source was detected in the centre of E+A 5 in the FIRST survey \citep{becker95} consistent with the presence of an AGN in this galaxy.

Since our spectra extend blue-ward of the [OII]$\lambda 3727$ line we can check whether the three galaxies with H$\alpha$ emission would have
been selected in the sample if the [OII]$\lambda 3727$ line was available and the normal [OII]$\lambda 3727>-2.5$\AA\, selection cut was applied. 
We measured the [OII]$\lambda 3727$ line equivalent width in a 3\,arcsecond aperture centred on the galaxy to match the SDSS fibre size. 
The [OII]$\lambda 3727$ equivalent width of E+A 2 is $\sim -1.9\pm 0.7$\,\AA\, satisfying the normal selection criterion. E+A 3 and E+A 5 have
[OII]$\lambda 3727$ equivalent widths of $\sim -2.7\pm 2.3$\AA\, and $-5.2\pm 1.3$\AA, respectively, both slightly above the threshold, although in the case
of E+A 2 it is well within the error. The rest of the sample have  [OII]$\lambda 3727 > -2.5$\,\AA. and consistent with zero at 1$\sigma$.

\subsection{Kinematics}
\begin{figure*}
  \begin{center}
    \begin{minipage}{0.95\textwidth}
      \includegraphics[height=3.2cm, angle=0, trim=0 0 0 0]{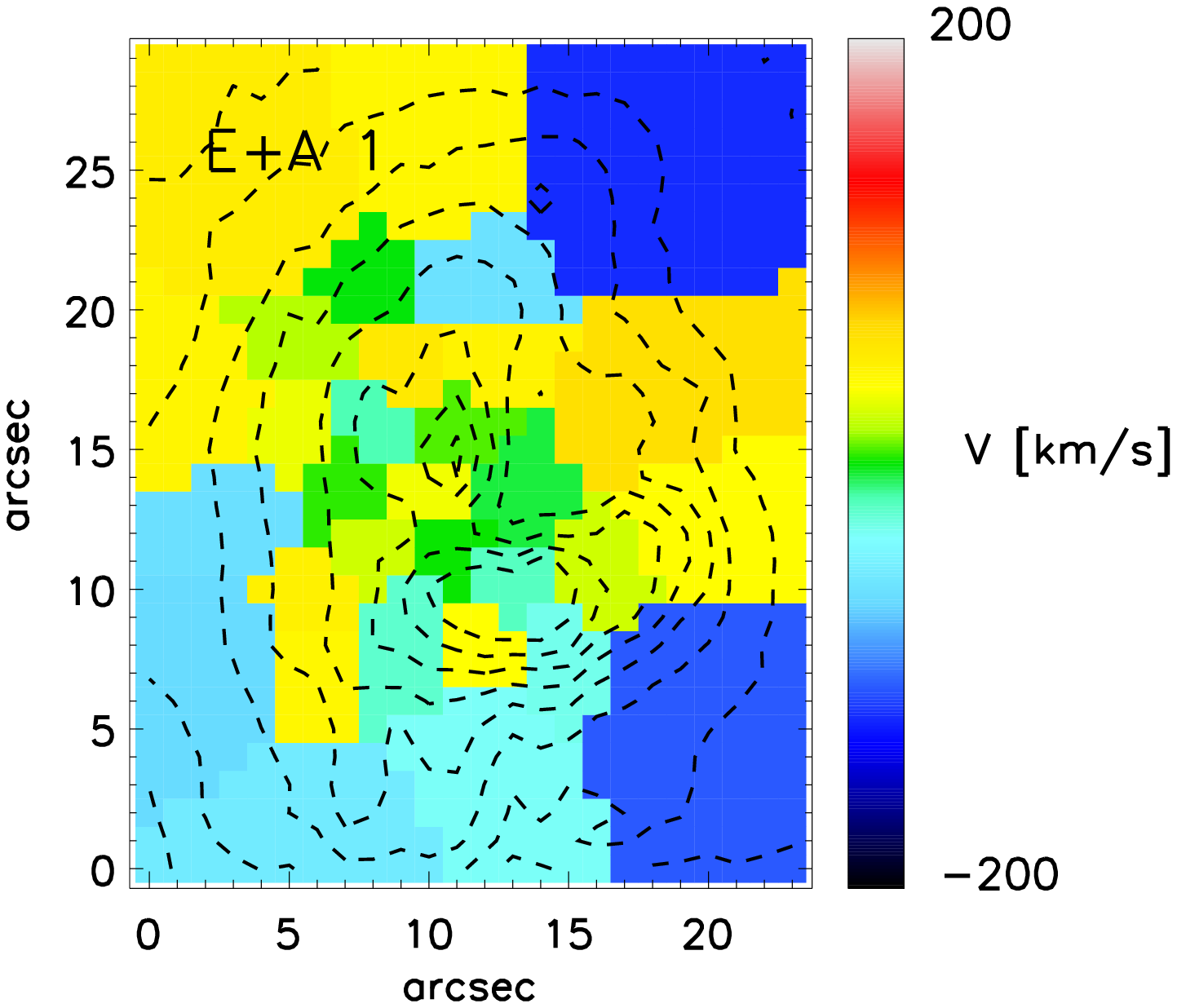}
      \includegraphics[height=3.2cm, angle=0, trim=100 0 0 0]{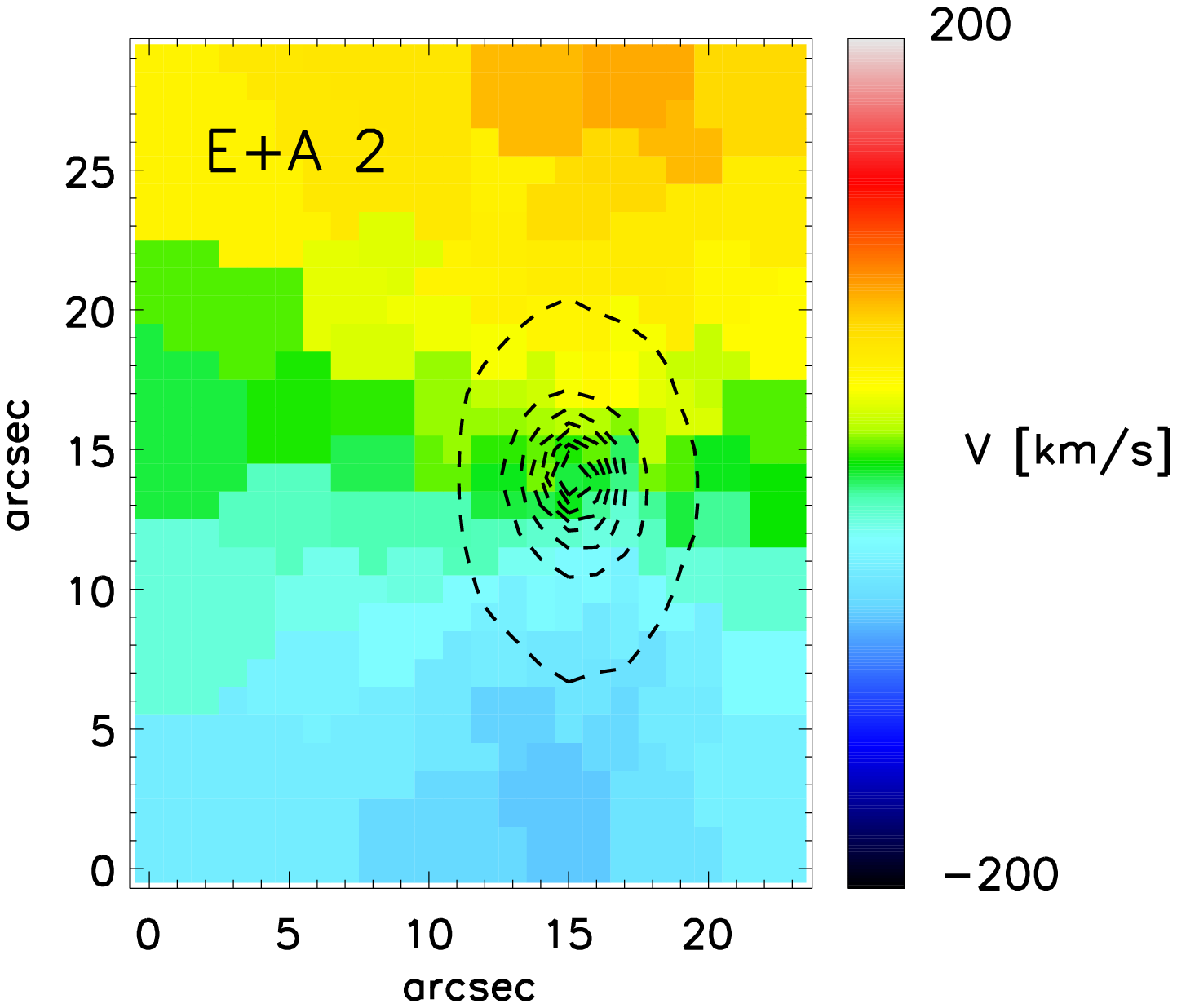}
      \includegraphics[height=3.2cm, angle=0, trim=100 0 0 0]{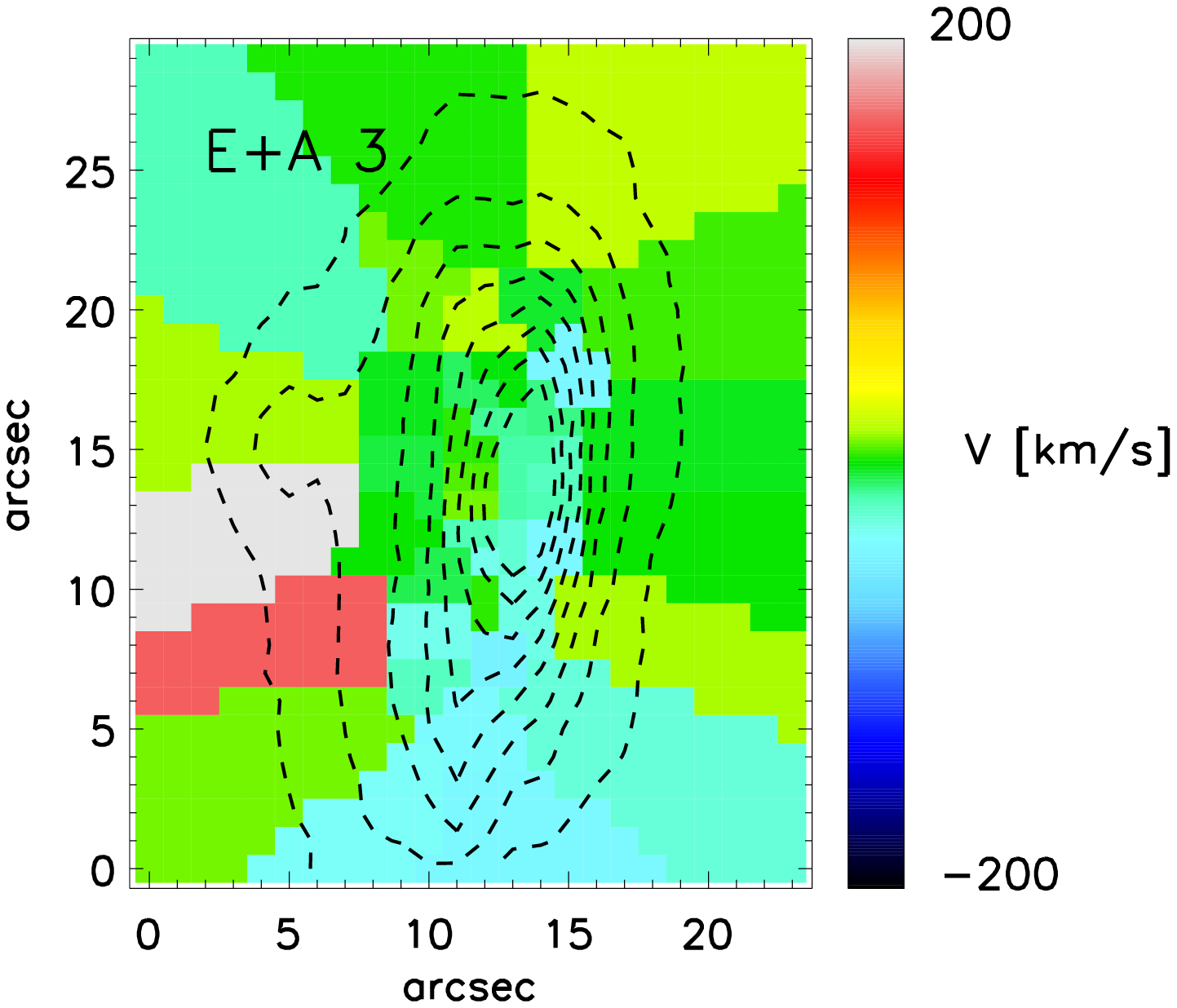}
       \includegraphics[height=3.2cm, angle=0, trim=100 0 0 0]{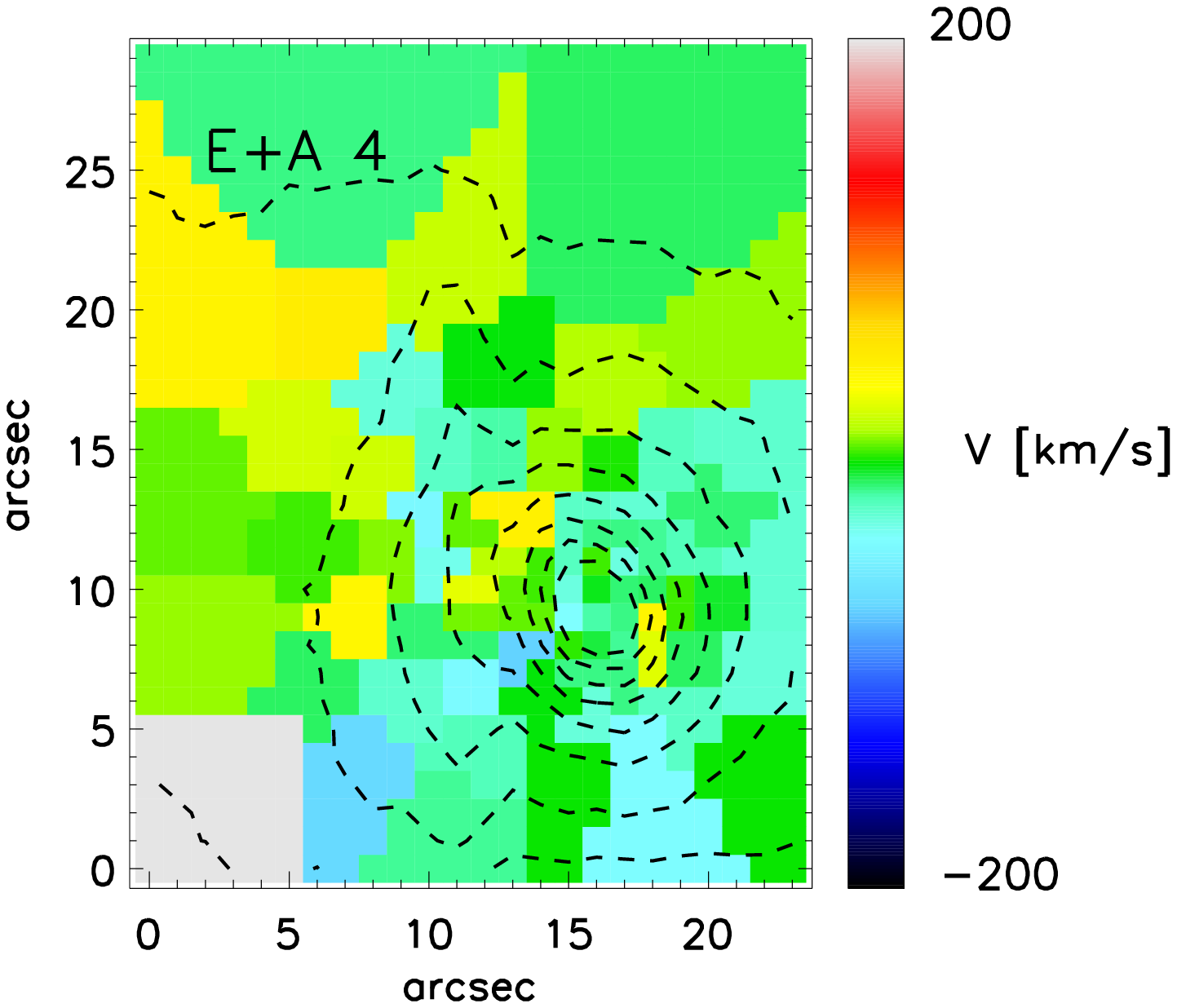}
    \end{minipage}
    \begin{minipage}{0.95\textwidth}
      \includegraphics[height=3.2cm, angle=0, trim=0 0 0 0]{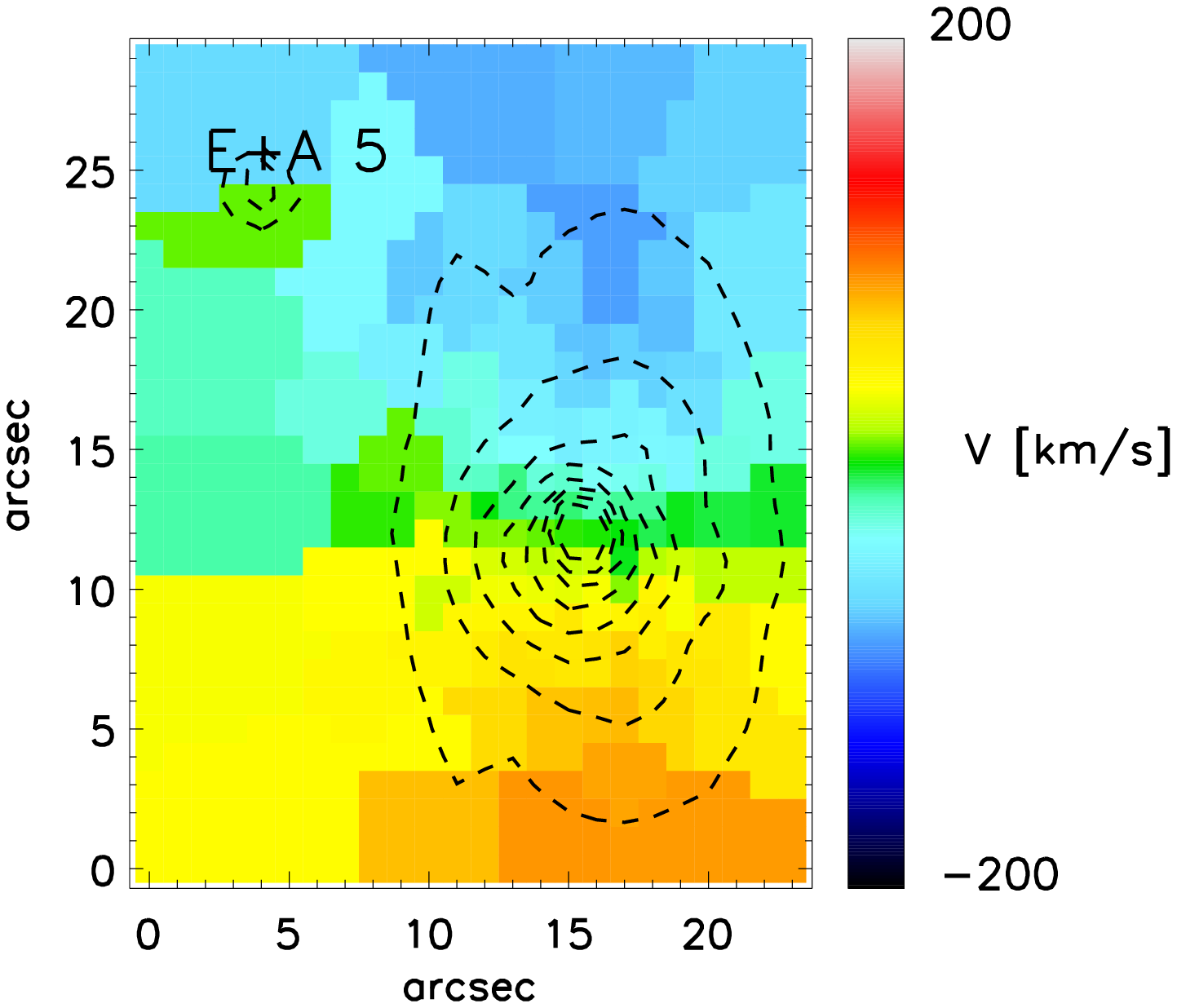}
      \includegraphics[height=3.2cm, angle=0, trim=100 0 0 0]{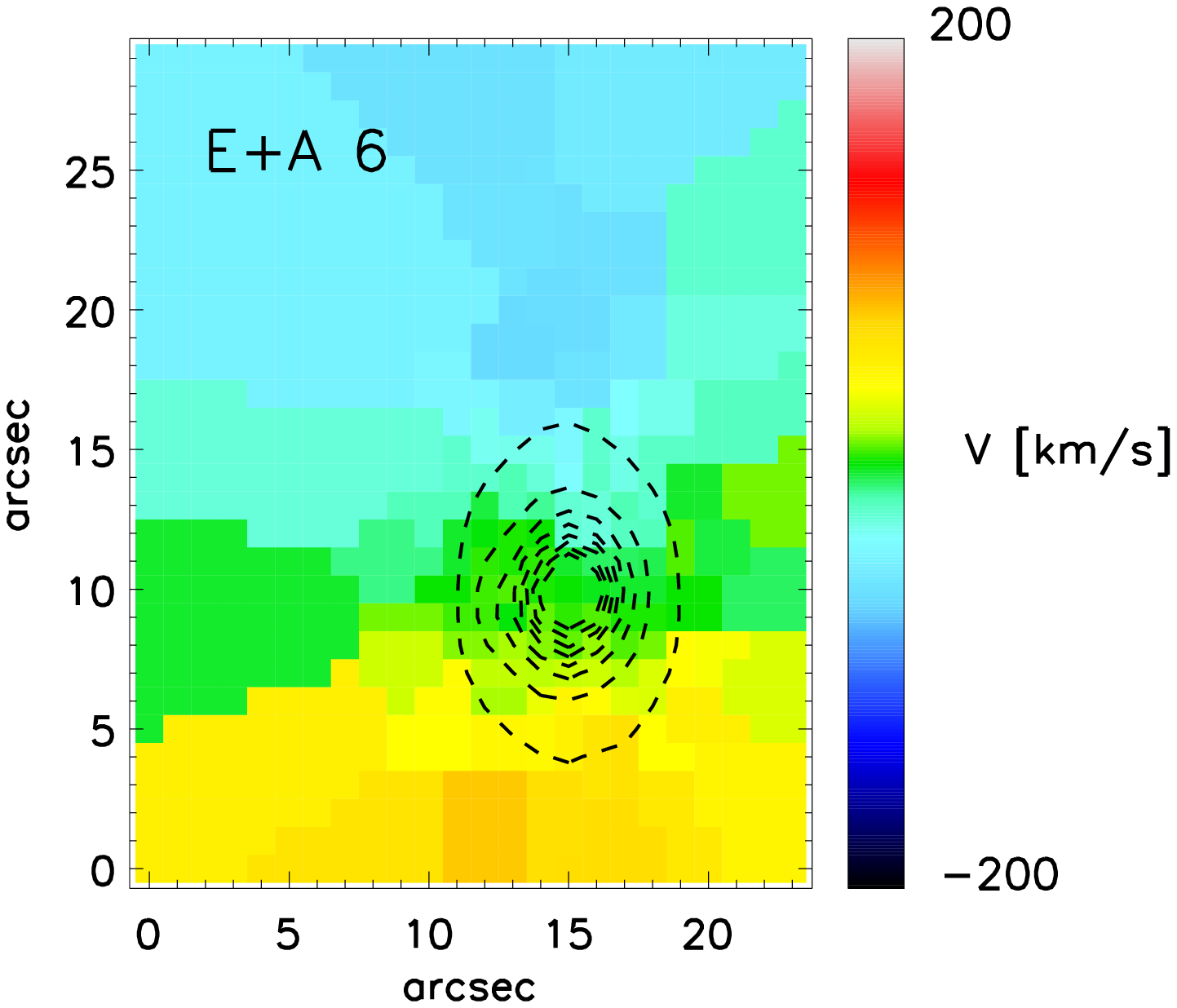}
      \includegraphics[height=3.2cm, angle=0, trim=100 0 0 0]{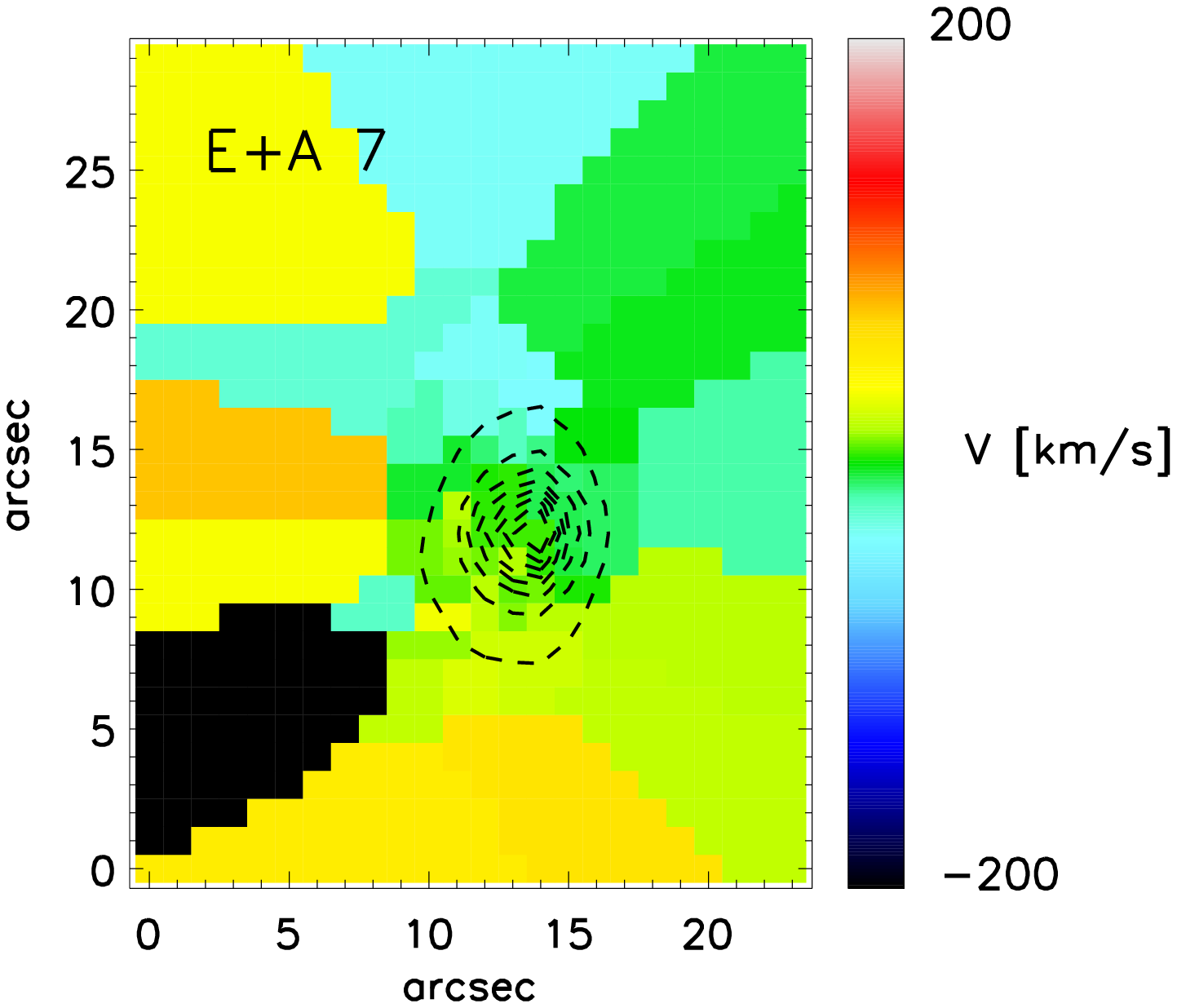}
     \end{minipage}
\end{center}
\vspace{-0.3cm}
    \caption{Shows the stellar streaming velocity maps for our sample. E+A 1 and E+A 3 have no detectable rotation and E+A 4 has only marginal evidence for rotation.
The remaining galaxies all show clear rotation fields. The black contours overlaid are the galaxy integrated light profiles.}
    \label{fig:streaming}
\end{figure*}

Two dimensional streaming velocity maps are shown in Fig. \ref{fig:streaming} and all the galaxies except E+A 1 and E+A 3 show rotation at some level.
In the case of E+A 3 this is surprising since it is a disk galaxy viewed somewhat edge on and rotation is to be expected. E+A 3 is the faintest and highest redshift
galaxy in our sample and was observed in very poor seeing and it maybe the case that the data quality is insufficient to detect the velocity field. 
Using a parameter, $\lambda_{R}$, derived from two--dimensional spectroscopy, early--type galaxies can be separated into two distinct 
kinematic categories: fast and slow rotators \citep{emsellem07}. The $\lambda_{R}$ parameter involves a luminosity weighted average 
over the kinematic field obtained from IFU spectroscopy. \citet{emsellem07,emsellem11} found that approximately three quarters of the SAURON 
sample were fast rotators. Likewise, in a sample of eight E+A galaxies selected from
the 2dFGRS and followed up with integral field spectroscopy seven were found to be fast rotators \citep{pracy09}. The prevalence within the E+A population of fast rotators
argues that unequal mass mergers rather than major mergers were the dominant progenitors to the E+A galaxies since the probability 
of rotating remnants increases with increasing merger mass ratio \citep{bournaud08}. Although major mergers can still produce fast rotating remnants \citep{bois11}.

\begin{figure}
      \includegraphics[width=5.8cm, angle=90, trim=0 0 0 0]{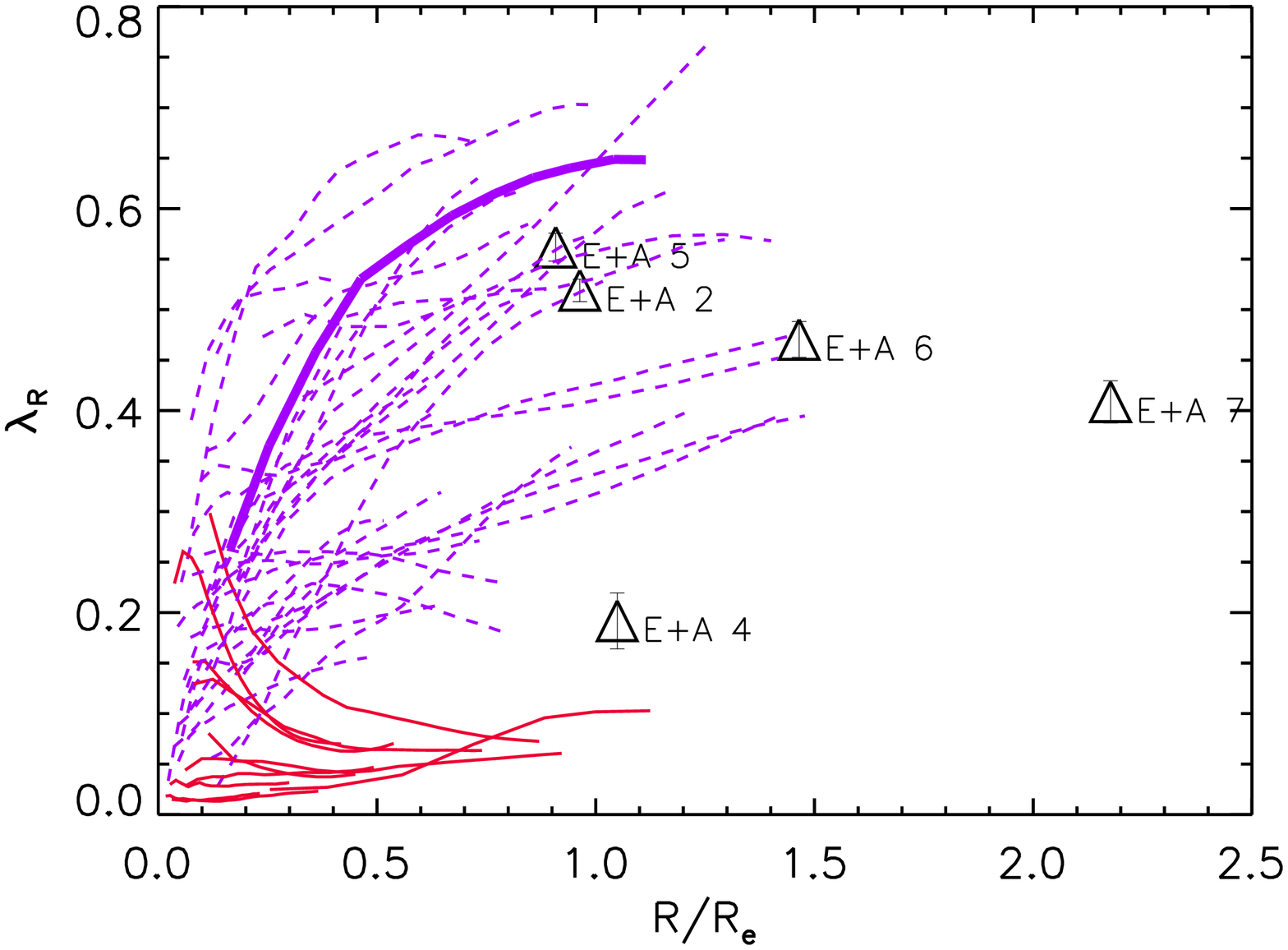}
     \includegraphics[width=5.8cm, angle=90, trim=0 0 0 0]{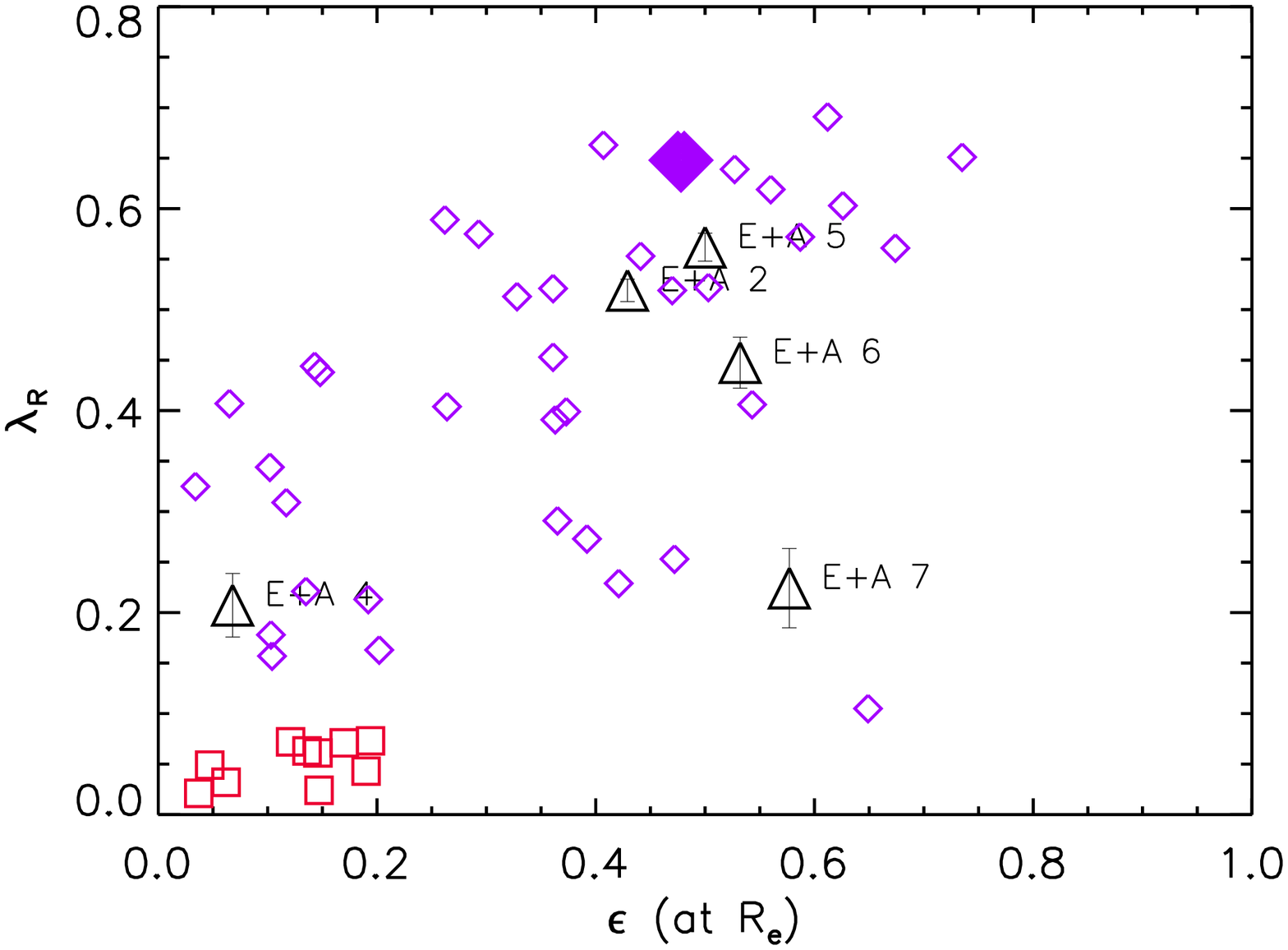}
\caption{\label{fig:lambdar} Top panel: the $\lambda_{R}$ parameter versus the radius over which it was measured for 
the early type galaxies in our sample. We measured $\lambda_{R}$ over a circular aperture which is 85\% covered by the IFU data and
correct the radius to an effective circular radius which has the same areal coverage. 
The tracks for $\lambda_{R}$ from the SAURON sample \citep{emsellem07} are overlaid as {\it blue dashed lines} for the fast rotators
and as {\it red solid lines} for the slow rotators. Bottom panel: the $\lambda_{R}$ parameter versus ellipticity measured at one effective radius. Again the SAURON data are overlaid
for fast ({\it blue diamonds}) and slow ({\it red squares}) rotators. The SAURON data for NGC 3156 (E+A 2) are plotted as a {\it thick line} (top panel) 
and {\it filled large diamond} (bottom panel)}
\end{figure}

In Fig. \ref{fig:lambdar} we plot the values of $\lambda_{R}$ for the early
type galaxies in our sample (not including E+A 1 and E+A 3 which are irregular and late type disk systems) versus the radius over which it is measured (top panel)
and against ellipticity  at one effective radius (bottom panel). 
The streaming velocities and velocity dispersions used to calculate $\lambda_{R}$ were derived from the spectral fitting described in Section 3.4.
We overlay in Fig. \ref{fig:lambdar} the data from the SAURON sample for 
fast rotators ({\it blue lines and symbols}) and slow rotators ({\it red lines and symbols}). In all five cases our data overlap the fast rotator 
regions of these plots consistent with previous findings \citep{pracy09}. The $\lambda_{R}$ values for NGC 3156 (E+A 2) from the SAURON sample are 
shown as a {\it thick line} and {\it large filled diamond} in the top and bottom panels respectively. These can be directly compared with the
value for E+A 2 from our measurements; the derived values of $\lambda_{R}$ are similar with the SAURON study value $\sim$0.1 larger. 
 
There is a minimum accurately measurable value for $\lambda_{R}$ \citep{emsellem07}
which depends on the quality of the data. Since $\lambda_{R}$ depends on the square of the velocity of each spaxel, even for data with zero true rotation, 
random noise will induce a non-zero positive value of $\lambda_{R}$. To determine if our derived values of $\lambda_{R}$ are reliable we simulate this effect
by setting the velocity values in each spaxel to zero and adding a random velocity drawn from a normal distribution with a width determined by the velocity
error for that spaxel. We use 10000 such simulations for each galaxy to estimate the minimum value of $\lambda_{R}$ that can be reliably measured.  
For E+A 2 and E+A 5 this value is $\lesssim$15\,per cent of the measured value and these galaxies can be confidently classified as fast rotators. 
For E+A 7 this value is $\sim$40\, per cent and in the cases of E+A 4 and E+A 6 the minimum measurable values are of order 60\,per cent of the measured values.
In a similar way we calculated the errors on the $\lambda_{R}$ values by randomly varying the streaming velocity and velocity dispersion of each spaxel and 
using the scatter in the resulting values of $\lambda_{R}$ as an estimate of the error. 

\section{Discussion}
By selecting a sample of E+A galaxies that are considerably closer, but less massive, than previously studied systems, we have
been able to obtain spatially resolved spectroscopy of the core (1\,kpc) region of E+A galaxies for the first time. 
Models that form E+A galaxies through tidal interactions or mergers suggest that starbursts occur in the core of 
the galaxies due to tidal torques funneling gas into the core \citep{barnes96,mihos94,hopkins09,snyder11}.
This region should be compact, on scales of $\sim$1~kpc,  \citep{barnes96,bekki05,hopkins09} 
and contain strong stellar population gradients observable as Balmer line gradients \citep{pracy05,bekki05,snyder11}.
The primary motivation of our observations was to test these predictions by acquiring spatially resolved spectroscopy that resolved this region. 

Previous studies, that used colours from imaging  \citep{yang08} or spectroscopy that does not resolve the core \citep{norton01,chilingarian09}, have suggested
that the post-starburst region is centrally concentrated but not necessarily confined to the galaxy core. Other groups have been unable
to detect any evidence for a central concentration in their observations  \citep{yagi06,pracy09}.
The most distant object in our sample has a  redshift of $z\sim$0.009  corresponding to a physical to angular scale of $\sim$0.18~kpc per arcsecond
which allowed us (for this worst case object) to easily spatially resolve the central 1\,kpc region even in moderate seeing conditions.
The resulting Balmer line equivalent width maps confirm that the post starburst region is centrally concentrated and contained within $\lesssim$1\,kpc
(see third column of Fig. \ref{fig:images}) consistent with model expectations. The radial profiles of the Balmer line absorption maps which
reveal a steep negative Balmer line gradient are consistent with the model predictions for E+A galaxies produced from mergers or tidal interactions \citep{pracy05}. The exception
is E+A 1 which has uniformly strong Balmer line absorption and no radial gradient. E+A 1 differs from the rest of the sample in other respects: it is the only
galaxy with an irregular morphology, it is the bluest galaxy in the sample by a considerable margin and has the strongest Balmer line equivalent width values
implying a younger post--starburst age. After emission line filling is taken into account it has stronger residual H$\alpha$ emission than the rest of the sample
indicating more ongoing star-formation which is distributed over the entire galaxy. E+A 1 may represent an earlier evolutionary stage 
than the other galaxies in the sample or a different class of object altogether. 

It has been pointed out by \citet{snyder11} that the presence of strong radial Balmer gradients in E+A galaxies
means that the galaxies selected by a given equivalent width criterion depend on the galaxy distance
since a fixed diameter fibre will subtend a greater physical distance at higher redshift. 
For centrally--concentrated Balmer gradients this implies that for a given selection criterion a more extreme population will be selected
at higher redshift. This is due to the aperture at higher redshifts including more light from larger galacto-centric radii where the Balmer line
absorption is weaker. At the median redshift of our sample, the 3\,arcsecond fibre of the SDSS subtends just $\sim$300\,pc whereas
at the median redshift of the SDSS ($z\sim$0.1) a 3\,arcsecond fibre subtends $\sim$5\,kpc. 
For our galaxies the strong radial gradients and strong `E+A levels' of Balmer line absorption are contained within the central few hundred parsecs.
Similar galaxies at higher redshift would not necessarily be classified as E+A galaxies. Selecting galaxies based on measurements from a central
fibre also biases us against selecting galaxies which have positive H$\delta$ gradients (central deficits) which are expected from the sudden truncation of 
star formation in a disk \citep{pracy05}. Our sample contains on average less luminous galaxies and also objects with less extreme overall 
absorption line strengths than many of the previous E+A samples in the literature. 

The morphologies of the galaxies in our sample, like E+A samples in general, are mostly early types with the exceptions being one late--type 
irregular galaxy and one late--type disk galaxy. Most of the early type galaxies are isolated systems with no interacting neighbor present
-- see Table \ref{tab:targets} for a list of morphologies and environments for the sample. This isolation rules out ongoing interactions as the cause
of the post--starburst signature and argues in favor of these galaxies being merger remnants where the progenitors have already coalesced and relaxed.
The kinematics of the early types in our sample are all consistent with being fast rotators as defined by \citet{emsellem07}. 
This suggests that the mergers producing the E+A galaxy are often unequal mass or minor mergers since mergers with mass
ratios around unity will more commonly result in a non-rotating remnant \citep{bournaud08}, although, this does not strictly rule out major mergers \citep{bois11}.
This same conclusion was reached by \citet{pracy09} for E+A galaxies selected from the 2dFGRS
and is in agreement with the recent findings of \citet{crockett11}. The two late type galaxies in our sample (E+A 1 and E+A 3) both have near neigbours which
are very close in the case of E+A 1 ($\sim 5$\,kpc) or much more massive in the case of E+A 3 (see Table \ref{tab:targets} for details) 
and for these two galaxies the E+A signature could be explained by tidal interactions rather than merging.

There are detectable emission lines in two of the early type galaxies as well as in the late type disk galaxy. The
emission is centrally concentrated and absorption--corrected line ratios imply that in the early type galaxies
this emission is, at least in part, due to AGN activity. The line ratios in the disk galaxy are more consistent
with pure star formation. There are other known examples of low power AGN in E+A galaxies \citep{liu07,chilingarian09} and
it has been proposed that nuclear activity, in addition to supernova feedback, contributes to the truncation of star-formation
in E+As \citep{kaviraj07}. However, simulations have shown that the effect on the post--starburst phase from AGN feedback
is weak \citep{wild09,snyder11}, with the dominant effect being an increase in the observed post--starburst signature as
a result of the expulsion of obscuring dust.

\section{Conclusions}
We have selected a new local sample of seven E+A galaxies from the SDSS with a redshift limit of $z=0.01$ and obtained
integral field spectroscopy with the WiFeS instrument on the ANU 2.3-m telescope. The low redshift of our sample allowed
for much better physical scale resolution in the core region of the E+As than previous studies. This allowed us to unequivocally
verify that the post starburst regions are centrally concentrated, consistent with expectations. 
In summary our main conclusions
are:

\begin{list}{$\bullet$}{\itemsep=0.1cm}

\item Six of the seven E+A galaxies and all galaxies with regular morphologies have a centrally--concentrated young 
stellar population and negative Balmer line gradients on scales of $\lesssim$1\,kpc. The exception is the only irregular
galaxy in the sample which shows uniform Balmer absorption strength and stronger overall absorption compared to the other galaxies. 

\item The sample is dominated by isolated early type galaxies. All the early type galaxies show some level of rotation and
are consistent with the fast rotator population.

\item In the early type galaxies the combination of a centralized young population and isolation argues in favor of a merger origin where the progenitors have
already coalesced.

\item  The two late type galaxies in the sample have nearby companions and their spectral signature could be explained by tidal interaction.

\end{list}

\section{Acknowledgments}
We acknowledge the financial support of the Australian Research Council throughout the course of this work. 
This research has made use of the NASA/IPAC Extragalactic Database (NED) which is operated by the Jet Propulsion 
Laboratory, California Institute of Technology, under contract with the National Aeronautics and Space Administration.
M.B.P would like to thank the School of Physics at the University of New South Wales for their hospitality. We thank
the referee for helpful comments that improved this paper.


\bsp

\bibliographystyle{mn2e}
\bibliography{wifes2}

\label{lastpage}

\end{document}